\documentclass{emulateapj}
\usepackage{lscape}
\usepackage{graphicx}
\usepackage{natbib}
\usepackage{amssymb}
\usepackage{float}
\usepackage{makeidx}
\usepackage{amsmath}
\usepackage{rotating}
\usepackage{longtable}
\usepackage{subfigure}
\usepackage{subfloat}
\usepackage{captcont}

\nocite{*}

\newcommand{\etal}{et~al.\/}

\newcommand{\Hline}[1]{\mbox{H{\footnotesize {#1}}}}
\newcommand{\Halpha}{\Hline{\mbox{$\alpha$}}}

\newcommand{\Msun}{\mbox{${\cal M}_\odot$}}

\newcommand{\cm}{cm$^{-2}$}

\newcommand{\oii}{\hbox{[O\,{\sc ii}}]}

\shortauthors{Werk \etal}
\shorttitle{Rogue Metals}

\begin{document} 
\slugcomment{accepted to ApJ; April 18 2011}

\title{Metal Transport to the Gaseous Outskirts of Galaxies}

\author{J. K.\ Werk\altaffilmark{1, 2, 3},
M. E.\ Putman\altaffilmark{3},
 G. R.\ Meurer\altaffilmark{4},
 \& N.\ Santiago-Figueroa\altaffilmark{3,5}
}
\altaffiltext{1}{present address: Department of Astronomy and Astrophysics \& UCO/Lick
Observatory, University of California, 1156 High Street, Santa
Cruz, CA 95064; $jwerk@ucolick.org$}
\altaffiltext{2}{Department of Astronomy, University of Michigan, 500 Church St., 
		Ann Arbor, MI 48109}
\altaffiltext{3}{Department of Astronomy, Columbia University, 550 West 120th Street, New York, NY 10027, USA}
\altaffiltext{4}{ICRAR/The University of Western Australia, 35 Stirling Highway, Crawley, WA 6009, Australia}
\altaffiltext{5}{present address: Department of Physics, Fisk University, Nashville, TN 37208, USA}
\begin{abstract}
We present a search for outlying HII regions in the extended gaseous outskirts of nearby (D $<$ 40 Mpc) galaxies, and subsequent multi-slit spectroscopy used to obtain the HII region nebular oxygen abundances. The galaxies in our sample have extended HI disks and/or interaction-related HI features that extend well beyond their primary stellar components. We report oxygen abundance gradients out to 2.5 times the optical radius for these galaxies which span a range of morphologies and masses. We analyze the underlying stellar and neutral HI gas distributions in the vicinity of the HII regions to understand the physical processes that give rise to the observed metal distributions in galaxies. These measurements, for the first time, convincingly show flat abundance distributions out to large radii in a wide variety of systems, and have broad implications for galaxy chemodynamical evolution. 

\end{abstract}


\section{Introduction}
\label{sec:rogues:intro}

Through nuclear fusion and well-studied stellar evolution processes, stars with M$_{*}$ $>$ 8 \Msun~generate metals in their cores, and return them to the local interstellar medium (ISM) near the end of their lifetimes. Since galaxies are neither static nor isolated, these metals do not simply stay put. Instead, they disperse throughout the galaxy, and sometimes flow into and out of the galaxy via galactic winds and galaxy interactions \citep{oey03b, tremonti07, opp09}.

 The transport mechanisms that govern how the metals are distributed throughout and/or ejected from galaxies are widely debated, and remain poorly understood.  At the heart of the matter is the myriad of gas-phase metal distributions determined from observational data. Metals are evenly distributed in some spiral galaxies, while others exhibit steeply declining radial abundance gradients \citep{vilacostas92, oey93}. By comparison, dwarf galaxies tend to show shallow, if not flat,  abundance gradients \citep{croxall09}. Very recent work by \cite{kewley10} indicates that close galaxy pairs also tend to have flatter metallicity gradients than those observed in typical isolated spiral galaxies. While there is a steep inner-galaxy abundance gradient in M83, there is an observed flattening of the radial abundance gradient at large radii (r $>$ R$_{25}$; Bresolin et al. 2009\nocite{bresolin09a}). It is most likely a combination of astrophysical processes that leads to the observed metal distributions in galaxies of various properties. For example, disk ``viscosity"  and metal-mixing \citep{zaritsky92, ferguson01}, a radial dependence of star-forming activity \citep{phillips91}, large-scale metal blowout \citep{tremonti04}, and merger-induced gas flows \citep{rupke10} have all been put forth as potential causes for the observed metal distributions in star-forming, dwarf,  and/or interacting galaxies. 

Characterizing the metal distribution in galaxies of all morphological types and masses over a wide range of galactocentric distances is essential for determining the relative importance of various metal-mixing processes. These processes, in turn, have significant bearing on our understanding of galaxy evolution, galaxy feedback, and star formation. With regard to observational studies of galaxy-wide metallicity gradients, there are two key environments that have yet to be fully investigated: interacting or disturbed galaxies, and the far outer reaches of galaxies. It is in their gaseous outskirts that galaxies may reveal how they build themselves up over cosmic time, through interactions and gas flows.  In this paper, we probe the chemical abundances of the far outer regions of disturbed and/or extended-gas, gas-rich galaxies, and therefore extend the study of metallicity distributions in galaxies to a new and important locale. To do this study, we use HII regions that span a wide range of galactocentric distances: within the optical extent (r $<$ R$_{25}$)  and beyond (r $>$ R$_{25}$; referred to as ``outlying HII regions"; Werk et al. 2010 \nocite{werk10}). We search for these HII regions in narrow-band images of galaxies in the HI Rogues Catalog \citep{rogues}. 

The HI Rogues catalog consists of 189 HI synthesis maps of nearby  galaxies with extended, unusual, or disturbed HI morphologies, including:  galaxies with extended (many times the optical extent) or warped HI distributions, interacting groups, minor and major mergers, intergalactic debris with no optical counterparts, and early-type galaxies with HI inside and/or outside their optical bodies. As the host galaxies of outlying HII regions are strongly correlated with disturbances, companions, extended gas, and interactions, this collection of ``rogue" galaxies provides a potentially rich hunting ground for outlying, massive star formation \citep{werk10}. Another benefit of the HI rogues sample is that all galaxies have readily available HI synthesis maps, providing crucial information about the gaseous environments of the outlying stars.  In these HI Rogue systems, the oxygen abundances of the extended and/or stripped outer gaseous material can place much-needed constraints on metal transport in gas-rich galaxies. 

Extragalactic HII regions have long been used as reliable indicators of the current metal abundance of the gas in their vicinity. At the low electron densities typical of HII regions (n$_e$ $<$ $10^{4}$ cm$^{-3}$), forbidden emission lines, such as [OIII] $\lambda\lambda$4959, 5007 and [OII] $\lambda\lambda$3727, are characteristic features of HII region emission-line spectra, along with the radiative recombination lines of the hydrogen Balmer and Paschen series (e.g. H$\alpha$). Because the strength of every emission line from every element present in the HII region depends on its  chemical abundance in the nascent ISM gas, HII regions are valuable metallicity indicators \citep{osterbrock89}. As such, they trace the star formation history of the gas and act as chemical tags that can aid in determining the origin of the gas. 

 The practice of determining chemical abundance gradients using HII region emission-line spectra was initiated by \cite{searle71}, who inferred a declining Galactic O/H abundance gradient from the observed increase in HII region excitation and temperature with radius (with [OIII]$\lambda$4959,$\lambda$5007 /[OII] $\lambda$3727 and [OIII]/H$\beta$ ratios). \cite{shields76} confirmed the anti-correlation between T$_{eff}$ and metallicity, and attributed the radial gradients to a metallicity dependent upper mass limit for star formation such that M$_{upper}$ $\propto$ Z$^{-1/2}$, a scaling theoretically determined by \cite{kahn74} who modeled massive star formation with a radiation ``cocoon."  Since then, numerous detailed studies have confirmed the overall trend for the decrease in oxygen abundance with galactocentric radii (e.g. Oey \& Kennicutt 1993; Zaritsky 1994; Bresolin et al. 1999; Kennicutt et al. 2003\nocite{oey93,zaritsky94,bresolin99, kennicutt03a}) in a large number of spiral (barred + unbarred) and irregular star-forming galaxies, although the underlying physical cause is still a subject of debate. The work of \cite{zaritsky94} finds the radial oxygen abundance gradients of nearby spiral galaxies to range from $-$0.03 dex/kpc$^{-1}$ to $-$0.15 dex/kpc$^{-1}$. 


Thus far, only a handful of measurements of gas-phase abundances at large optical radii (r $>$ R$_{25}$) have emerged \citep{ferguson98a, bresolin09a}, and many questions remain about the metal content in the outermost regions of galaxies. Our study aims to address those questions by determining the oxygen abundances of numerous inner and outlying HII regions in 13 HI Rogue galaxies. The morphologically diverse nature of our sample of galaxies allows us to consider the role played by galaxy interactions in the observed metal distributions. 

This paper proceeds as follows: Section \ref{sec:rogues:obs} describes imaging and spectroscopic data obtained at various telescopes over the course of several years; Section \ref{sec:rogues:anal} reviews our methods for obtaining strong-line oxygen abundances from HII region emission-line spectra; Section \ref{sec:rogues:results} presents radial oxygen abundance gradients reaching beyond R$_{25}$ for 13 HI rogues with outlying HII regions; Section \ref{sec:rogues:disc} discusses these results and comments on the physical mechanisms that could be responsible for the metal distributions; and Section \ref{sec:rogues:sum} summarizes and concludes. Our Gemini multi-slit data allow the largest to-date observational study of the metal content of outlying gas (beyond R$_{25}$), and how it compares to the metal content of gas within the galaxy.

\section{Optical Imaging and HI data}
\label{sec:rogues:obs}
\subsection{H$\alpha$ Imaging}
\label{sec:rogues:ha}

In April 2005, over the course of four clear nights, we imaged 27 galaxies from the HI Rogues Catalog \citep{rogues} in H$\alpha$ and R$-$band continuum with the MDM 2.4-m Hiltner telescope. The MDM 2.4-m telescope direct images, with the Echelle CCD, have a field of view of  9.5\arcmin~ $\times$ 9.5\arcmin~ and a plate scale of 0.275\arcsec~ per pixel.  The average seeing over the course of these four nights was 1.2\arcsec. These observations were made in a similar fashion to the H$\alpha$ imaging of the SINGG survey \citep{SINGG}. We used four of the SINGG narrow$-$band filters with bandpasses that corresponded to the H$\alpha$ emission line at the target galaxy's velocity. The central wavelengths, in angstroms, of the filters used are 6568, 6596, 6605, and 6628, each with a FWHM of $\sim30$ \AA~(Meurer et al. 2006 contains a full description of the SINGG filters).  Exposure times were 3$\times$ 600 s in H$\alpha$ and 3$\times$ 120 s in R. Targets were selected from the full HI Rogues catalog on the basis of their visibility from MDM observatory and their recessional velocities, such that the redshifted H$\alpha$ emission line is encompassed by the SINGG H$\alpha$ filters' narrow bandpasses. Galaxies were also selected to have angular optical and HI extents such that they were contained in the 9.5\arcmin~ field of view. Observations of several different spectrophotometric standards were used to flux calibrate the data. 

We performed a basic reduction of the data using IRAF CCD processing routines for bias subtraction and  flat field division. Images were then combined and astrometry was performed using the SINGG IDL pipeline routines (see Meurer et al. 2006) modified specifically for the MDM 2.4-m telescope. Photometry was performed on the images based on observations of spectrophotometric standards calibrated to the AB magnitude system. We followed the basic SINGG procedure detailed in Appendix A of Meurer et al. 2006.  The average 5$\sigma$ limiting H$\alpha$ flux for a point source is $\sim$1 $\times$ 10$^{-16}$ ergs cm$^{-2}$ s$^{-1}$ and the average 5$\sigma$ limiting R$-$band magnitude is $\sim$24.

\subsection{Searching Images for Outlying HII Regions }
\label{sec:rogues:eldot}

We searched the reduced, calibrated MDM H$\alpha$ and R$-$band images for ELdots (emission-line dots; see Werk et al. 2010\nocite{werk10}) using the basic selection criteria outlined in \cite{werk10}, with the exception of the requirement that the ELdot lie beyond 2 $\times$ R$_{25}$. For reference, ELdots are defined to be strong emission-line point sources in narrow-band images that  lie well outside the broadband optical emission of nearby galaxies \citep{werk10}. Generally, ELdots can be outlying H$\alpha$ emitting HII regions at a similar velocity to the target galaxy or they can be background galaxies emitting a different line ([OIII] $\lambda$5007, [OII] $\lambda$$\lambda$3727, or H$\beta$) that is redshifted into the narrow filter passband used for the observations. For this study, we selected by-eye sources that appeared to lie beyond the main R$-$band optical extents of the potential host galaxies. The reason for relaxing our projected distance requirement is that R$_{25}$ is not well-defined for merging, interacting, and/or irregular, asymmetric galaxies. For comparison to the sample in \cite{werk10}, and an estimate of the ELdots' projected distances from the main optical extents of their host galaxies, we did calculate an approximate R$_{25}$ at a later time. In order to obtain  the approximate location of R$_{25}$  in these messy systems, we used the IRAF task ELLIPSE found in the STSDAS package, fitting elliptical isophotes to the target rogue galaxies in the 2.4-m MDM R$-$band images.  In one case (NGC 3227), two merging galaxies had to be fit with a single ellipse. In other cases, incidences of tails, plumes, and warps caused the fixed-center elliptical isophotes to be significantly larger than they otherwise would have been. The laxity of the R$_{25}$ fitting should be noted, and the quantity itself should be treated as an optical radial reference point when discussing oxygen abundance gradients. All of the rogue ELdots, originally identified by-eye, lie beyond this optical radius. 

13 of the 27 HI rogues imaged with the MDM 2.4-m were found to have over 40 ELdots in total. We began the process of spectroscopically confirming their association with the target galaxy, with longslit spectroscopy at the MDM 2.4-m telescope (Section \ref{sec:rogues:long}). Various difficulties with blind offsets and large observational overhead  forced us to discontinue the follow-up longslit spectroscopy before it was complete. Yet, of the dozen ELdot spectra that were obtained, all were confirmed as outlying HII regions (the H$\alpha$ emission line, instead of any of the redshifted strong blue emission lines, falls in the SINGG narrow bandpass). This high confirmation rate led us to pursue deeper multi-slit spectroscopy with the Gemini 8-m telescope, as described in Section \ref{sec:rogues:spec}. 

\subsection{Properties of the Sample of 13 HI Rogues}

Figure \ref{fig:slitsandgas} shows the  H$\alpha$ continuum-subtracted images for the 13 HI rogue fields in which we found ELdots. For nine of these galaxies, we also show HI column density contours generated from reduced HI data cubes kindly provided by several project PIs (see Table 1).  These images highlight the inner and outer-galaxy star formation and often do not show the full MDM field of view or the full extent of the HI distribution. We show R$_{25}$ on the images, in addition to all of the HII regions (outer + inner) for which we were able to obtain GMOS multi-slit spectra (see next section).  Regarding the few galaxies for which HI data cubes were not available to us, we refer the reader to the online HI rogues catalog and references therein \citep{rogues}.  Table \ref{tab:c5t1}  lists the optical, HI, and derived properties of this sample.  Column 7 provides an estimate of the total oxygen abundance for each galaxy from the literature. In most cases, the oxygen abundance measurement was made with various strong emission-line ratios for HII regions in the central parts  (within R$_{25}$ in all cases) of each galaxy.  Column 8 lists the HI morphology code from the HI rogues catalog for each target. The codes are defined in the caption of Table \ref{tab:c5t1}, and the HI morphology of each target is described in greater detail in Section \ref{sec:rogues:results}. 

	For the calculations we perform in Section \ref{sec:rogues:disc}, we derive the total baryonic (stellar + gas) and global star-formation rate (SFR) for each galaxy in our sample. To obtain a value of the stellar mass, we use the published total integrated broad-band magnitudes in B and V$-$bands available on NED in conjunction with the stellar mass-luminosity/color relation from \cite{bell03}. Most of the total B$-$band and V$-$band apparent magnitude estimates come from the 1991 De Vaucoleurs Catalog, using a combination of new and previously published data, and are corrected for internal and Milky Way extinction. UGC 5288 is an exception in this regard (not included in the De Vaucoleurs Catalog), and therefore we used the broadband optical B and R$-$band photometry from \cite{vanzee00}. To obtain the total stellar mass from the B$-$band luminosity and the B$-$V color (sometimes B$-$R), we use Table 7 of \cite{bell03} which gives coefficients for their derived stellar mass-luminosity/color relation for SDSS galaxies. These stellar masses are accurate to within 50\%, and are given in column 11 of Table \ref{tab:c5t1}. In some cases, we are able to compare our B$-$V results with other calibrations of the \cite{bell03} mass-luminosity/color relations using different broadband filters, which give values that are consistent within the errors. The HI masses in column 12 of Table \ref{tab:c5t1} are obtained from the literature, and references are given in column 10. 
	
	We calculate SFRs using the relation SFR (\Msun~yr$^{-1}$) = 5.3 $\times$ 10$^{-42}$ L(H$\alpha$) (ergs s$^{-1}$) \citep{calzetti08} and obtain total H$\alpha$ luminosities from the MDM 2.4-m images within an elliptical aperture at 2 $\times$ R$_{25}$. These total H$\alpha$ luminosities are corrected for Galactic extinction \citep{schlegel98} and for the contribution of [NII]$\lambda$6583 using the [NII]/H$\alpha$ emission-line ratios from the literature (listed in column 9 of Table \ref{tab:c5t1}) and convolving this contamination with the SINGG narrow-band filter transmission curve (see Appendix A of Meurer et al. 2006 for details). This [NII]/H$\alpha$ emission-line ratio for each galaxy represents a global average (from integrated spectra or averages of individual HII regions in the galaxy) over its optical extent. The only exception is NGC 3227, in which only one measurement of an HII region at $\sim$ 0.8 $\times$ R$_{25}$ was available. We did not attempt to add in the potential effects from metallicity gradients across the galaxy since such effects would be negligible in light of our rather high photometric errors. Considering errors in the continuum subtraction,  occasional H$\alpha$ filter reflections/diffuse patterns, and uncertainties in the extinction correction, we estimate that these total H$\alpha$ luminosities, and thus the total SFRs,  are accurate to within $\sim$20\%. 
	
Table \ref{tab:c5none} lists the remaining 14 targets in which we did not find any ELdots, along with their HI morphologies, distance, and R$_{25}$. Although their properties are not fundamentally different from those of the HI rogues with ELdots, we do note that the only three optical early-type elliptical galaxies with HI that we imaged with the MDM 2.4-m appear in this sample. 

\subsection{Comparison with an Unbiased Sample of HI-Selected Galaxies}
\label{sec:rogues:sample}

The HI Rogues sample is neither complete nor representative. It was compiled as a sample largely dominated by interacting or otherwise ``weird-featured" gas-rich galaxies. Of the 13 galaxies (or galaxy pairs) studied here, 10 are mergers or remnants of mergers with extended gaseous features. Both NGC 3359 and UGC 5288 are galaxies with visible star formation contained in a much larger, more extended concentration of HI gas which may or may not be the result of some past interaction despite very regular kinematic structures. NGC 3718 exhibits a large warp in its HI distribution of unknown origin. As we discuss in Section 5, each individual system is at a different stage in the interaction process and the nature of the interaction ranges from minor to major. 

To compare our galaxies with a sample of ``HI-normal" galaxies, we use the HI-selected sample
of the Survey of Ionization in Neutral Gas Galaxies (SINGG; \nocite{SINGG} Meurer et al. 2006).  
The SINGG sample was selected from the HI Parkes All-Sky Survey (HIPASS; Barnes et al. 2001\nocite{barnes01})
purely on the basis of HI mass and recessional velocity. The primary goal of SINGG is to uniformly survey the star
formation properties of HI-selected galaxies across the entire HI mass function sampled by HIPASS (Log(M$_{HI}$/M$_{\odot}$) $\sim$ 8.0 - 10.6) in a way that is blind to
previously known optical properties of the sources.

\cite{hanish06} used the SINGG Release 1 data to
calculate the luminosity densities in \Halpha\ and the $R$ band of the
local universe for HI-selected galaxies in the SINGG sample.  They
then derive the fractional contribution to these luminosity densities as
a function of various addtional parameters (see their figure ~5 and
table~5).  Compared to our minimum ${\cal M}_{\rm HI}$, about 12\%\ and
22\%\ of the $R$ and \Halpha\ luminosity density, respectively, comes
from galaxies with lower ${\cal M}_{\rm HI}$ values, while about 6\%\ of
both luminosity densities comes from galaxies having ${\cal M}_{\rm HI}$
larger than our maximum value.  The median ${\cal M}_{\rm HI} =
5.3\times 10^9\,{\cal M}_\odot$ is slightly higher than the value ${\cal
M}_{\rm HI}\approx 3.6\times 10^9\,{\cal M}_\odot$ corresponding to the
midpoint contributing to both these luminosity densities. Hence, the
${\cal M}_{\rm HI}$ values for our sample corresponds well to that of
SINGG, bur are skewed perhaps to slightly larger values.  We can also
compare ${\cal M}_\star$ values, if we convert the SINGG R band
luminosities to stellar mass.  For this rough calculation, we assume ${\cal M}_\star/L_R
\approx 1$ and adopt $M_R = 4.61$ ABmag for the sun's absolute magnitude.
Then we find that at most 3\%\ of the local $R$ and \Halpha\ luminosity
densities have ${\cal M}_\star$ lower than our minimum value, while
30\%\ of the $R$ luminosity density and 13\%\ of the \Halpha\ luminosity
density comes from galaxies more massive than our largest ${\cal
M}_\star$.  The median ${\cal M}_\star = 1.31\times 10^{10}\, {\cal
M}_\odot$ for our sample, while the midpoint for the $R$ and \Halpha\
luminosity densities is ${\cal M}_\star = 3.3\times10^{10}\,{\cal
M}_\odot$ and $1.4 \times10^{10}\,{\cal M}_\odot$, respectively.  Hence
our sample covers the ${\cal M}_\star$ range of the SINGG sample fairly
well, but may be deficient in the highest ${\cal M}_\star$ galaxies.


\begin{figure*} [hp]
\centering 
\subfigure[NGC 2146]{ 
\label{fig:slitsandgas:a} 
\includegraphics[width=0.48\linewidth]{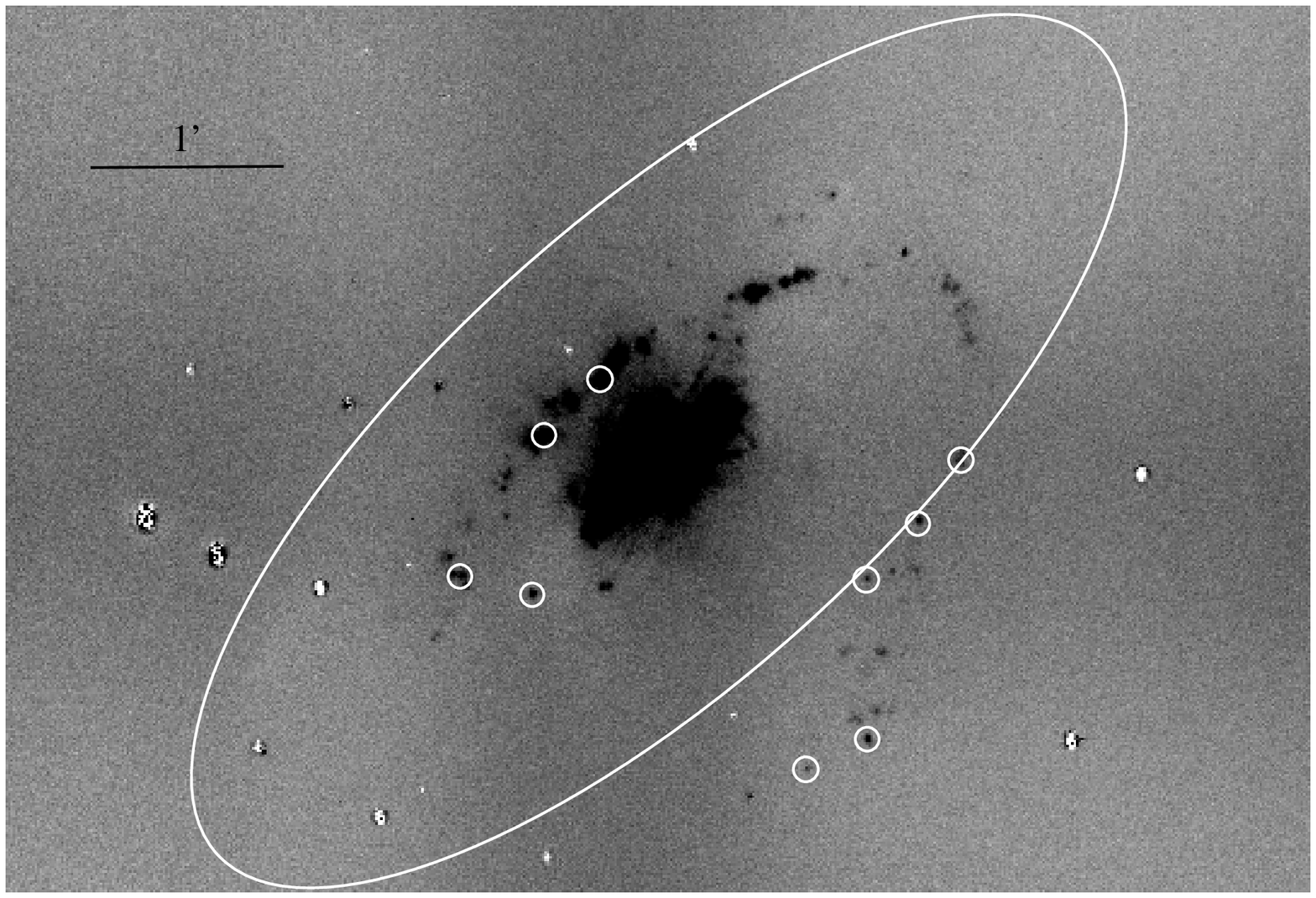}} 
\subfigure[NGC 2782; with HI contours at  2.5 to 10.5 M$_{\odot}$ pc$^{-2}$]{ 
\label{fig:slitsandgas:b} 
\includegraphics[width=0.48\linewidth]{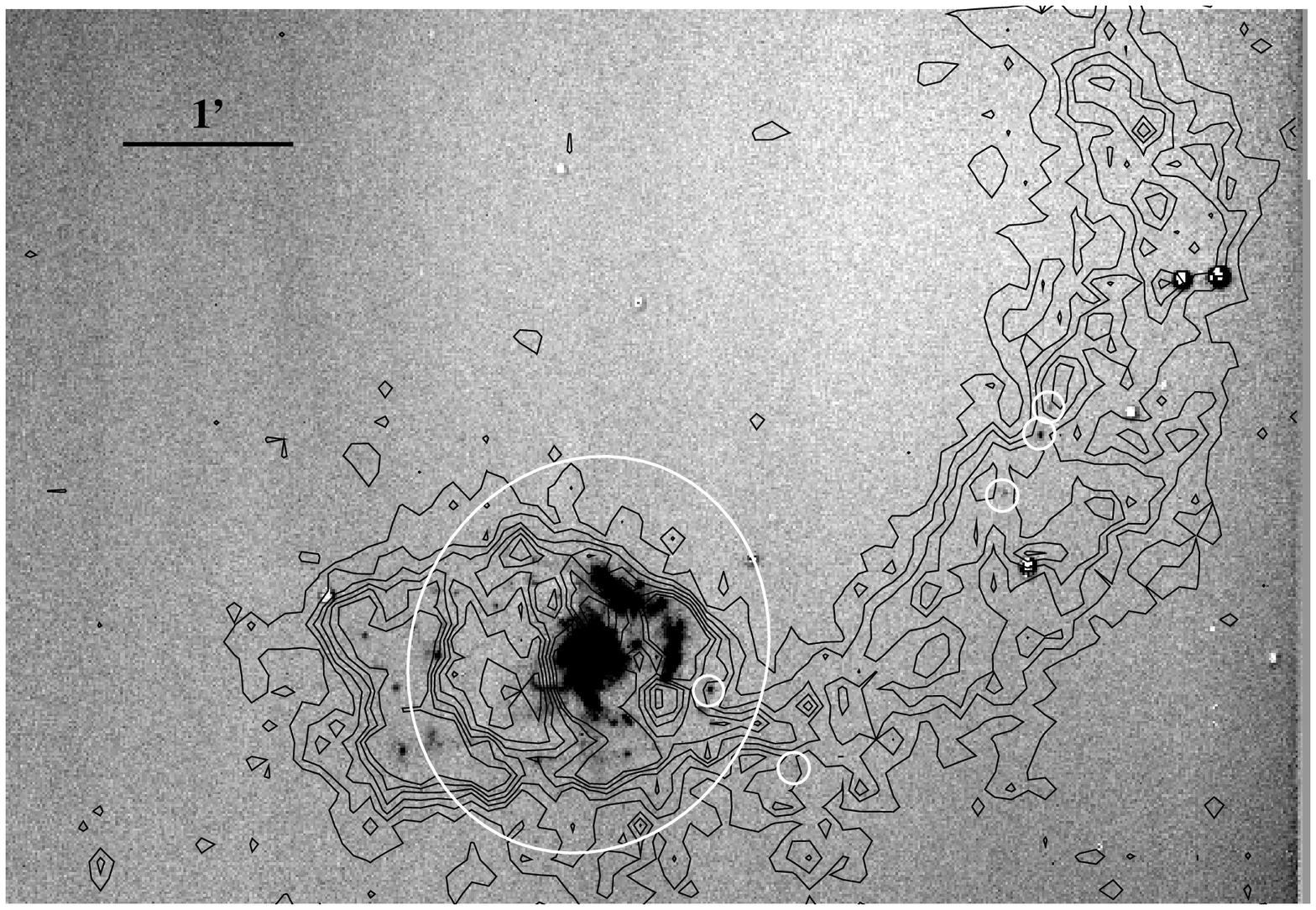}} 
\subfigure[NGC 3227; with HI contours at 1.5 to 8.5 M$_{\odot}$ pc$^{-2}$]{ 
\label{fig:slitsandgas:c} 
\includegraphics[width=0.48\linewidth]{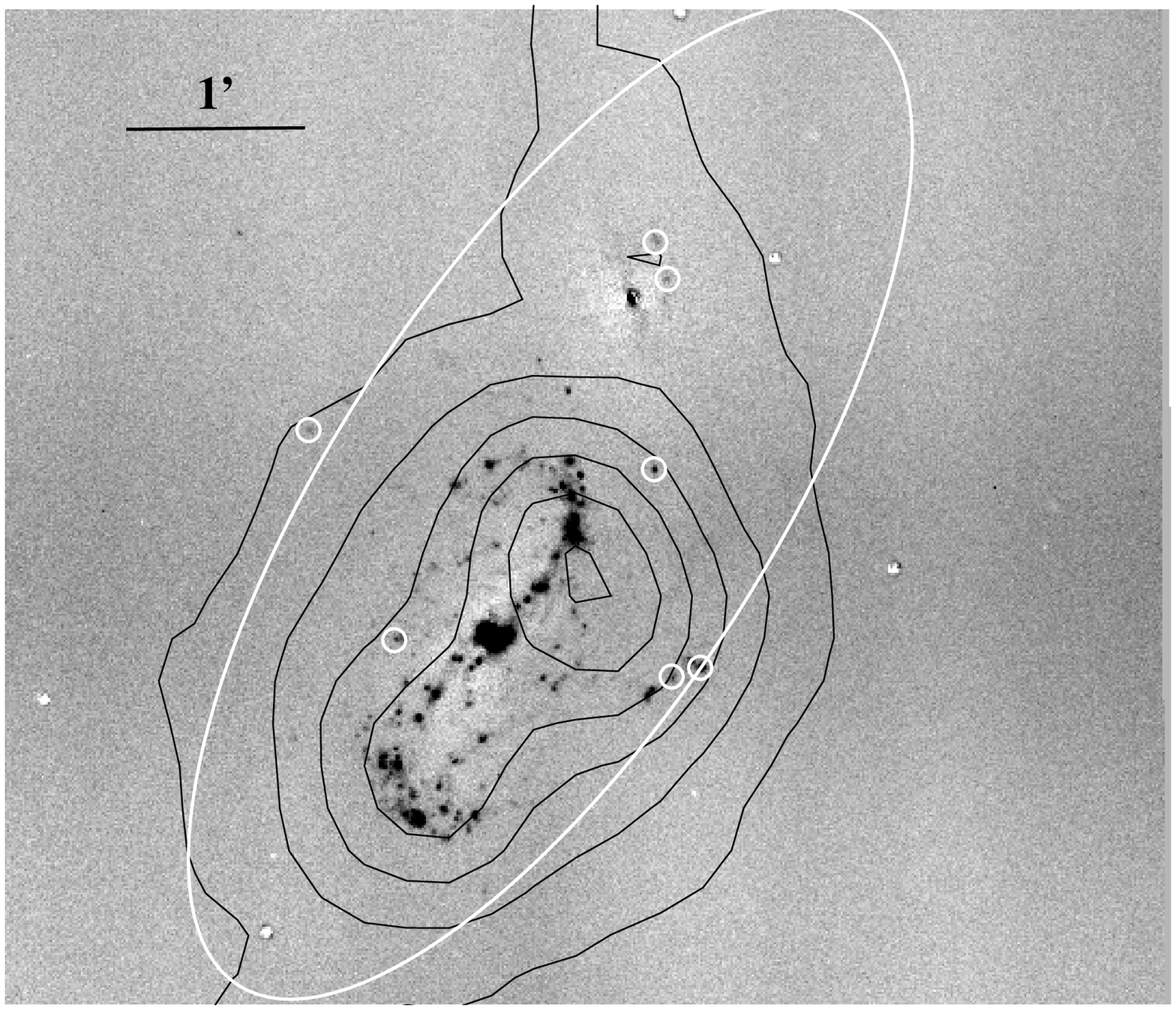}} 
\subfigure[NGC 3239; with HI contours at 0.5 to 8.5 M$_{\odot}$ pc$^{-2}$]{ 
\label{fig:slitsandgas:d} 
\includegraphics[width=0.48\linewidth]{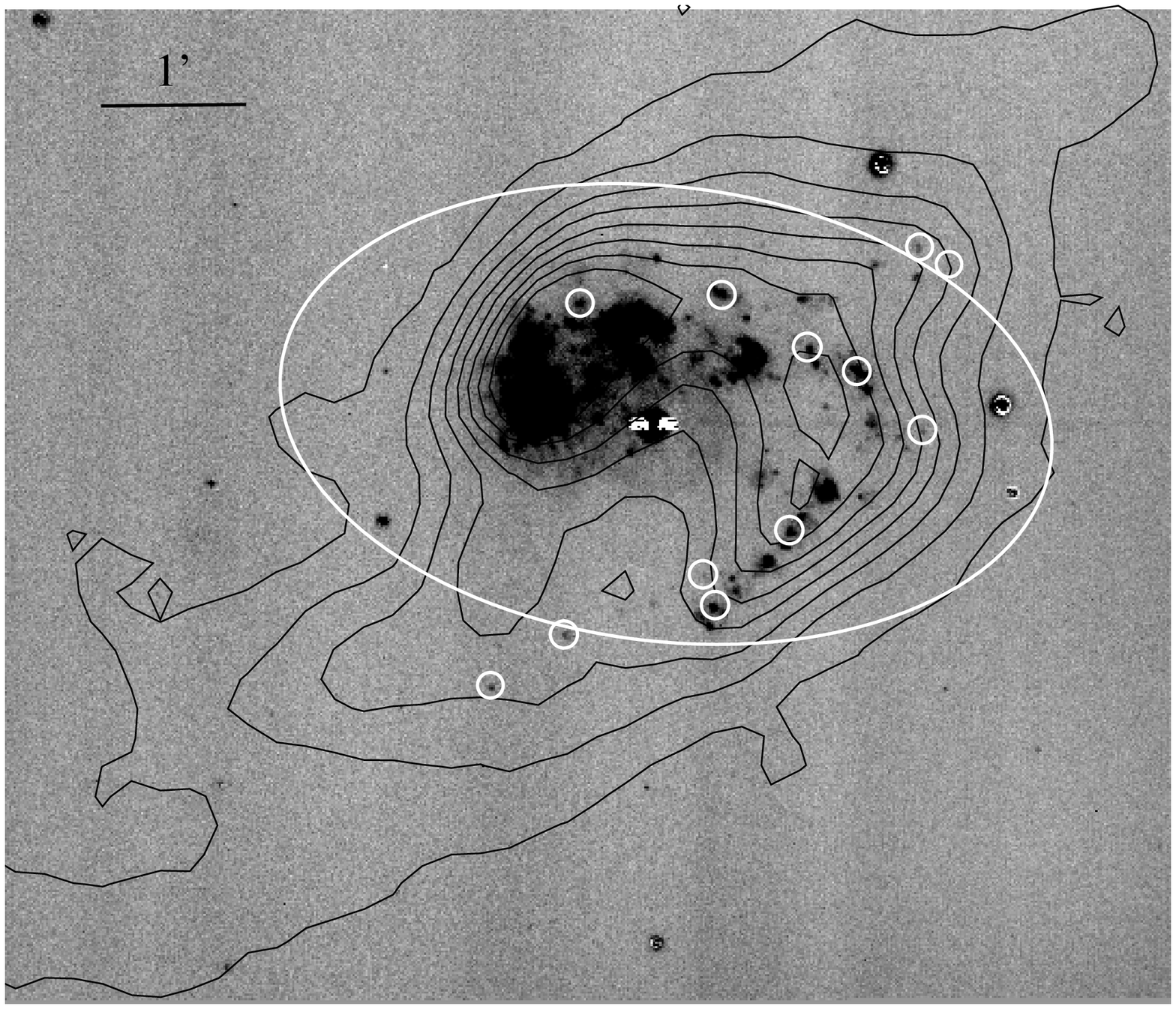}} 
\captcont{The continuum-subtracted H$\alpha$ emission images from the MDM 2.4-m telescope, as described in Section \ref{sec:rogues:obs}, overlaid with HI column density contours, where available. Original references and additional details on the HI data (beam size, etc.) can be found in the online HI Rogues catalog at http://www.nrao.edu/astrores/HIrogues/ and in Section \ref{sec:rogues:results}. The white ellipse shows the location of the elliptical R-band 25$^{th}$ magnitude isophote (R$_{25}$) for the galaxy, while the smaller white circles point to the locations of the HII regions for which we obtained GMOS-N multi-slit spectra. Images are aligned to a world-coordinate system such that north is upwards, and east is to the left. Continued.} 
\end{figure*} 
\begin{figure*}[hp!] 
\centering 
\subfigure[NGC 3310]{ 
\label{fig:slitsandgas:e} 
\includegraphics[width=0.48\linewidth]{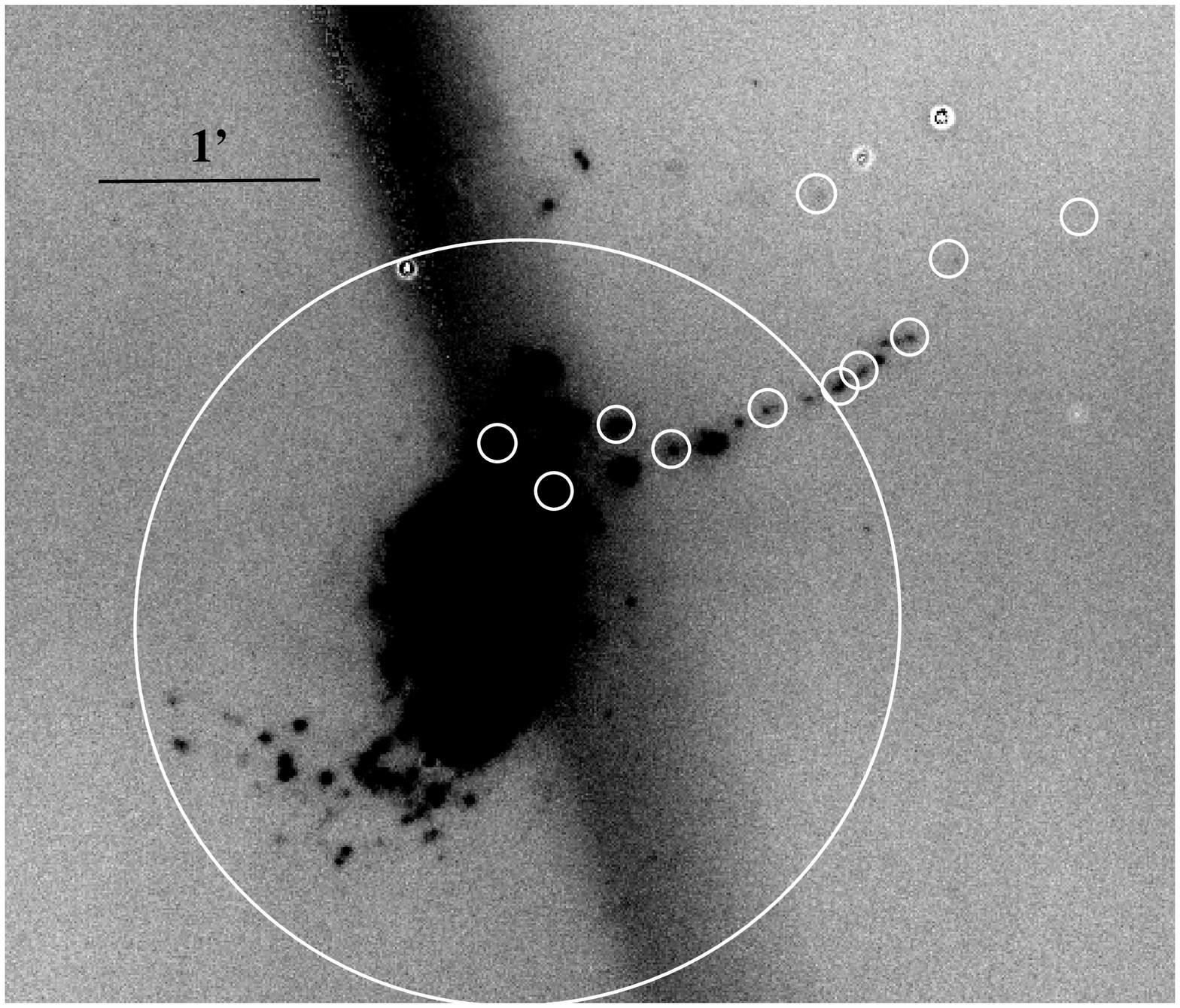}} 
\subfigure[NGC 3359]{ 
\label{fig:slitsandgas:f} 
\includegraphics[width=0.48\linewidth]{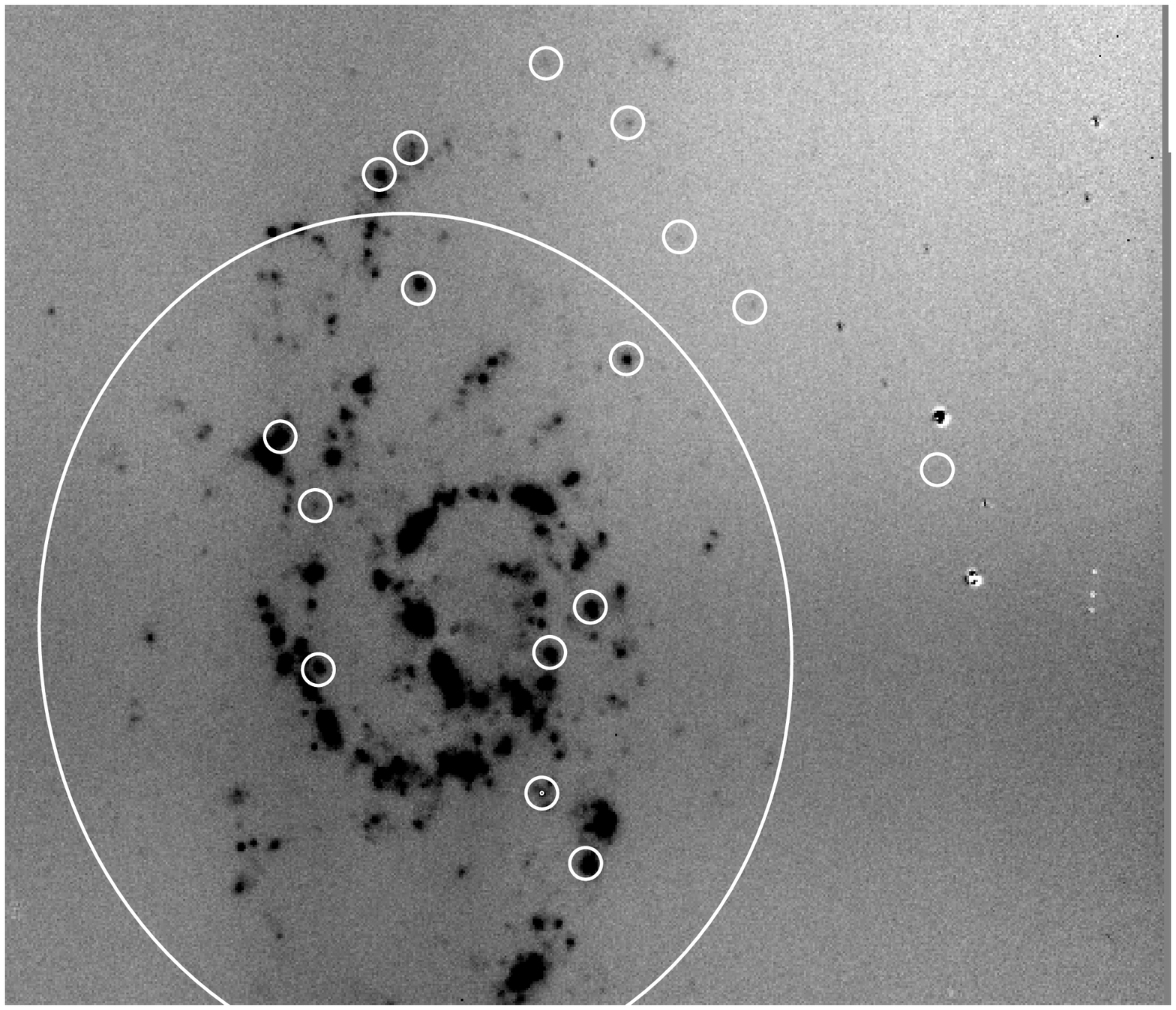}} 
\subfigure[NGC 3432; with HI contours at 6.5 to 60 M$_{\odot}$ pc$^{-2}$]{ 
\label{fig:slitsandgas:g} 
\includegraphics[width=0.48\linewidth]{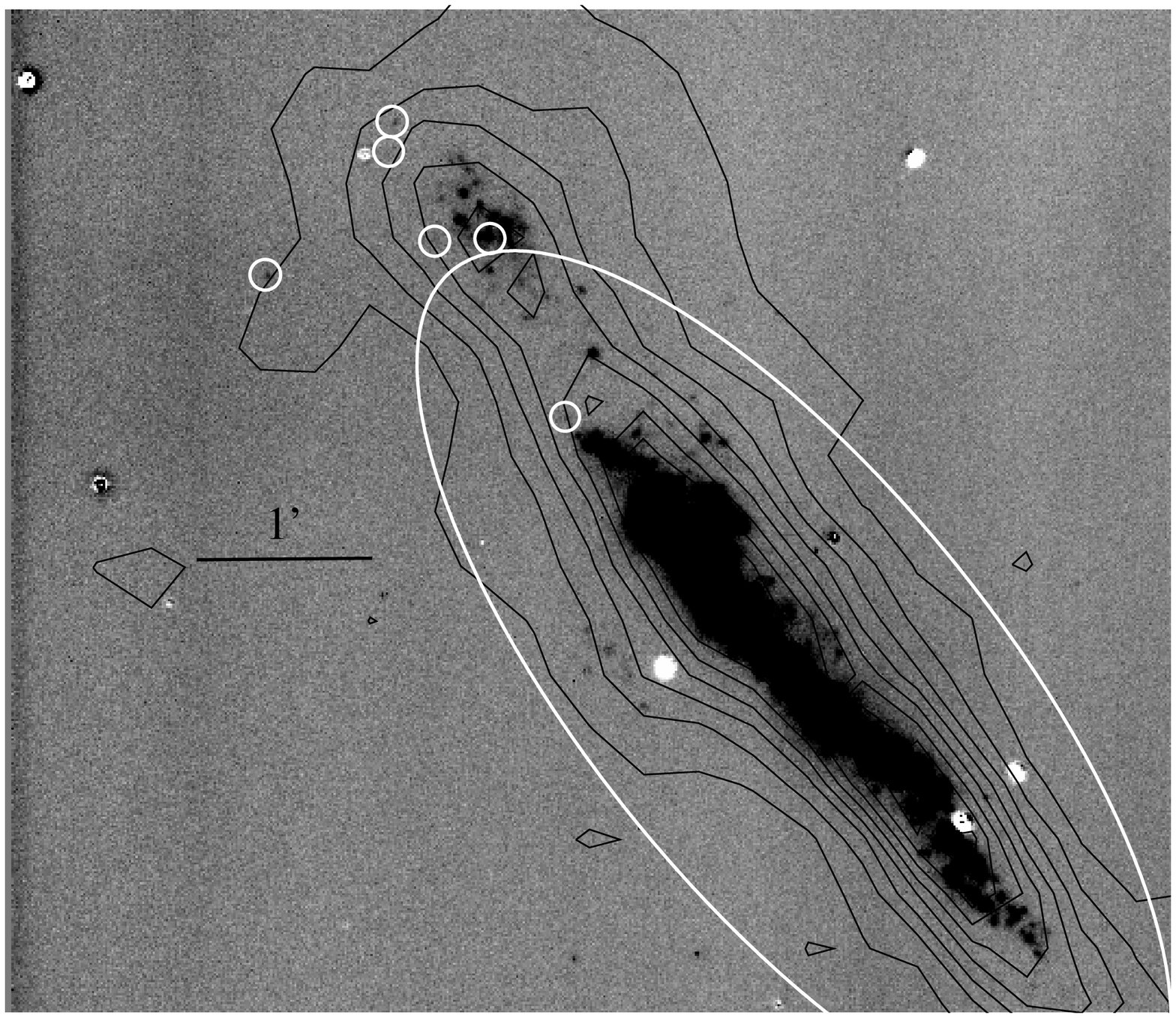}} 
\subfigure[NGC 3718; with HI contours at 0.5 to 20 M$_{\odot}$ pc$^{-2}$]{ 
\label{fig:slitsandgas:h} 
\includegraphics[width=0.48\linewidth]{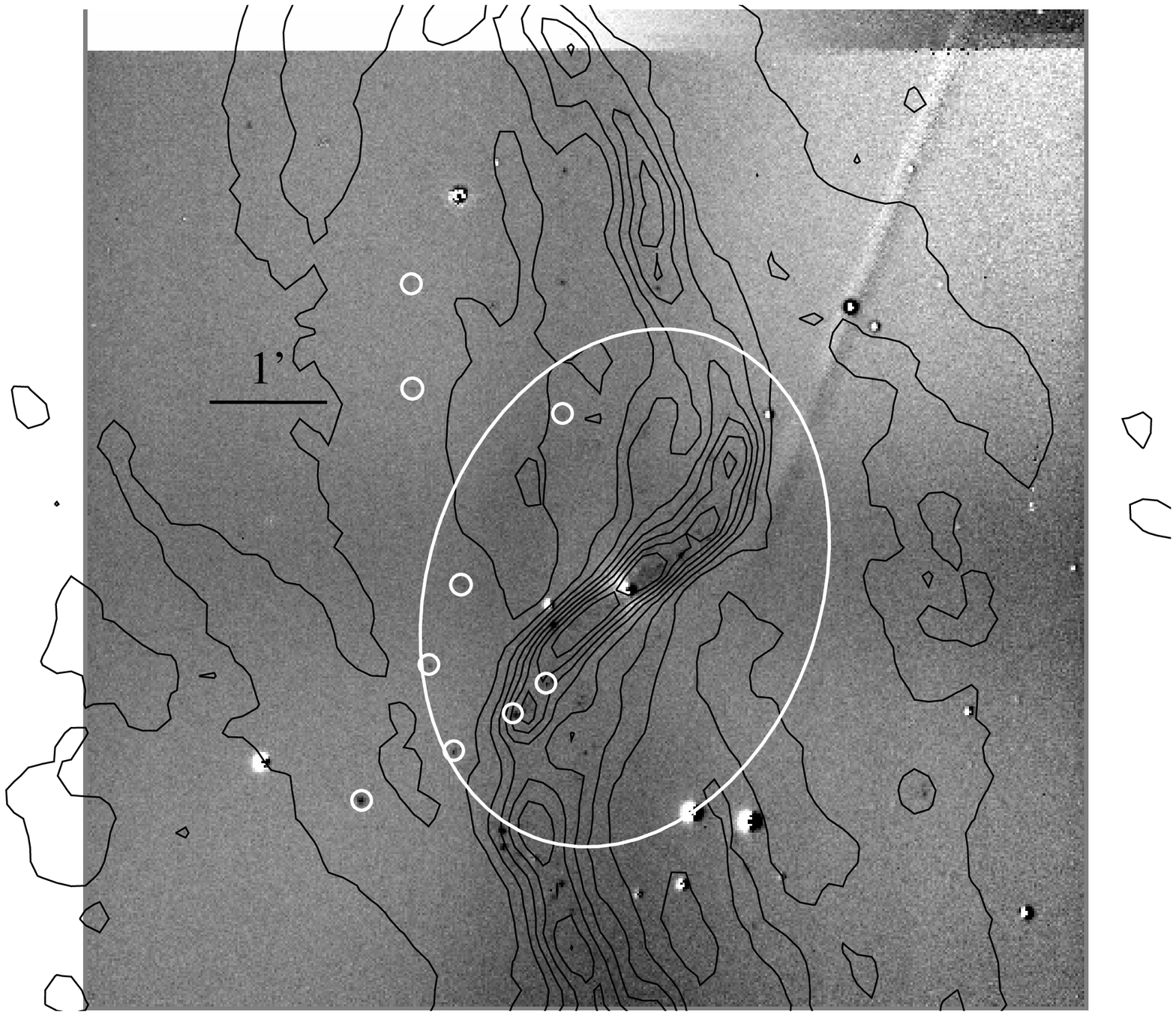}} 
\captcont{cont'd} 
\end{figure*} 
\begin{figure*} [hp!]
\centering 

\subfigure[NGC 3893; with HI contours at 3.5 to 17.0 M$_{\odot}$ pc$^{-2}$]{ 
\label{fig:slitsandgas:i} 
\includegraphics[width=0.40\linewidth]{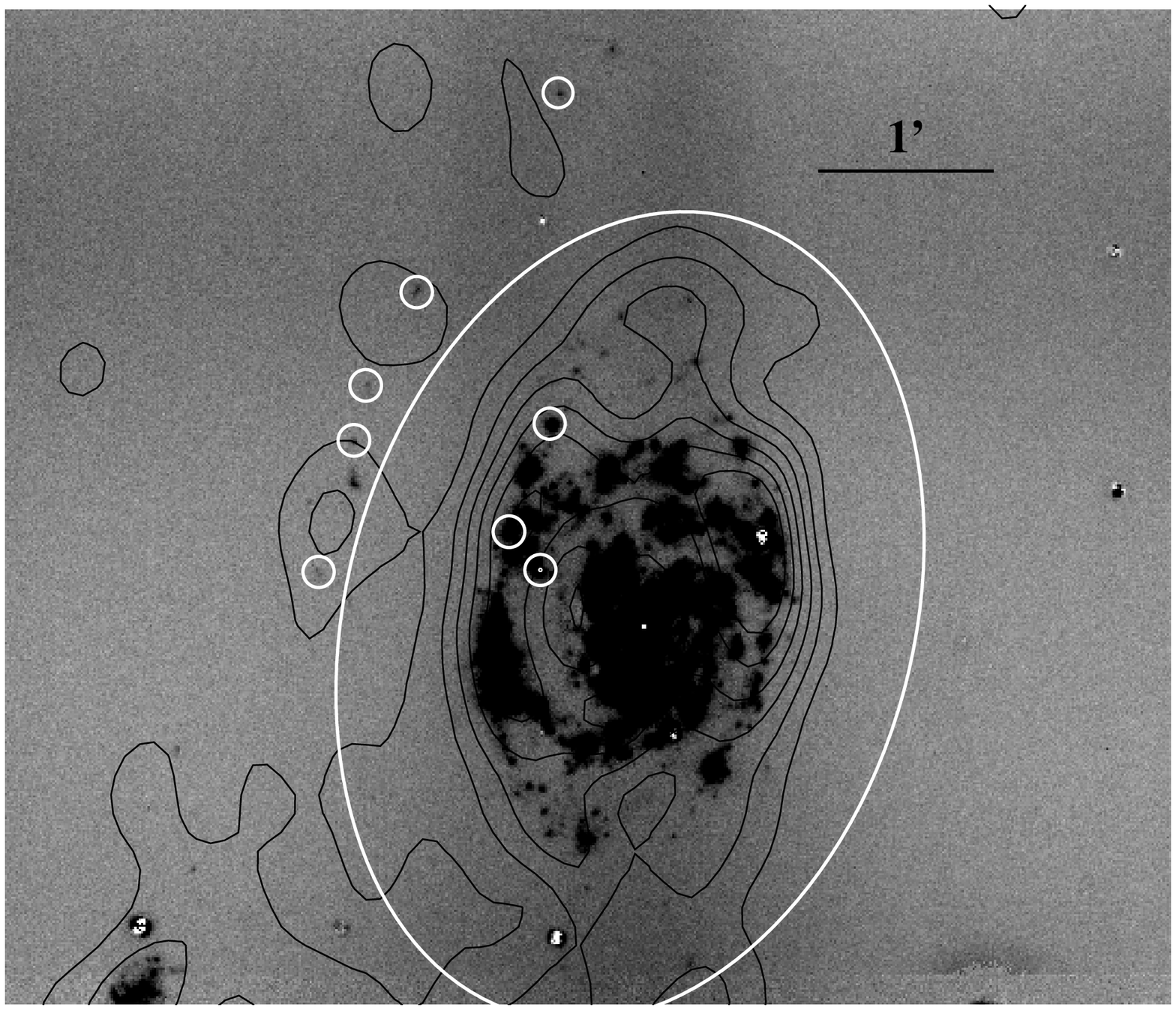}} 
\subfigure[NGC 5774/5; with HI contours at  2.8 to 25 M$_{\odot}$ pc$^{-2}$]{ 
\label{fig:slitsandgas:j} 
\includegraphics[width=0.40\linewidth]{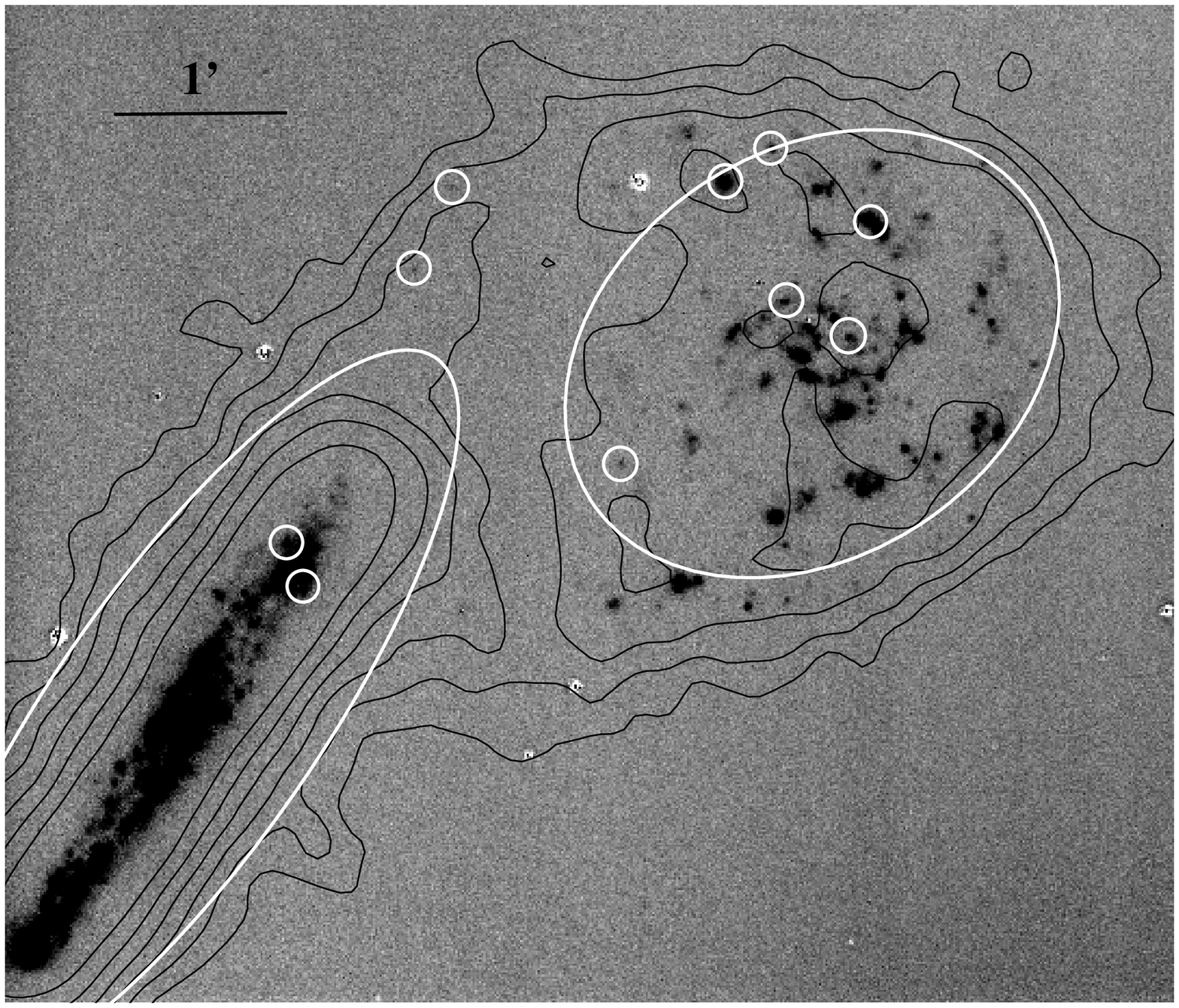}} 
\subfigure[NGC 6239; with HI contours at 2.0 to 25 M$_{\odot}$ pc$^{-2}$]{ 
\label{fig:slitsandgas:k} 
\includegraphics[width=0.40\linewidth]{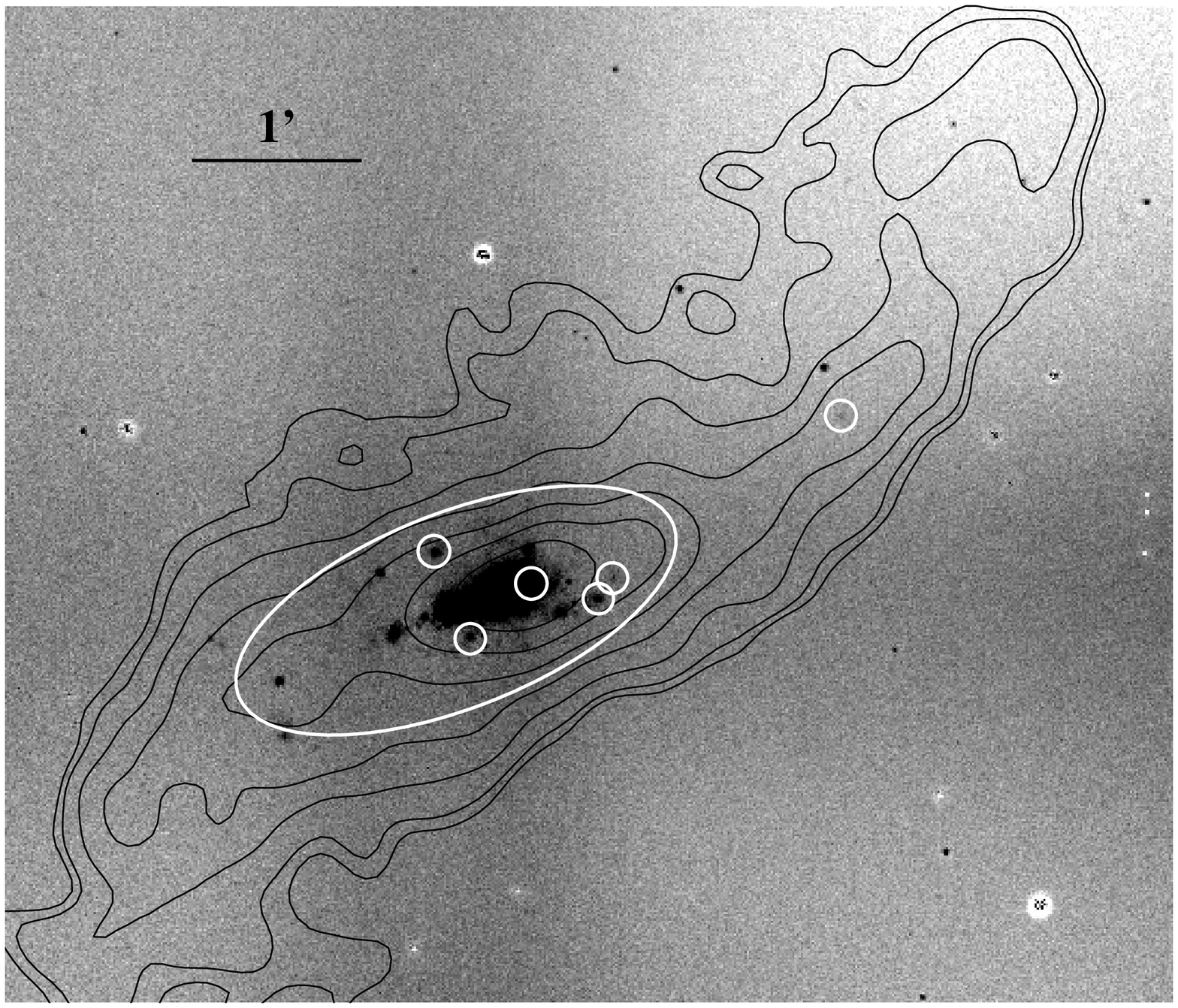}} 
\subfigure[UGC 5288; with HI contours at 0.25 to 6.0 M$_{\odot}$ pc$^{-2}$]{ 
\label{fig:slitsandgas:l} 
\includegraphics[width=0.40\linewidth]{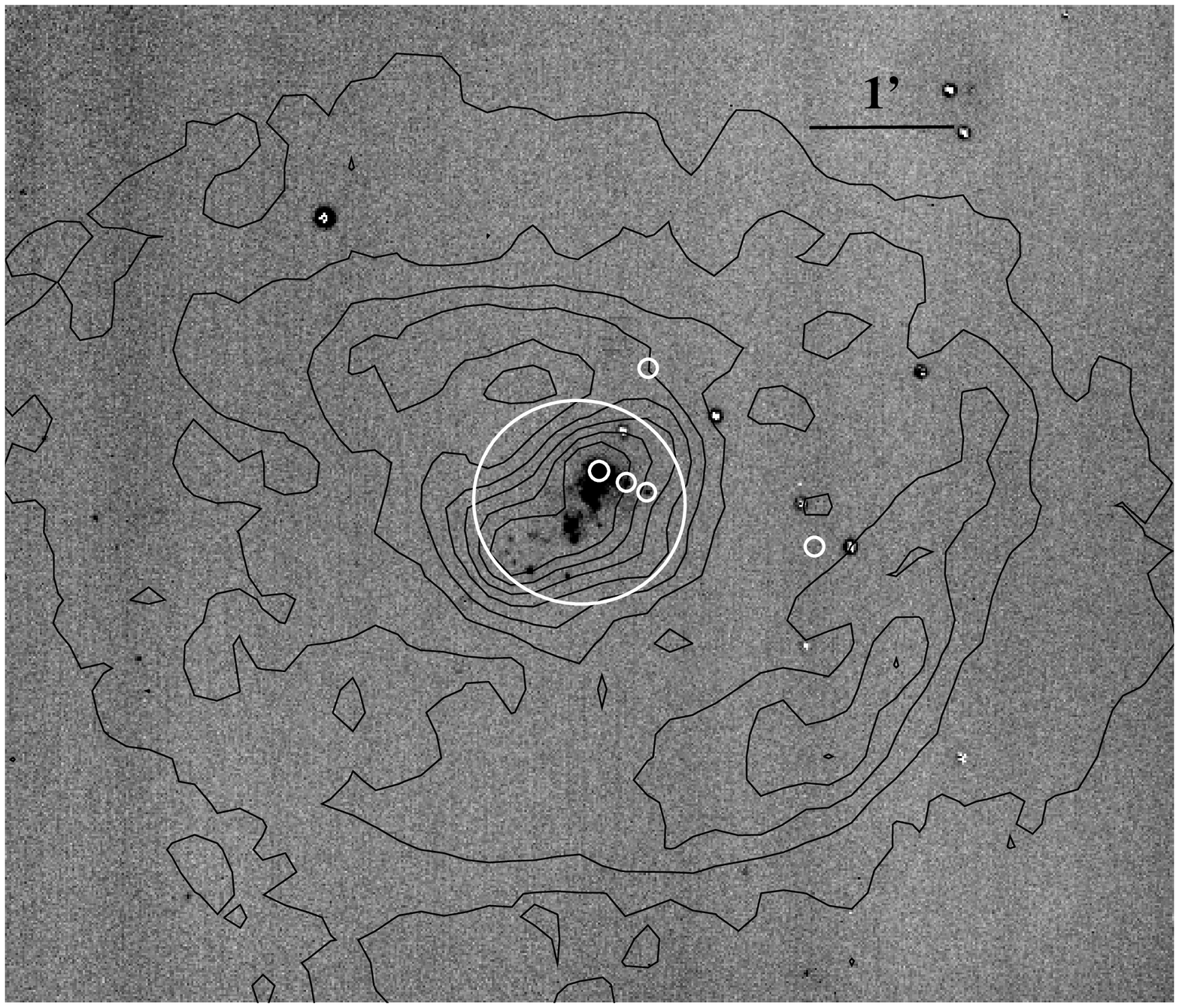}} 
\subfigure[UGC 9562]{ 
\label{fig:slitsandgas:m} 
\includegraphics[width=0.40\linewidth]{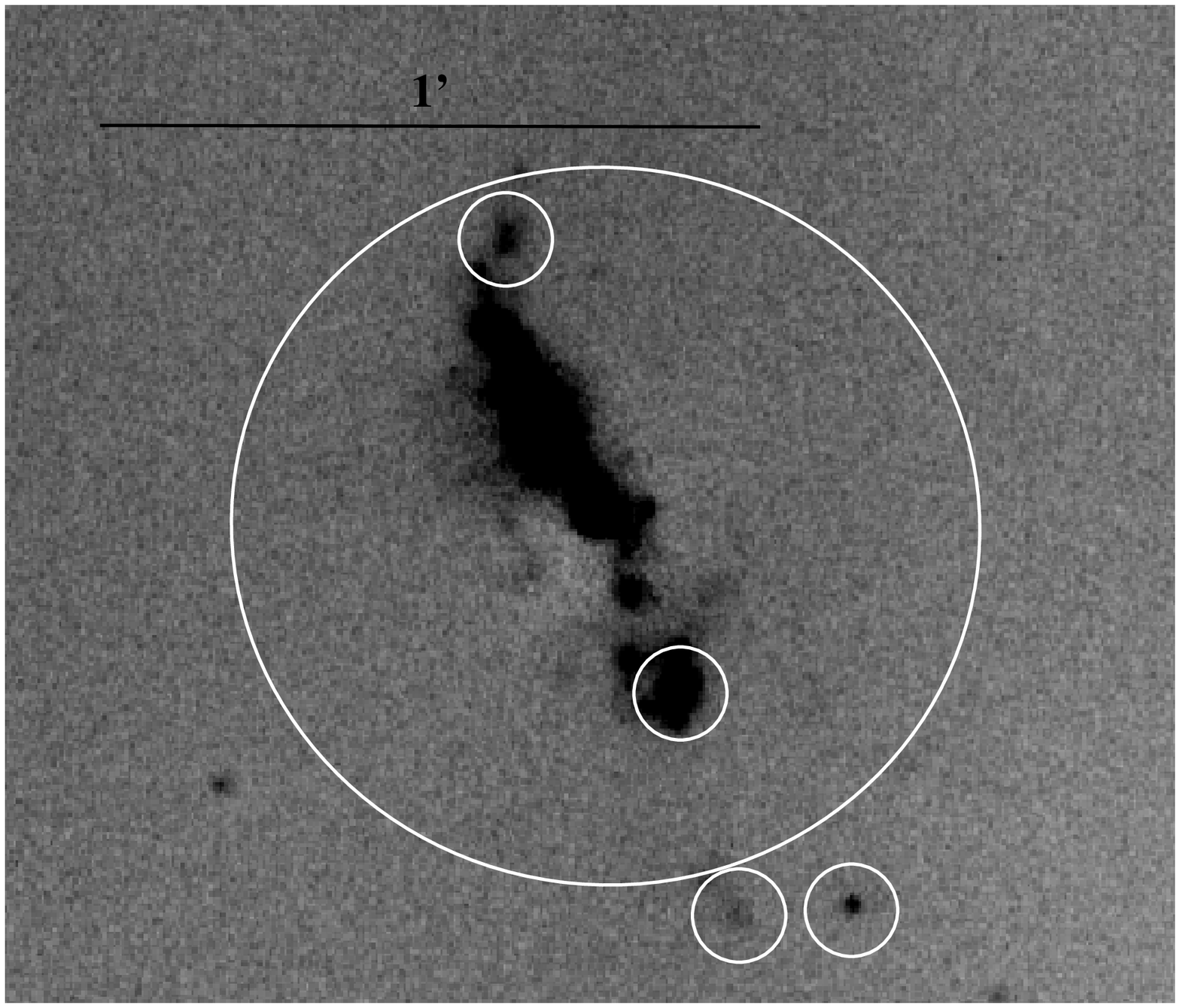}} 
\caption{ cont'd.} 
\label{fig:slitsandgas} 
\end{figure*}


\begin{sidewaystable*}[p]\centering \scriptsize
\vspace{4.5in}
\begin{tabular*}{0.90\textwidth}{lccccccccccccc}
\hline
Name & RA & dec & d & R$_{25}$ & E(B$-$V) & Abun & [NII]/H$\alpha$ & HI morph & Beam Size & HI Reference &M$_{*}$ & M$_{HI}$ & SFR$_{tot}$\\
\hline
(1) & (2) & (3) & (4) & (5) & (6) & (7) & (8) & (9) & (10) & (11) & (12)& (13) & (14) \\
\hline
\hline
 NGC 2146 &  06 18 38.1  & +78 21 26.3 & 22.45 & 22.15 & 0.096 &  8.68$^{1}$ & 0.47$^{1}$& mMR & 77 $\times$ 49 & \cite{taramopoulos01} &6.02 &   2.10  &   2.2\\
NGC 2782 &  09 14 05.1 & +40 06 48.9 & 37.30 & 12.93 & 0.016 &  8.80$^{1}$ & 0.47$^{1}$& mMR & 9.4 $\times$ 8.7 & \cite{smith94}&1.55  & 0.42   &  1.2 \\
 NGC 3227  &  10 23 29.0 &  +19 52 42.3 & 20.85 & 20.57 & 0.023 &  8.60$^{2}$ & 0.30$^{2}$& E$-$Sp & 80 $\times$ 47 &\cite{mundell01}& 5.17   & 0.20  &  0.47 \\
NGC 3239  &  10 25 04.3 &  +17 09 11.5 &  8.10 &  6.26 & 0.032 &  8.19$^{1}$ & 0.07$^{1}$& Sp$^{+}-$Sp$^{+}$& 64.3 $\times$ 54.6 & \cite{iyer01}& 0.122  & 0.12  &  0.29 \\
NGC 3310 &  10 38 44.3  & +53 30 07.9 & 18.10 &  9.17 & 0.022 &  8.40$^{1}$ &0.23$^{1}$& mMR & 20 $\times$ 20 & \cite{iyer01}&0.798  & 0.52  &  3.0 \\
 NGC 3359 &  10 46 37.7  & +63 13 25.5 & 17.80 & 13.05 & 0.008 &  8.60$^{3}$ &0.15$^{3}$& HIc & 30 $\times$ 30 &  \cite{boonyasait01}&1.31  & 0.45  &  0.89 \\
NGC 3432  &  10 52 31.2 &  +36 37 06.9 & 12.98 & 12.46 & 0.013 &  8.30$^{1}$ &0.12$^{1}$& mM & 30 $\times$ 30 &  \cite{swaters02}&0.313  & 0.60  &  0.50\\
NGC 3718  &  11 32 35.0  & +53 04 05.2 & 17.00 & 12.24 & 0.014 &  8.48$^{1}$ &0.22$^{1}$& WARP & 30 $\times$ 30& \cite{verheijen01}&2.94  & 1.00   &  0.25 \\
 NGC 3893 &  11 48 38.6  & +48 42 40.8 & 18.13 & 12.81 & 0.021 &  8.60$^{1}$ &0.33$^{1}$& mM & 30 $\times$ 30 &\cite{verheijen01} &2.45  & 0.60  &  1.60\\
 NGC 5774 &  14 53 42.5 & +03 34 56.8 & 26.80 & 12.18 & 0.042 &  8.55$^{4}$ &0.26$^{4}$& Sp$^{0}-$Sp$^{0}$& 29 $\times$ 23&\cite{irwin94}&1.10 &  0.54 &   0.47\\
 NGC 5775  &  14 53 57.2 & +03 32 48.5 & 26.80 & 20.04 & 0.042 &  8.70$^{4}$ &0.36$^{4}$& Sp$^{0}-$Sp$^{0}$ & 29 $\times$ 23 &\cite{irwin94}&1.36   &0.91  &   0.70 \\
 NGC 6239  &  16 50 05.9 & +42 44 18.9 & 22.30 &  8.92 & 0.018 &  8.40$^{4}$ &0.15$^{4}$& mM &18.2 $\times$ 16.7& \cite{hogg01}&0.403 &  0.70 &  0.94\\
 UGC 5288 &  09 51 17.2 & +07 49 37.1 &  6.03 &  1.37 & 0.034 &  8.08$^{5}$ &0.07$^{5}$& ENVL & 20.2 $\times$ 16.9 & \cite{vanzee01}&0.00852 & 0.023  &   0.0063\\
 UGC 9562 &  14 51 14.4 & +35 32 32.0 & 23.80 &  3.81 & 0.013 &  8.64$^{6}$ &0.22$^{6}$& Sp$^{0}-$Sp$^{0}$& 22.5 $\times$ 20.0&\cite{cox01}&0.095  & 0.082  &  0.15\\
     \hline
\end{tabular*}
\caption[Properties of the HI Rogue galaxies with outlying HII regions ]{Properties of the galaxies in the HI Rogues sample. (1)  NGC name of the galaxies. (2) RA (J2000) in hms. (3) dec (J2000) in dms. (4) Redshift-independent distances from NED, in Mpc. (5) Length of the 25$^{th}$ magnitude elliptical isophote (R$_{25}$) semi-major axis, in kpc. (6) Color excess E(B$-$V) from \cite{schlegel98}. (7) and (8) Oxygen abundances (average or integrated) as 12 + Log(O/H) and [NII]/H$\alpha$ ratios from various sources: 1. \cite{moustakas06} $-$ strong-line R23 abundances and [NII]/H$\alpha$ ratios from integrated spectroscopy 2. \cite{lisenfeld08} $-$ strong-line R23 abundances and [NII]/H$\alpha$ ratios from one HII region at 0.8 $\times$ R$_{25}$. 3. \cite{martin95} $-$ average of R23 oxygen abundances and [NII]/H$\alpha$ ratios for disk HII regions located between 4$-$ 9.5 kpc in projected distance from the optical center 4. \cite{marquez02} $-$  average of strong-line  [NII]/H$\alpha$ abundances and ratios (Pettini \& Pagel, 2004) for a number of  disk HII regions. 5. \cite{vanzee06} $-$ average strong-line R23 oxygen abundances and [NII]/H$\alpha$ ratios for  6 HII regions within R$_{25}$  6. \cite{shi05} $-$ average strong-line R23 oxygen abundances and [NII]/H$\alpha$ ratios for HII regions within R$_{25}$.  (9) HI morphology code from the online HI Rogues Catalog: mMR $-$ minor merger remnants; E$-$Sp $-$ interacting doubles, only 1 with HI;  Sp$^{+}-$Sp$^{+}$ $-$ interacting doubles, both with HI, both with tails; HIc $-$ detached HI clouds; mM $-$ minor mergers; WARP $-$ galaxies with two-sided warps; Sp$^{0}-$Sp$^{0}$ $-$ interacting doubles, both with HI, no tails; ENVL$-$ galaxies with extended HI envelopes. (10) Beam size of HI observations, in arcseconds. (11) Reference for the HI data. (12) Stellar masses derived from B, V, and sometimes R-band photometry and the luminosity/color and stellar mass functions of \cite{bell03} in units of 10$^{10}$ \Msun. (13) HI-masses for the galaxy in units of 10$^{10}$ \Msun, obtained from the literature (see text for the original sources) and corrected for the galaxy distances we use in column 5 of this table.  (14) Derived SFRs in \Msun~yr$^{-1}$ from the total H$\alpha$ luminosity obtained from the MDM 2.4-m images.\label{tab:c5t1} }
\end{sidewaystable*}

\begin{table}[h!]\centering \scriptsize
\begin{tabular*}{0.32\textwidth}{lccc}
\hline
Galaxy & d & R$_{25}$ & HI morph \\
\hline
(1) & (2) & (3) & (4) \\
\hline

NGC 2768 & 20.06 & 25.63& EpecH\\
NGC 2777    &18.61& 2.27 & mM \\
NGC 3396   &28.00* & 15.18 & Sp$^{+}-$Sp$^{+}$ \\
NGC 3471  &28.64& 9.0 & mM\\
NGC 3769   &15.50& 13.24 & mM\\
NGC 3998  &17.85& 9.00 & EpecH\\
NGC 4088  &15.80& 12.32 & mM-3\\
NGC 4111 &16.00& 16.70 & EpecH\\
NGC 4532  &15.80& 8.97 & IGHI\\
NGC 4639  &22.34& 10.08 & EXT\\
NGC 4789A  &4.14& 1.98 & ENVL\\
NGC 5916 &30.60& 14.24& Trip+\\
UGC 8201 &4.29& 2.18 & MISC\\
VIRGOHI 21 &14.00* & -- & IGHI\\

\hline
\hline

  \hline
\end{tabular*}
\caption[Galaxies imaged at MDM found to have no outlying HII regions.]{ Galaxies imaged at MDM found to have no outlying HII regions.  (1) Common name for the galaxy. (2) Redshift-independent distances from NED, in Mpc. *no red-shift independent distances available, Hubble-flow distance. (3) Length of the 25$^{th}$ magnitude elliptical isophote (R$_{25}$) semi-major axis, in kpc. (4) HI morphology code from the online HI Rogues Catalog: EpecH$-$ Normal early-type galaxies with peculiar HI; mM-3 $-$ Minor mergers, 3 body-encounters; IGHI $-$ intergalactic HI with no optical counterpart. To clarify, VIRGOHI 21 is one of these intergalactic clouds, while NGC 4532 is an optically-detected galaxy surrounded by such HI debris; Trip+ $-$  Interacting triples; MISC $-$ miscellaneous. The caption of Table 5.1 contains additional descriptions of morphology codes. \label{tab:c5none}}
\end{table}

\section{Optical Spectroscopy}
\subsection{Multi-slit Spectroscopy}
\label{sec:rogues:spec}

In the spring of 2008, we obtained multi-slit spectra of numerous HII regions associated with the 13 HI Rogues using the Gemini-North Multi-Object Spectrograph (GMOS-N) in order to make metallicity measurements of outer-disk and/or stripped gas. We were somewhat limited by the 5 \arcmin$\times$ 5 \arcmin~ field of view of GMOS, which restricted us to partial coverage of the rogue galaxies and their outskirts. The generation of the GMOS slit mask required that we obtain pre-imaging with GMOS-N  in queue mode prior to our spectroscopic observations. We used a 90 \AA-wide H$\alpha$ filter for GMOS to image the 13 fields in our sample, and in concert with the R$-$band continuum-subtracted MDM images, were then able to pick out and obtain precise instrumental positions for outlying and galactic HII regions. In only a few cases, we selected potential outlying HII regions from the Gemini pre-imaging that were not detected in the MDM images. Four to five alignment star holes per-field, and slitlet lengths of at least 10\arcsec~allowed us to fit $\sim25$ science slitlets per mask. In practice, however, we had an average of 8 usable science slitlets per mask (minimum = 4; maximum = 17) due to HII region position/alignment conflicts, very faint HII regions without multiple emission-line detections (discovered post-reduction), several blank ``sky slitlets" (see below), constraints on the instrument rotator angle due to available guide stars, and an increased slitlet length for the brightest central HII regions (for proper night-sky emission line subtraction).  

Our spectral observations were taken in queue mode under the following conditions: a grey lunar phase, an image quality of 1.5\arcsec~(occurring 85\%~of the time),  and some cloud cover (better than that occurring 50\% of the time). We used the B600-G5303 600 l/mm grating with various grating angles optimized for each individual slit mask so that key emission lines would not fall in either of the 2 CCD chip gaps. Central wavelengths range between 4100 and 4800 \AA.  The above grating configuration results in a spectral coverage of 2760 \AA, and was chosen so that we could detect emission lines between $\sim3700 -5007$ \AA~ for the vast majority of the HII regions on our masks. Several HII regions falling close to the image edges did not achieve this range, and, for a few of those HII regions, we were able to detect the H$\alpha$ and [NII] $\lambda6583$ emission lines. However, we generally did not obtain spectral data above 5500\AA~ for two reasons: 1. The low-resolution grating (R150, which would have given a larger spectral coverage) is not sensitive enough in the blue to obtain a detection of \oii~$\lambda$$\lambda$3727 for our faint objects  2. An additional grating configuration for red spectral coverage would have doubled our observing time, and therefore would have reduced the number of targets we could observe with Gemini. Since we opted for only blue spectral coverage, we used the R23 method for obtaining abundances. We use additional long-slit data with red spectral coverage to break the degeneracy of the R23 relation (see Section \ref{sec:rogues:r23} for details).  1.0\arcsec~wide slitlets and  4 (spatial) $\times$ 2 (spectral) binning yields a spectral resolution of 1.8 \AA~per pixel, and a spatial resolution of 0.28 arcsecond per pixel. Our science exposure times were 3$\times$ 20 minutes for each field. 

Arc-lamp (CuAr) and spectral flat field calibrations were done between and after each of the science observations. Baseline Gemini calibrations additionally include the observation of one standard star per grating configuration, HZ-44 in our case, over the execution of the entire science program. The main dangers of using a single  standard not taken on the same night as the targets are that seeing variability combined with differential refraction will alter the effective sensitivity function, and that the atmospheric extinction curve will change.  Fortunately, in every case, our  targets (and the standard star) were observed close to zenith, minimizing the effect of seeing variability with differential refraction.  Additionally, the variation in atmospheric extinction (or overall throughput) is minimal on Mauna Kea at optical wavelengths \citep{extinctionvar}. Nonetheless, our single relative sensitivity function probably subjects our relative line fluxes to higher errors than if we had used at least one standard star on every night data were taken. 

For the reduction of the GMOS multislit data, we used the IRAF Gemini/GMOS software package, specifically designed for GMOS multislit data reduction. Once the basic reduction is performed (GPREPARE, GSFLAT, GSREDUCE), we apply the wavelength solution from GSWAVELENGTH, using GSTRANSFORM, the fits to which had a typical RMS of 0.8$-$1.0 \AA. The short slit lengths complicated night-sky line subtraction for several of the bright or more extended HII regions in our sample. Several blank sky slitlets on every mask allowed us to properly subtract the sky for these bright/extended objects whose emission lines occupied the entire length of the slitlet. We therefore did two passes of sky subtraction for each mask dataset, one using GSSKYSUBTRACT, and the other using the manual sky subtraction method. Using GEMCOMBINE and GSEXTRACT, we obtained 1-dimensional reduced spectra, which we flux-calibrated using GSCALIBRATE based on our sensitivity function derived from HZ-44. 

As a test of our manual sky-subtraction, we compared the two methods using calibrated 1-d spectra of the compact, fainter HII regions, and we found that line fluxes were identical.  Typical RMS noise in the final, reduced spectra range from $\sim$9 $\times$ 10$^{-18}$ ergs s$^{-1}$ cm$^{-2}$ at 3730 \AA, to $\sim$2 $\times$ 10$^{-18}$ ergs s$^{-1}$ cm$^{-2}$ at 5000 \AA, to $\sim$7 $\times$ 10$^{-19}$ ergs s$^{-1}$ cm$^{-2}$ at 6500 \AA~(when data were redder due to objects being on images edges). These values do vary somewhat for different slit masks/fields, and depend on the aperture size, and are noted only to illustrate that the emission-line Poisson noise is not a major contributor to the error. The RMS of the fitted sensitivity function was 6.5\%, and we estimate that the addition of flux calibration uncertainty, read noise, sky noise, and flat-fielding errors bring our total errors in line flux measurements up to $\sim$ 10\%. 
 
\subsection{Longslit Spectroscopy}
\label{sec:rogues:long}

Over the course of three observing runs, one at the MDM 2.4-m in the winter of 2007, one at the Keck 10-m in the spring of 2010, and one at the Shane 3-m telescope at Lick Observatory in the spring of 2011,  we obtained longslit, low-resolution spectra for outlying HII regions in our sample of HI rogues with extended star formation. We use these spectra in the subsequent analysis to obtain [NII] $\lambda$6583 and H$\alpha$ emission lines, since these lines were generally absent in the GMOS multislit data. At MDM, we used the low-resolution Mark III spectrograph with the Echelle CCD,  employing the 300/5400 grating and a 0.824 arcsecond slit width. The 300/5400 grating centered at 6500 \AA~ gives a spectral coverage of 3120 - 9879 \AA, a range containing all of the optical emission lines we could hope to detect, and a  resulting dispersion of 3.3 \AA~  per pixel.  Blind offsets from nearby stars were required for positioning the often very faint ELdots in the slit. As mentioned in Section \ref{sec:rogues:eldot}, the run on the MDM 2.4-m telescope was intended as a spectroscopic confirmation run, and we obtained spectra for only a dozen ELdots. Exposure times ranged from 300 s to 2 $\times$ 600 s, depending on the narrow-band flux of the ELdot. 

At Keck, we were able to use the low-resolution imaging spectrometer (LRIS) for longslit spectroscopy of 3 HI Rogues with ELdots, during an observing run which had several target list ``holes" at the very start and end of the nights. The data were taken with the 600 l/mm grism (blue-side) and 400 l/mm grating (red side) which gives a spectral coverage between 3000 and 9300 \AA. Binning the data 2 $\times$ 2 resulted in a dispersion of 1.2 and 2.3 \AA~ per pixel for the blue and red cameras, respectively. ELdots were positioned in the 1\arcsec~ slit precisely by aligning the offset position angle between it and a bright star with the slit angle. In one case, NGC 6239, we were able to align several other galactic HII regions with the slit position angle as well. A problem with an amplifier on the red camera results in an unusable lower CCD chip on the red side (each side has two chips). Exposure times were 2 $\times$ 600 s in the blue and 3 $\times$ 400 s in the red.

At Lick, we used the Kast spectrometer on the Shane 3-m telescope to observe 9 remaining outlying HII regions without red spectral coverage. We  used Grism 1(600/4310) on the blue side, the D55 dichroic, and the 600/7500 grating on the redside, which gave full spectral coverage between 3300 and 8800 \AA. Exposures were taken through light clouds in most cases, and exposure times ranged from 30 minutes to 1.5 hours. The same observing strategy of bright-star position angle alignment used at the Keck telescope was employed here, and our overall success rate of acquiring the faint outlying HII regions was 100\%.  

Basic reduction steps, which included bias-subtraction, flat-fielding, and flux calibration with a single spectrophotometric standard star (one for each telescope) were performed with the IRAF longslit package for MDM data, and with the LowRedux \footnote{http://www.ucolick.org/$\sim$xavier/LowRedux/index.html} IDL software package for the Keck and Lick data.  We use these longslit data only to obtain [NII]/H$\alpha$ ratios in order to break the R23 degeneracy (as described in \ref{sec:rogues:r23}) for our blue GMOS multislit data. Because these two emission lines ([NII] $\lambda6583$ and H$\alpha$, at $\lambda$ = 6563 \AA) are so close in wavelength, we did not require an accurate flux (or wavelength) calibration or reddening correction. Table \ref{tab:c5t2} lists the outlying HII regions for which we were able to observe the [NII]/H$\alpha$ ratio using the longslit data. Table \ref{tab:c5t2} also lists the outlying HII regions so far to the edge of the CCD in the GMOS data for NGC 2146 and NGC 2782 that we were able to obtain  [NII]/H$\alpha$ ratios.

\begin{table}[h!] \centering \scriptsize
\begin{tabular*}{0.45\textwidth}{lcccc}
\hline
Galaxy & Region ID & R/R$_{25}$ & Telescope & [NII]/H$\alpha$\\
\hline
(1) & (2) & (3) & (4) & (5) \\
\hline
\hline

NGC2146 & 14 & 0.5 & GMOS & 0.31$\pm$0.03 \\ 
NGC2146 & 13 & 0.4 & GMOS & 0.31$\pm$0.03\\ 
NGC2146 & 12 & 0.4 & GMOS & 0.27$\pm$0.03 \\
NGC 2782 & 2 & 2.5 & MDM & 0.24$\pm$0.02\\ 
NGC 2782 & 4 & 1.4 & GMOS & 0.27$\pm$0.02\\
NGC 2782 & 6 & 0.6 & GMOS & 0.21$\pm$0.02\\
NGC 3239 & 4 & 1.3 & Lick & 0.07$\pm$0.01\\
NGC 3310 & 7 & 1.5 & Lick & 0.31$\pm$0.04\\ 
NGC 3310 & 12 & 1.3 & Keck & 0.18$\pm$0.01\\ 
NGC 3359 & 11 & 1.4 & Lick & 0.08$\pm$0.01\\ 
NGC 3359 & 14 & 1.1 & Lick & 0.09$\pm$0.01\\ 
NGC 3432 & 14 & 1.0 & Lick & 0.11$\pm$0.01\\ 
NGC 3718 & 15 & 1.4 & MDM & 0.20$\pm$0.03 \\ 
NGC 3718 & 4 & 1.7 & Lick & 0.11$\pm$0.02 \\
NGC 3893 &  7 & 1.4 & MDM & 0.13$\pm$0.02 \\ 
NGC 5774/5 & 4 & 1.5 & MDM & 0.18$\pm$0.02 \\
NGC 5774/5 & 7 & 1.1 & Lick & 0.18$\pm$0.02 \\
NGC 5774/5 & 12 & 1.0 & Lick & 0.12$\pm$0.02 \\ 
NGC 6239 & 5 & 1.9 & Keck & 0.08$\pm$0.02 \\ 
NGC 6239 & ** & 0.6 & Keck & 0.11$\pm$0.02 \\  
NGC 6239 & ** & 0.4 & Keck & 0.22$\pm$0.01 \\ 
NGC 6239 & 17 & 0.3 & Keck & 0.12$\pm$0.01 \\  
NGC 6239 & ** & 0.2 & Keck & 0.20$\pm$0.02 \\  
NGC 6239 & ** & 0.1 & Keck & 0.10$\pm$0.01 \\
UGC 5288 & 1 & 1.4 & Keck & $<$0.08 \\
UGC 5288 & 10 & 0.4 & Keck & 0.06$\pm$0.01 \\
UGC 9562 & 7 & 1.2 & Keck & 0.08$\pm$0.01 \\ 
UGC 9562 & 14 & 0.5 & Keck & 0.08$\pm$0.01 \\

\hline

  \hline
\end{tabular*}
\caption[NII/H$\alpha$ for selected HII regions from longslit spectra obtained at the MDM, Keck, and Lick Observatories. ]{ [NII]/H$\alpha$ for selected HII regions from longslit spectra obtained at the MDM, Keck, and Lick Observatories. Values are also tabulated for those HII regions with emission lines red-ward of H$\alpha$ in the GMOS multi-slit data.  (1) Host galaxy name (2) Region name that corresponds to the GMOS multi-slit ID given in HII region properties tables, if applicable. (3) Projected distance from the 25$^{th}$ magnitude elliptical isophote. (4) The telescope at which the spectral data were obtained. (5) the ratio of [NII] $\lambda6583$ to H$\alpha$ with associated errors. In one case, this value is an upper limit. We use these ratios to break the R23 degeneracy for the Gemini data.  \label{tab:c5t2}}
\end{table}

\section{Analysis}
\label{sec:rogues:anal}

\subsection{Emission-line Measurements}

The HII regions in our final sample, and marked in Figure \ref{fig:slitsandgas} were included based on the presence of the relevant emission lines for R23 strong-line abundances.  A direct measurement of the HII region's oxygen abundance from the oxygen emission lines present in its spectrum requires a measurement of the temperature-sensitive line ratio [OIII] $\lambda\lambda$4959, 5007/$\lambda$4363 \citep{osterbrock89}. In these spectra, the critical emission line [OIII] $\lambda$4363 is not detected. Therefore we use the R$_{23}$ = ([OII] $\lambda$3727 + [OIII] $\lambda\lambda$4959, 5007) / H$\beta$ \citep{pagel79} calibration of \cite{mcgaugh91} to determine the oxygen abundance for numerous HII regions in 13 HI Rogue galaxies. The majority  of the multi-slit spectra contained H$\gamma$ as well, which we used to correct for interstellar reddening (see below). The highest signal-to-noise spectra on every mask often contained additional emission lines, including H$\delta$, HeI $\lambda$4026 and HeII $\lambda$4687, occasionally  [NeIII] $\lambda$3869 and $\lambda$3970, and rarely the very faint, temperature-sensitive [OIII] $\lambda$4363 auroral emission line which provides a direct oxygen abundance measure.  Using the IRAF task {\it{splot}}, and fitting gaussian profiles we obtain integrated line fluxes. For the inner HII regions with substantial underlying continuum, we corrected for absorption from underlying stellar populations by subtracting gaussian fits to the Balmer-line absorption. 

We apply a correction for interstellar reddening to all line measurements from the observed H$\gamma$ to H$\beta$ ratio for case B recombination where H$\gamma$/H$\beta$ = 0.459 at an effective temperature of 10,000 K and electron density of 100 cm$^{-3}$ \citep{hummer87}. These lines have been corrected for underlying Balmer absorption.  We use a reddening function normalized at H$\beta$ from the Galactic reddening law of \cite{ccm} assuming R$_{v}$ = A$_{v}$/E(B$-$V) = 3.1. In most cases, the measured E(B-V) from the Balmer decrement is similar to the value given in Table \ref{tab:c5t1}, or lower. We do note additionally that the reddening has very little impact on the calculated strong-line abundances, which have very large errors. Tables \ref{tab:prop2146} to \ref{tab:prop9562} present the reddening-corrected strong line measurements for the HII regions in all 13 rogue galaxies.  All listed line fluxes are relative to  H$\beta$ = 100. 

\subsection{Breaking the Degeneracy of the R23 Relation}
\label{sec:rogues:r23}

We use the \cite{mcgaugh91} calibration of the R23 relation (henceforth, M91) to obtain the oxygen abundances of all the HII regions in our sample of 13 rogue systems.  The R23 relation exhibits a well-known degeneracy, with a turn-over in the relation at Z $\sim$ 0.3 Z$_{\odot}$ (12$+$Log(O/H) $\sim$ 8.35). The most robust way to place an HII region on the upper or lower branch of the R23 relation is to use the [NII] $\lambda$6583 to [OII] $\lambda$$\lambda$3727 ratio. When log [NII]/[OII] $<$ $-$1.0, HII regions lie on the lower metallicity branch of the R23 relation \citep{kewley08}. Since our multislit data are bluer than 5500 \AA~ in most cases, we cannot use this method. Instead, we use the [NII]/H$\alpha$ ratio as described in  \cite{kewley08} from our longslit spectra and from the literature. The primary benefit of this line ratio is that it does not require an accurate (or absolute) flux calibration or reddening correction due to the proximity in wavelength of [NII] $\lambda$6583 and H$\alpha$. This benefit is important for our data redward of 5600 \AA, since we are using spectra from 4 different telescopes taken under varying conditions. 



 \cite{kewley08} discuss how to use the [NII]$\lambda$6583/H$\alpha$ ratio (N2 index) to break the degeneracy in the R23 relation. If Log N2 $<$ -1.3 ([NII]/H$\alpha$ $<$ 0.05), there is a high degree of certainty ($\sim$86\%) that the oxygen abundance is on the lower-branch of the R23 relation.  Whereas if Log N2 $>$ -1.1 ([NII]/H$\alpha$ $>$ 0.08), the oxygen abundance is on the upper branch. Between -1.1 and -1.3 (0.05 and 0.08), Log (N2) (([NII]/H$\alpha$) does not accurately discriminate between the upper and lower branches of R23 because the oxygen abundance is very close to the turnover at 12 + Log (O/H) = 8.3.  This method of breaking the R23 degeneracy works because the N2 index is sensitive to and monotonic with the oxygen abundance to within 0.3 dex accuracy at a 95\% confidence level up to 12 + Log(O/H) = 8.8 (Pettini and Pagel 2004, henceforth PP04\nocite{pettini04}).  The primary drawback of using this method, then,  is the large uncertainty in the calibration ($\sim$0.4 dex), and the limited range over which it is useful. It is apparent from the [NII]$\lambda$6583/H$\alpha$ ratios given in Tables 1 and 3 that most of the outlying and central HII regions lie on the upper branch of the R23 relation, except for a few cases which we discuss below.
 
   We do not have any measurements of the N2 index for the outlying HII regions in NGC 2146, and therefore cannot definitively place them on either branch. Furthermore, the integrated  [NII]$\lambda$6583/H$\alpha$ ratios of NGC 3239 and the average  [NII]$\lambda$6583/H$\alpha$ ratio of the HII regions in UGC 5288 put both galaxies in the ambiguous region of the N2 index. Fortunately, for UGC 5288 and NGC 3239, there are published values of the log [NII]/[OII] that allow us to clear the N2 index ambiguity in the central regions. The integrated spectroscopy of the inner R $=$ 3 kpc ($\sim$ 0.5 $\times$ R$_{25}$) of NGC 3239 indicates a Log [NII]/[OII] of -1.05 \citep{moustakas06}.  UGC 5288 has an average Log [NII]/[OII] of -1.16 within R$_{25}$ \citep{vanzee06}. Therefore, the inner HII regions of  both galaxies lie on the lower branch of the R23 relation.
   
    The outlying HII regions of NGC 3239 and UGC 5288 have N2 indices in the ambiguous region as well. In the case of UGC 5288, the upper-branch M91 R23 values for its outer HII regions would put them at super-solar oxygen abundance, which is highly unlikely given that it is a dwarf galaxy and its central regions have an average oxygen abundance of 0.1 solar.  We therefore assume that these outlying HII regions are on the lower branch of the R23 relation along with their more central counterparts. In the case of NGC 3239, the situation is less clear, since both upper and lower branch values of the oxygen abundance are near the R23 turnover region. We cannot make any assumptions for the outlying HII regions in this galaxy, and therefore report both upper and lower branch values. 

\section{Results: The Distribution of Oxygen in 13 HI Rogues} 
\label{sec:rogues:results}

\begin{figure*} 
\centering 
\subfigure{ 
\label{fig:radgrads:a} 
\includegraphics[width=0.48\linewidth]{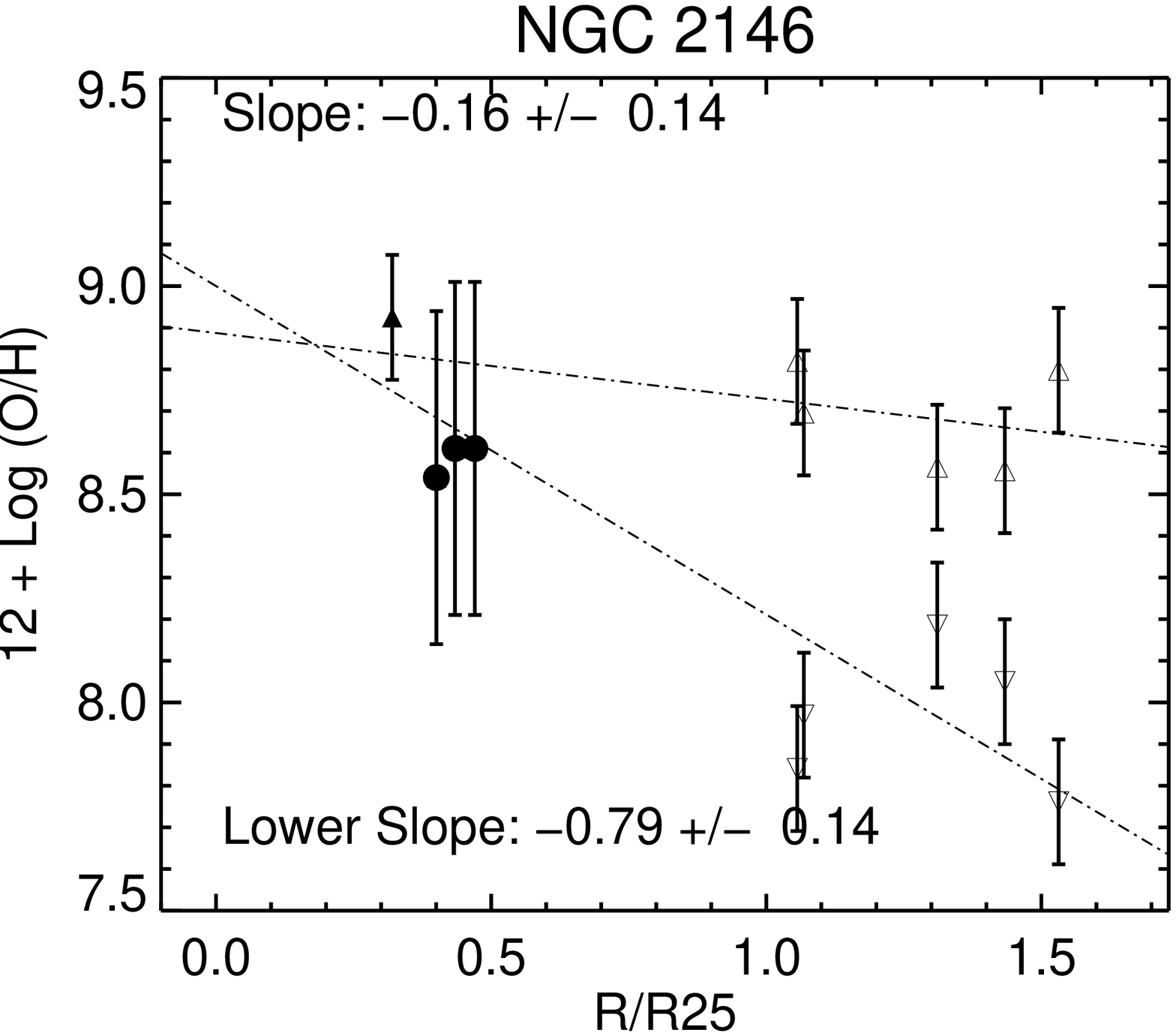}} 
\subfigure{ 
\label{fig:radgrads:b} 
\includegraphics[width=0.48\linewidth]{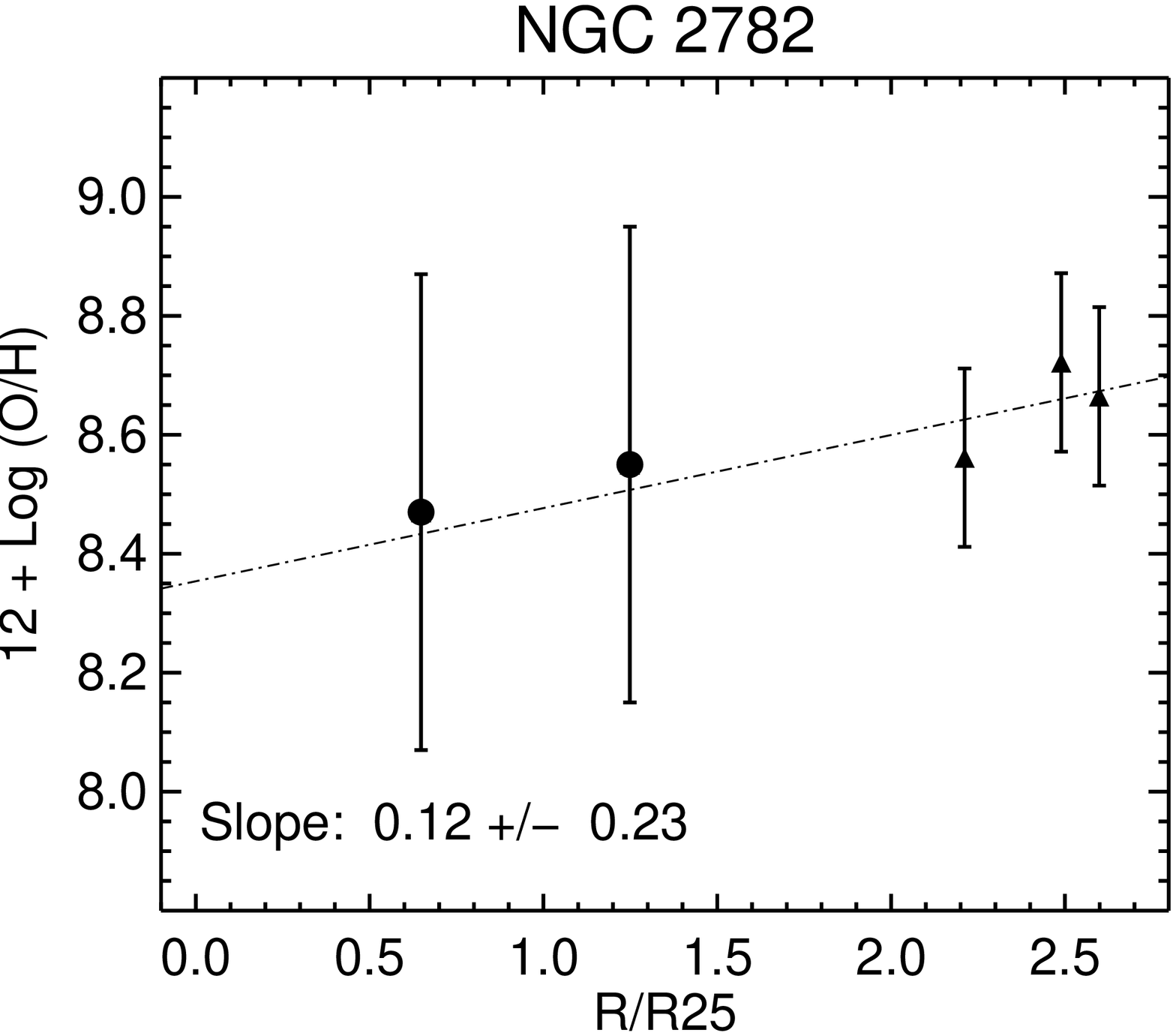}}
\subfigure{ 
\label{fig:radgrads:c} 
\includegraphics[width=0.48\linewidth]{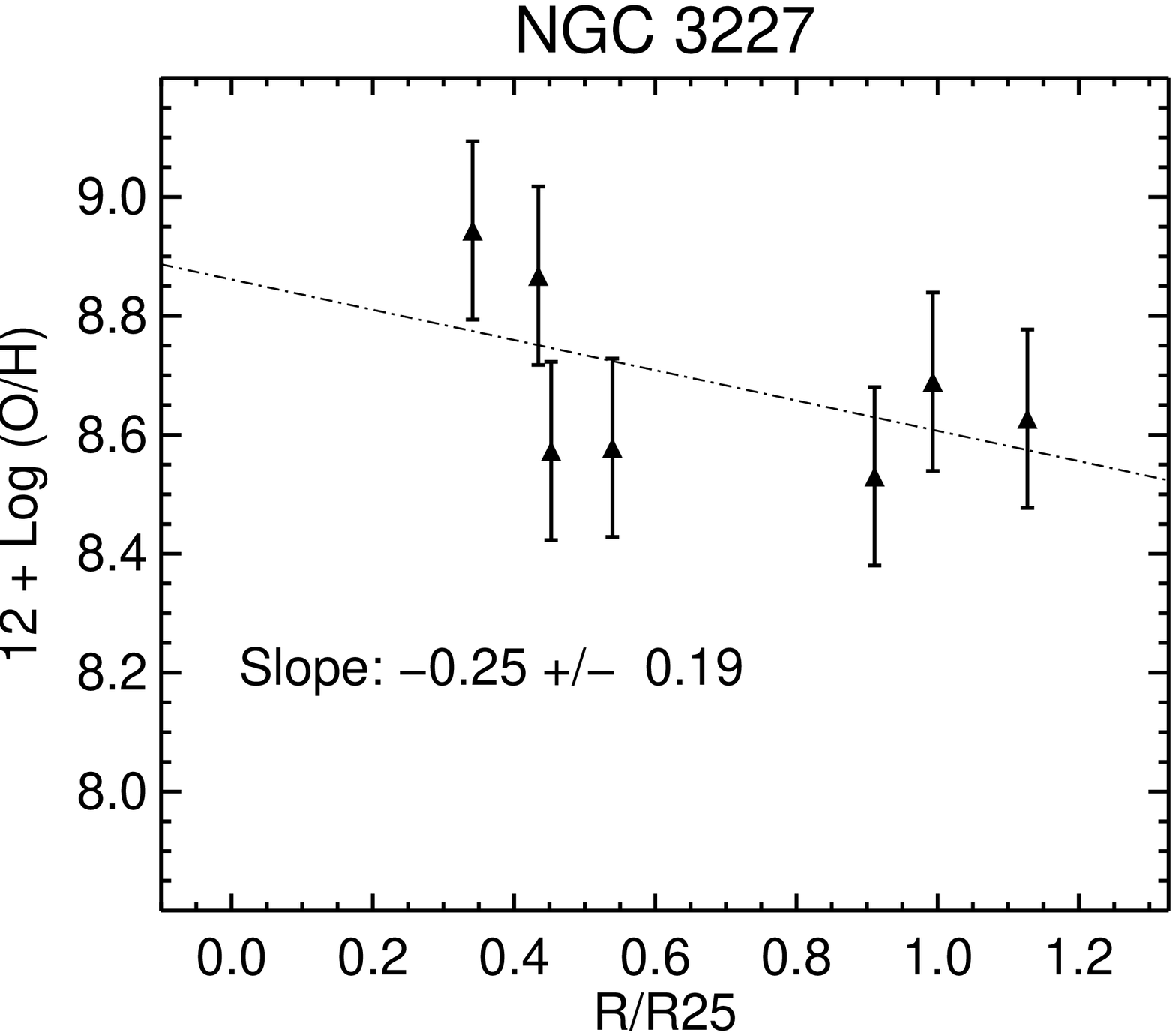}} 
\subfigure{ 
\label{fig:radgrads:d} 
\includegraphics[width=0.48\linewidth]{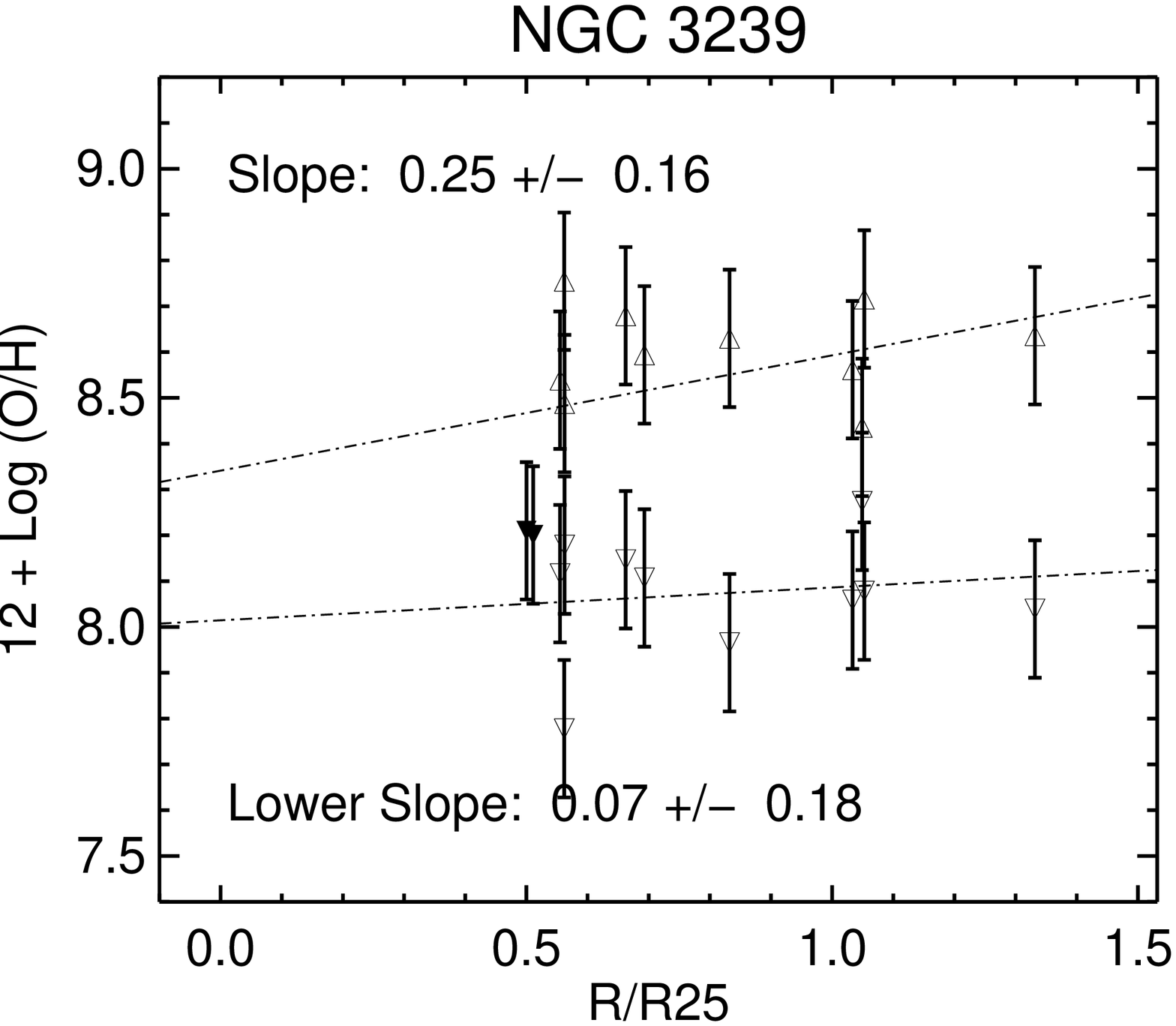}} 
\subfigure{ 
\label{fig:radgrads:e} 
\includegraphics[width=0.48\linewidth]{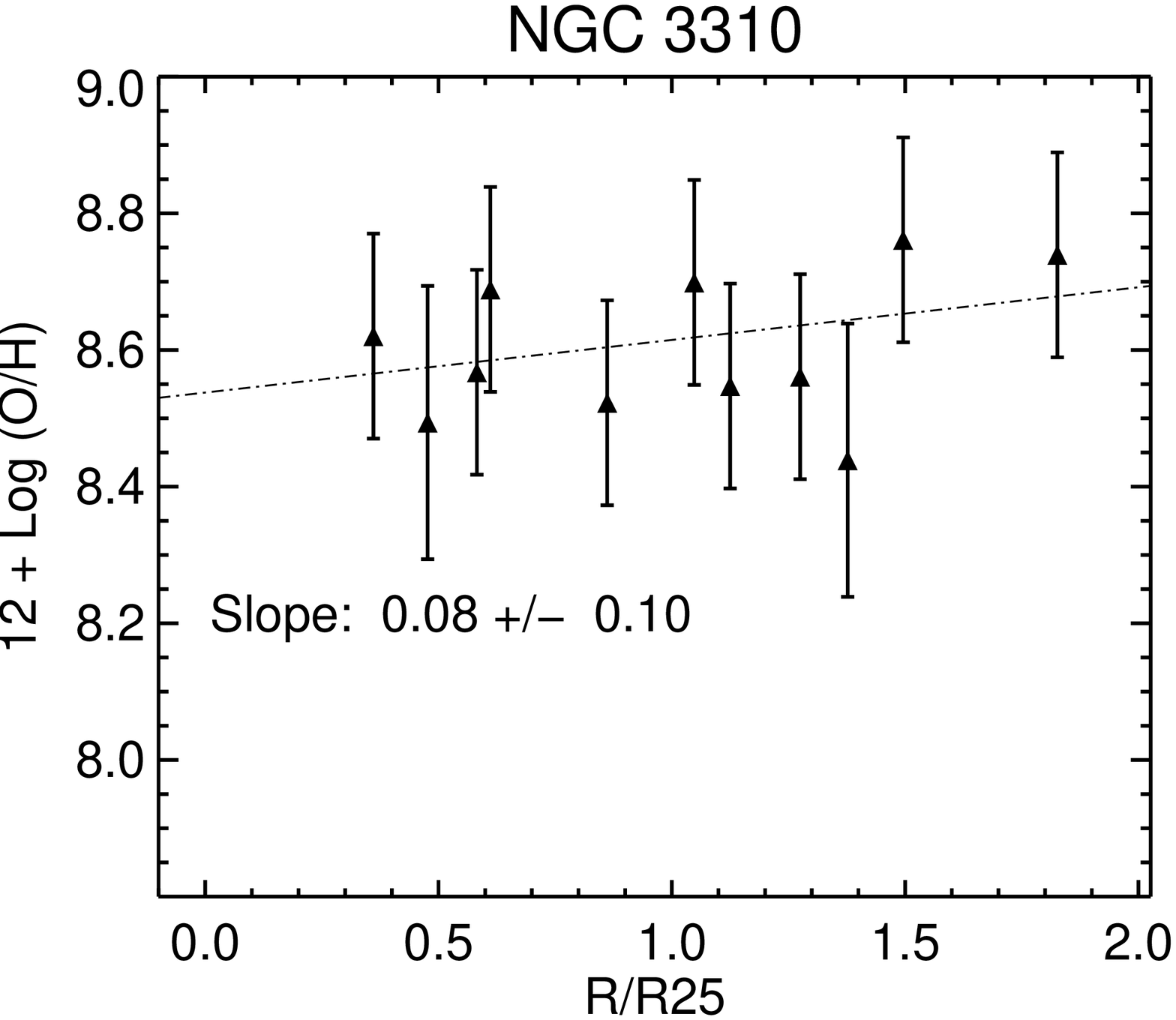}} 
\subfigure{ 
\label{fig:radgrads:f} 
\includegraphics[width=0.48\linewidth]{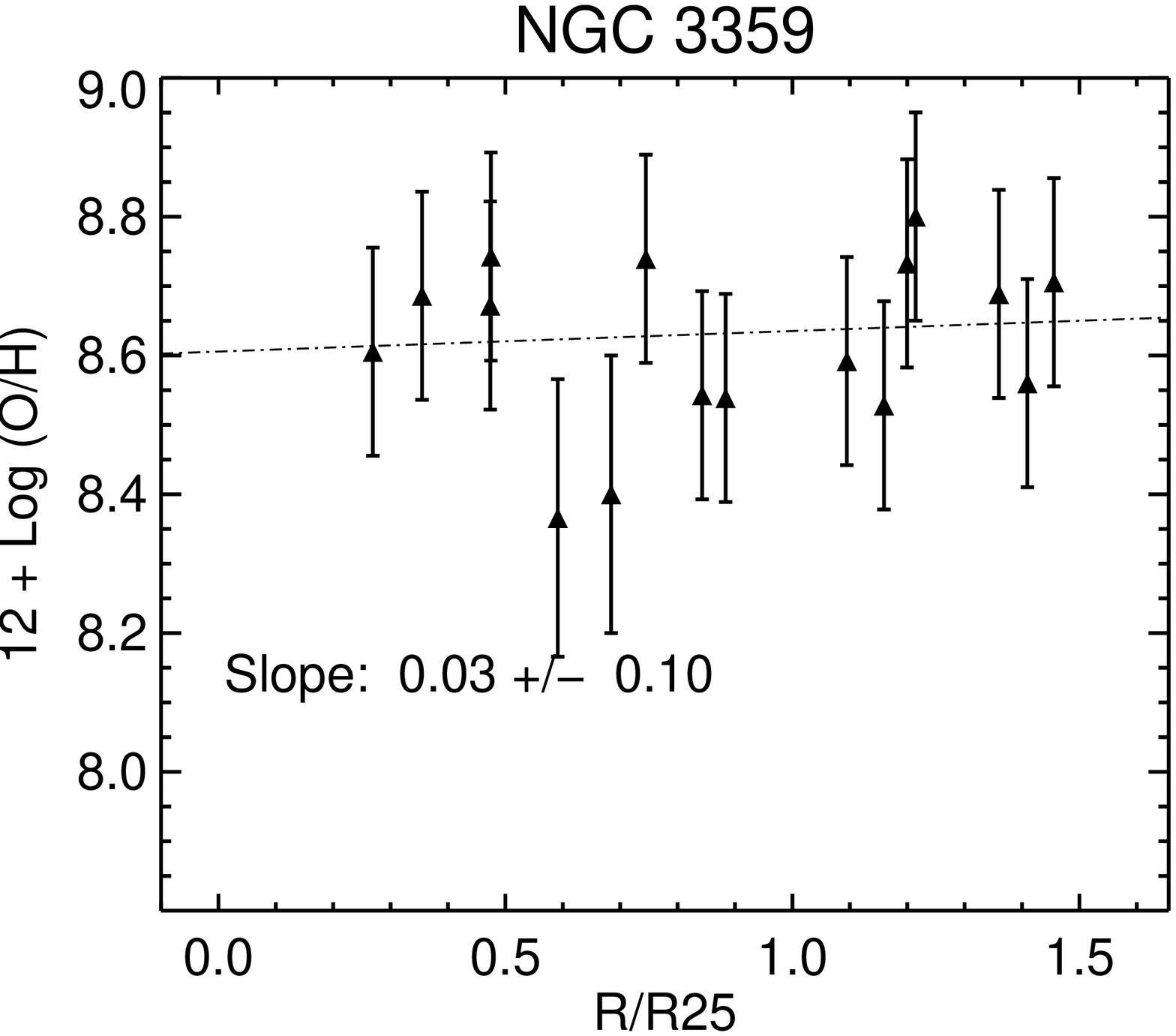}} 
\captcont{Radial oxygen abundance gradients, all in terms of R/R$_{25}$. The errors for the strong-line abundances range between 0.15 and 0.2 dex; the higher value is for abundances near the R23 turnover, 8.3 $<$ 12 $+$ Log $<$ 8.5 (O/H). (cont'd) } 
\end{figure*} 

\begin{figure*} [htp]
\centering 
\subfigure{ 
\label{fig:radgrads:g} 
\includegraphics[width=0.48\linewidth]{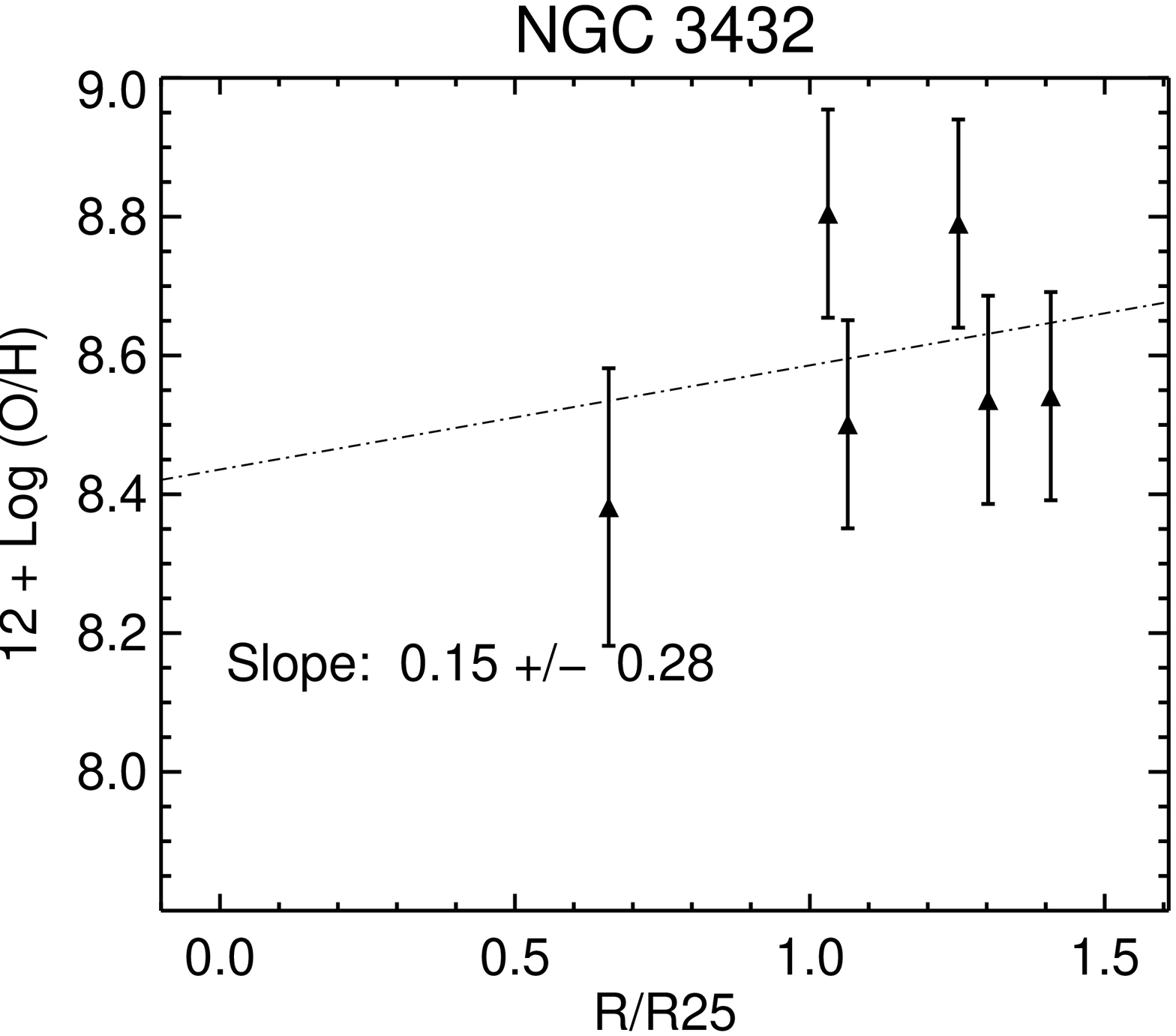}} 
\subfigure{ 
\label{fig:radgrads:h} 
\includegraphics[width=0.48\linewidth]{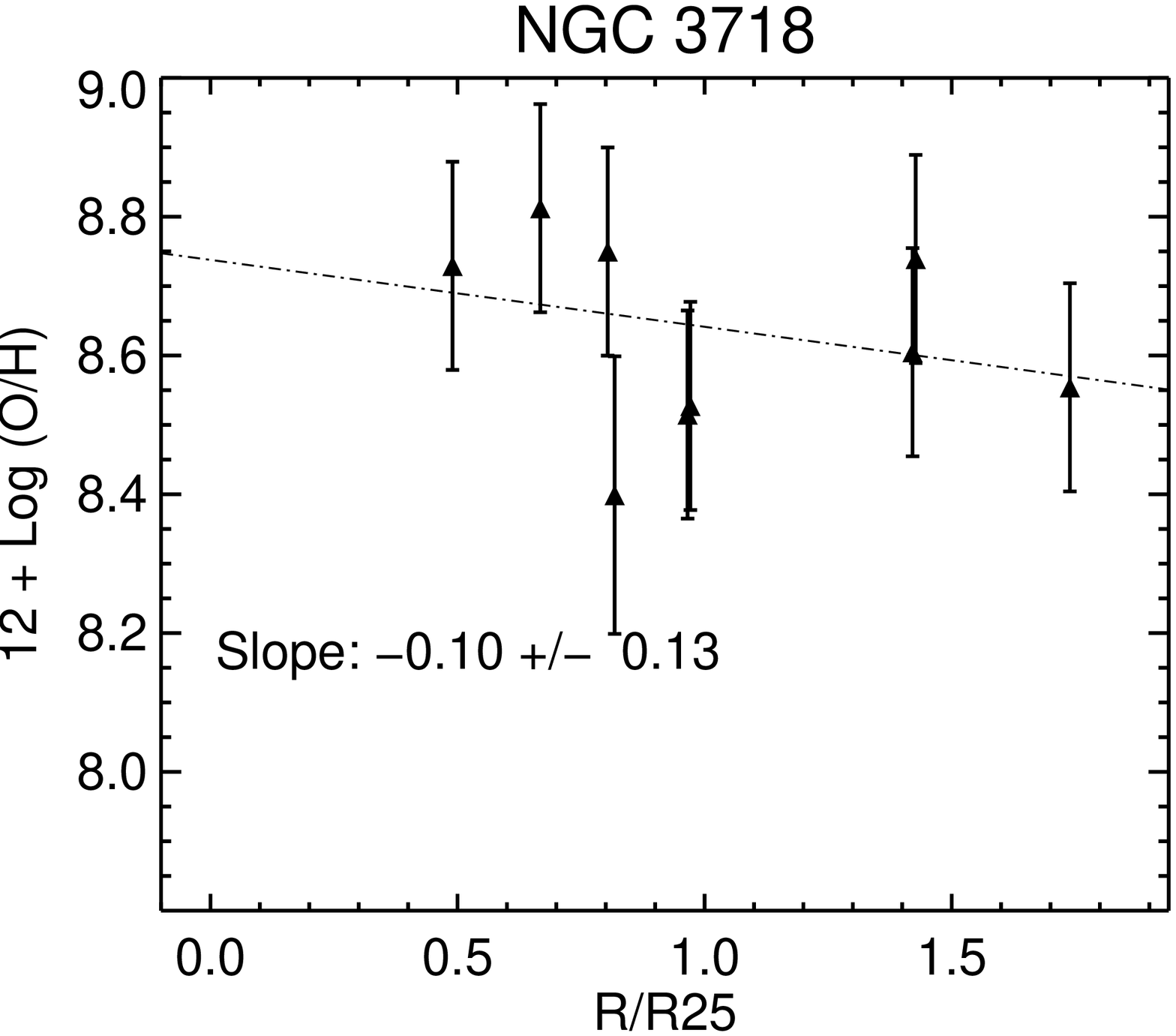}}
\subfigure{ 
\label{fig:radgrads:i} 
\includegraphics[width=0.48\linewidth]{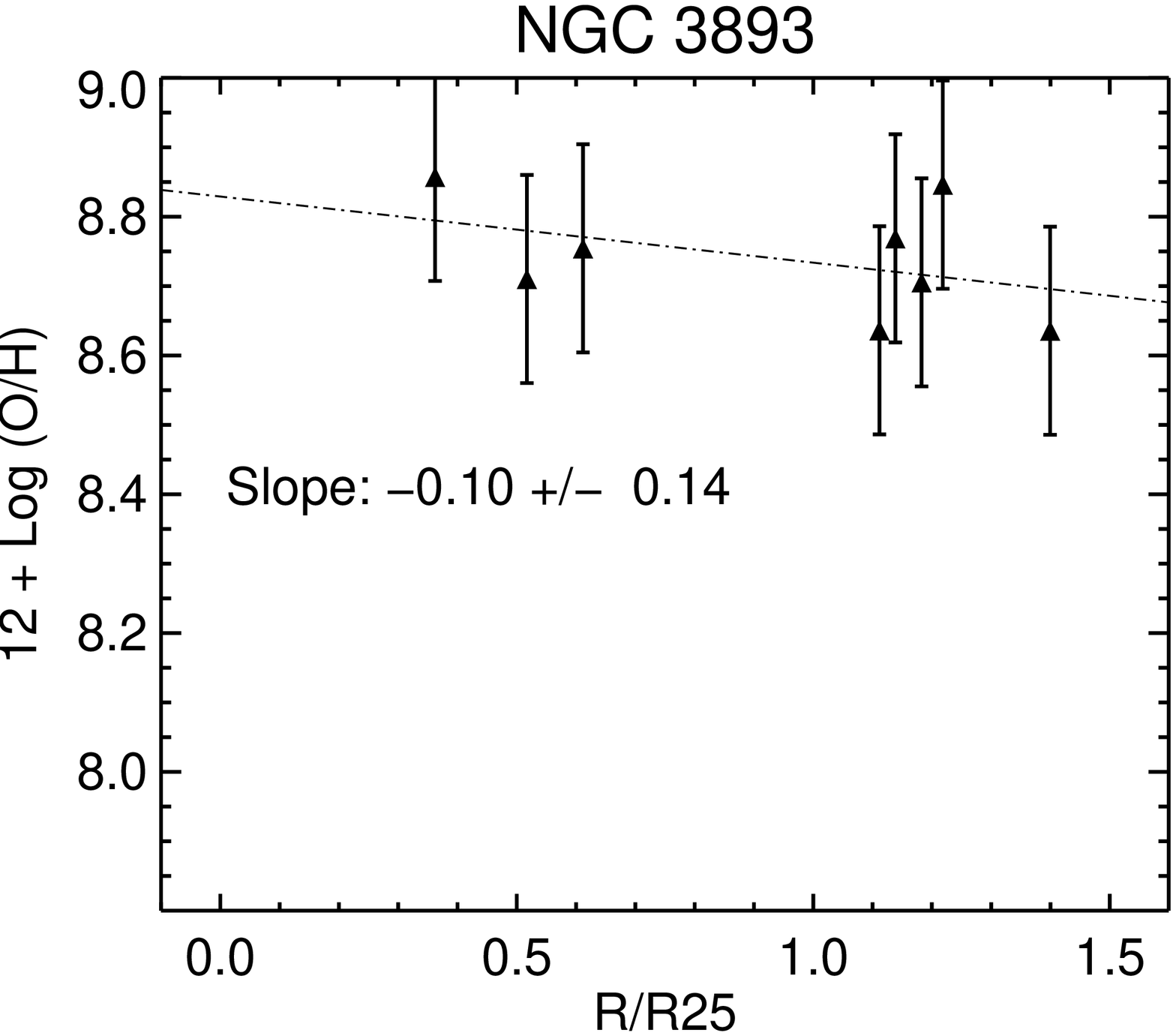}} 
\subfigure{ 
\label{fig:radgrads:j} 
\includegraphics[width=0.48\linewidth]{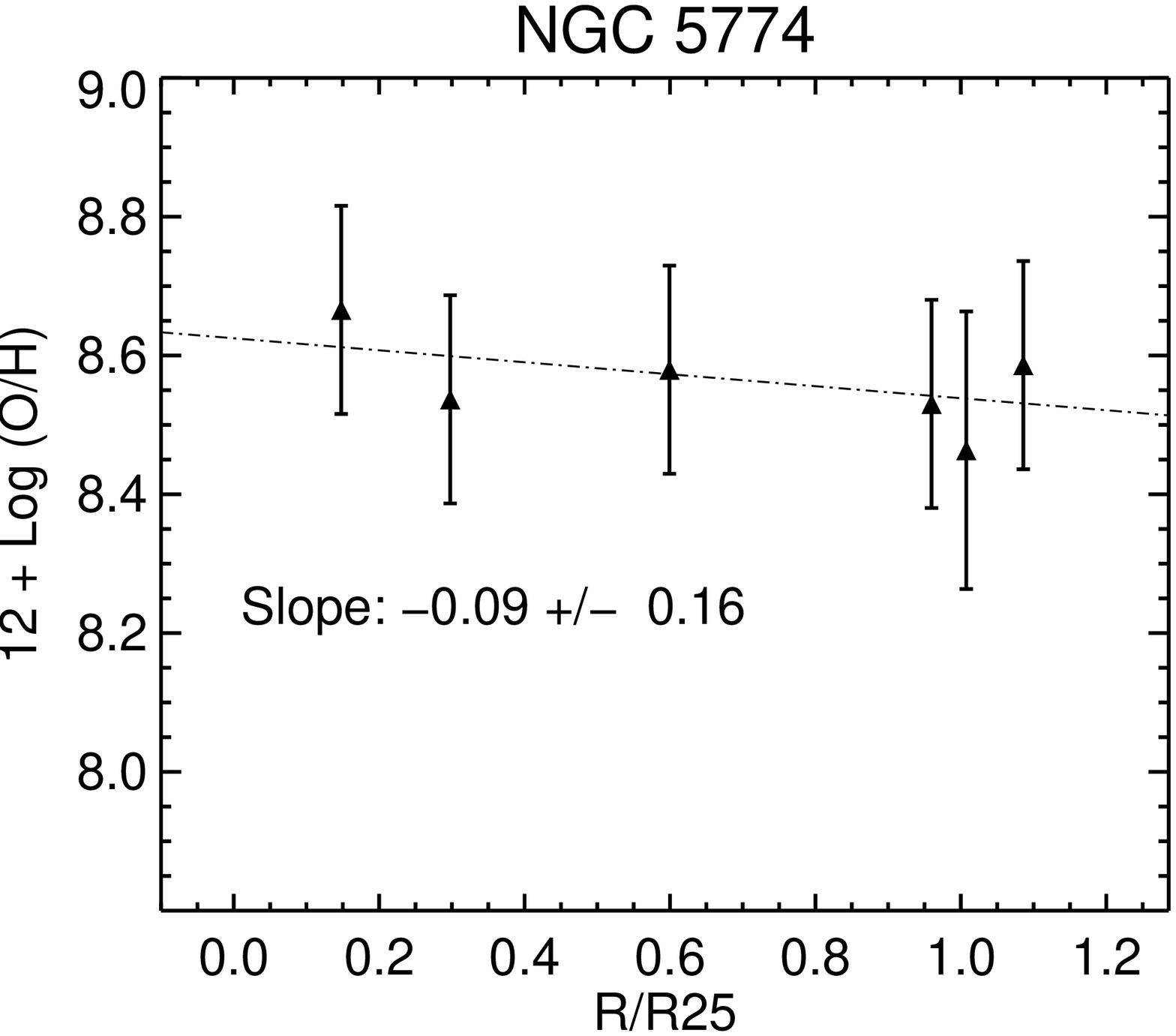}} 
\subfigure{ 
\label{fig:radgrads:k} 
\includegraphics[width=0.48\linewidth]{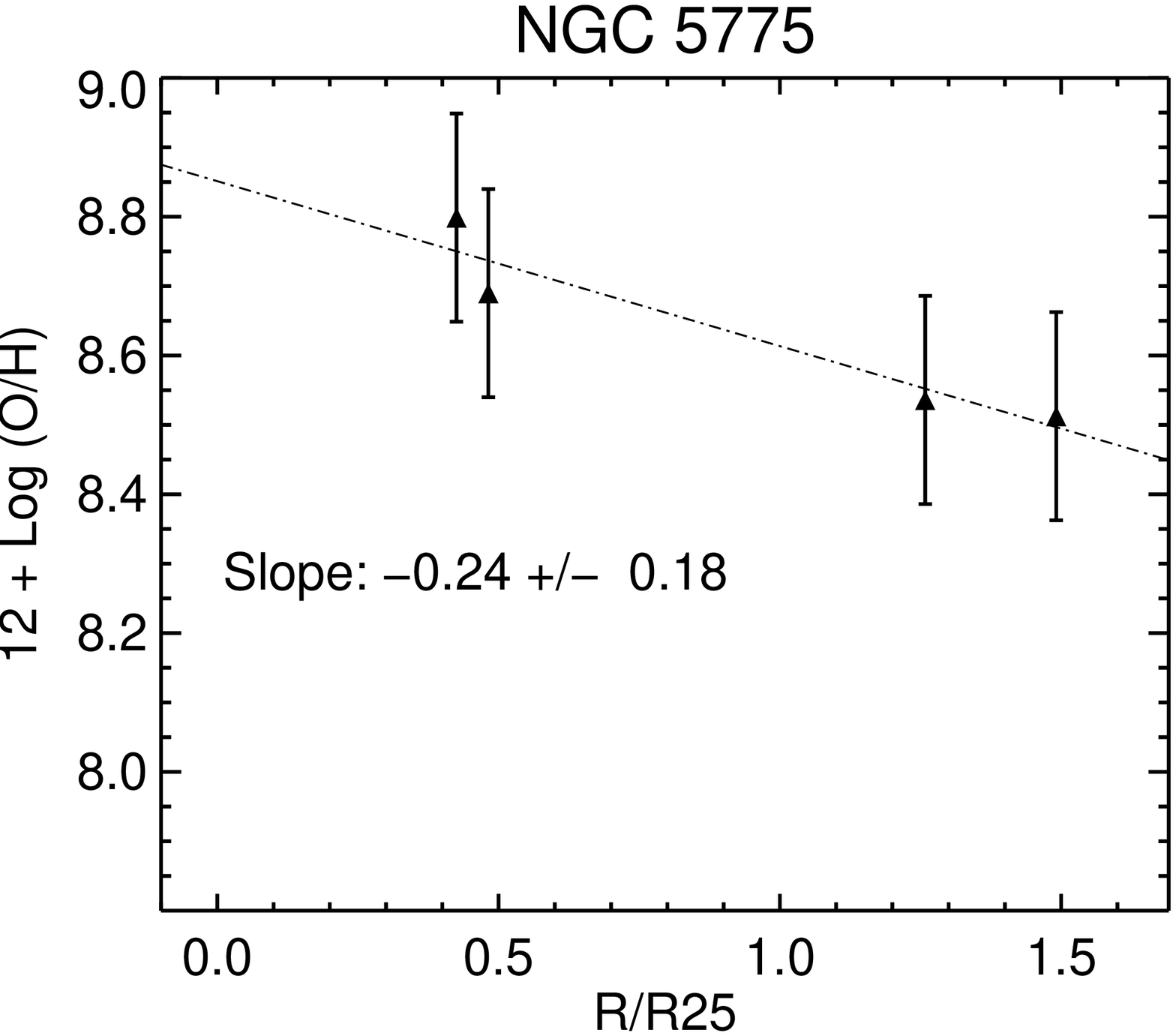}} 
\subfigure{ 
\label{fig:radgrads:l} 
\includegraphics[width=0.48\linewidth]{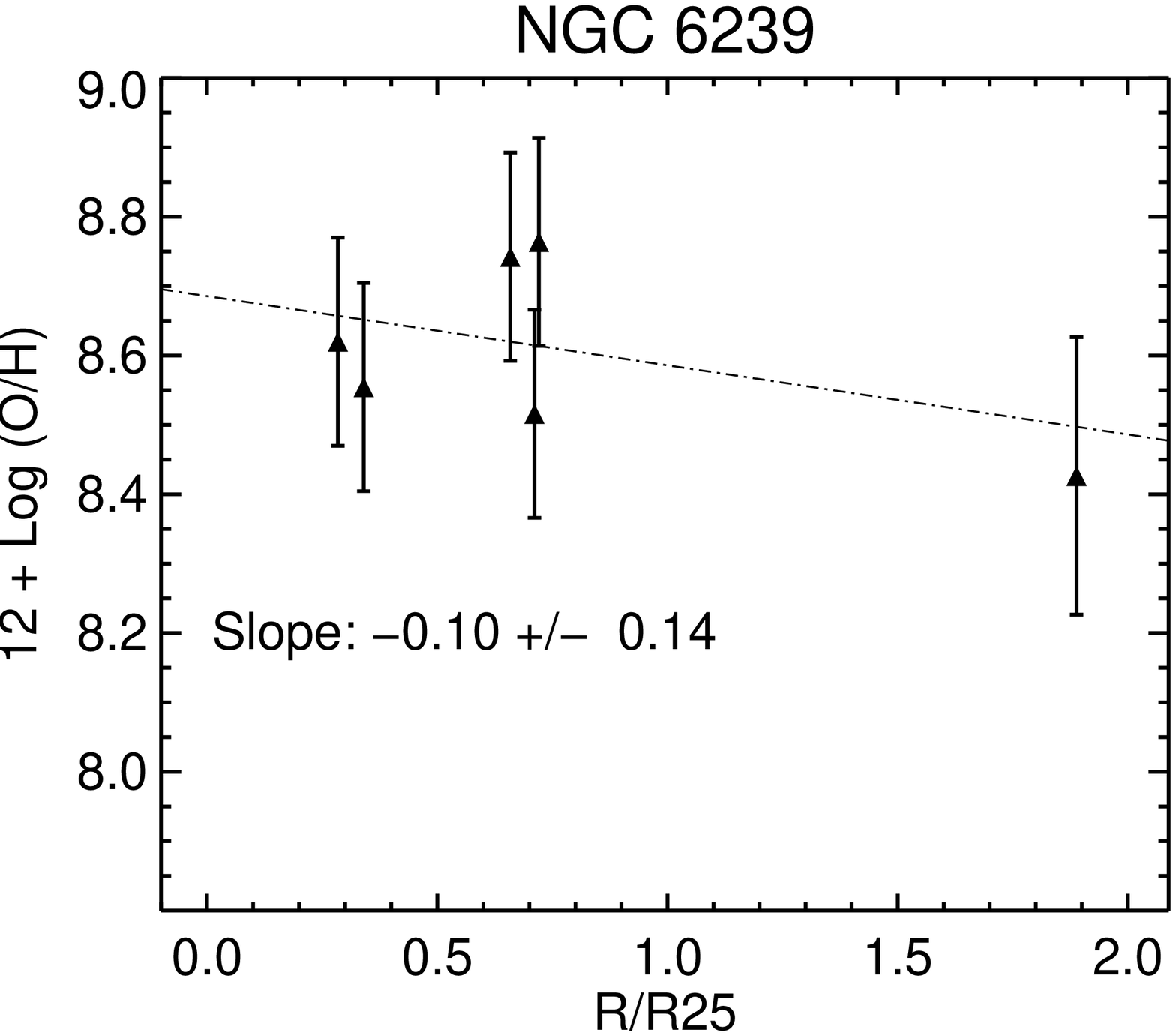}} 
\captcont{The dashed-dotted line represents a linear-least-squares fit to the abudance data, the slope (and its associated $\sigma$) of which is given on the plot. In two cases where we were unable to break the R23 degeneracy, NGC 2146 and NGC 3239, we plot fits to both upper (upward-facing, open triangles) and lower (downward-facing, open triangles) branch values.  In all cases except these two, we also put the gradients on the same y-axis for ease of comparison. (cont'd)} 
\end{figure*} 
\begin{figure*} [htp!]
\centering 
\subfigure{ 
\label{fig:radgrads:m} 
\includegraphics[width=0.48\linewidth]{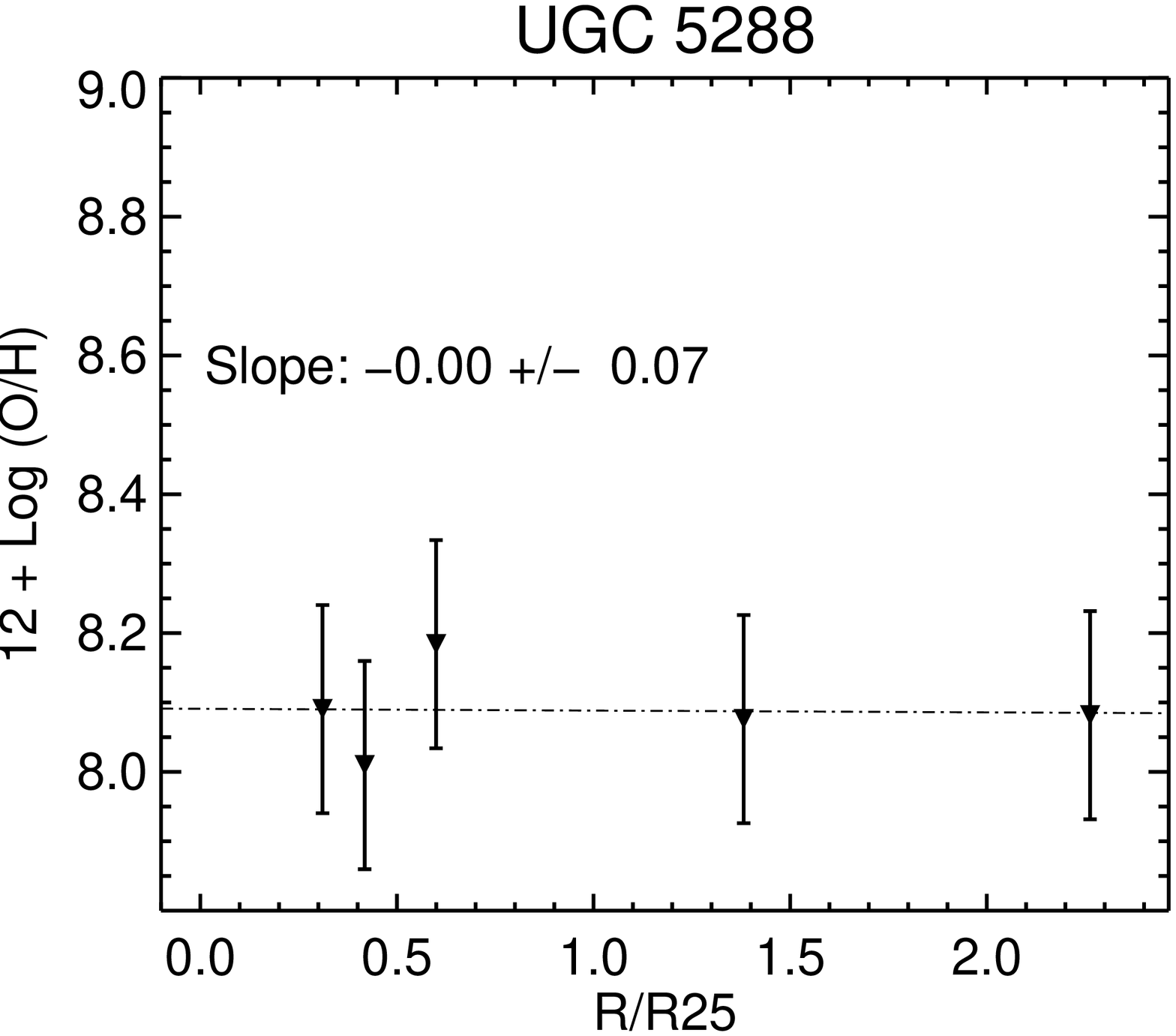}} 
\subfigure{ 
\label{fig:radgrads:n} 
\includegraphics[width=0.48\linewidth]{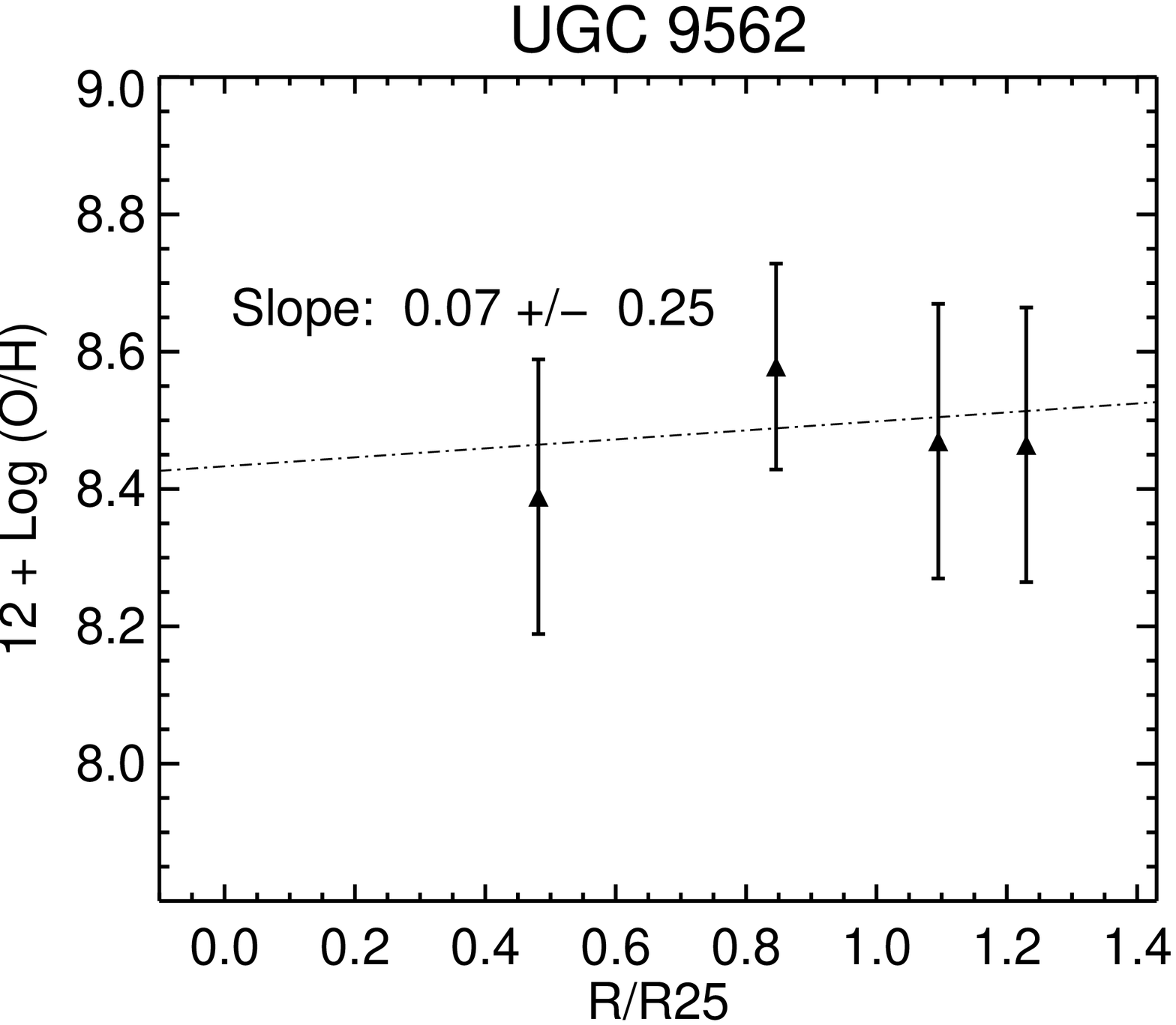}}
\caption{The filled circles with larger error bars on the plots of NGC 2146 and NGC 2782 show PP04-derived oxygen abundances for reference. We could determine only PP04 abundances for these few HII regions that fell on the very edge of the GMOS slit masks.}
\label{fig:radgrads} 

\end{figure*}


Here, we report radial oxygen abundance distributions for our sample of 13 HI Rogues with Gemini GMOS-N spectral data.  In this section, we consider each galaxy separately, and comment on its environment (level of interaction), HI morphology, and the distribution of star formation.  For reference, HI column density contours, where available, are shown in Figure \ref{fig:slitsandgas} for nine of the 13 galaxies discussed here. The HI maps for those galaxies for which we were not able to construct our own zeroth moment maps are available for viewing in the HI Rogues Gallery \citep{rogues}. 

Figure \ref{fig:radgrads} shows the M91 R23 oxygen abundance, 12 + Log (O/H), versus the galactocentric projected distances (in R/R$_{25}$) for the HII regions in each galaxy. \cite{mcgaugh91} reports several different values of systematic error associated with his strong line abundance calibration for {\emph{zero-age}} HII regions, depending on the resultant oxygen abundance. He finds the systematic errors to be $\sim$ 0.1 dex, 0.05 dex, and 0.20 dex on the upper branch, lower branch, and within 0.1 dex of the turnover, respectively.  Additional sources of error come from age effects, which can change the ionization parameter and shape of the ionizing spectrum \citep{stasinska96}, and geometrical effects, which give rise to a large dispersion in ionization parameters \citep{ercolano07}. The overall impact of these effects increases the systematic errors of the M91 R23 calibration to 0.1 $-$ 0.3 dex, on average, regardless of branch, in the worst-case scenario. We therefore adopt 0.15 dex as an average systematic error in our abundance measurements, unless a value is within 0.1 dex of the turnover, where it is then estimated to be 0.2 dex. 

The dash-dotted line on each plot represents an error-weighted linear-least-squares fit to the data. For reference, the solar oxygen abundance used in the M91 grid is 12 + Log (O/H)$_{\odot}$ $=$ 8.92, and the turnover of the R23 relation occurs at 12 + Log (O/H) = 8.35 \citep{mcgaugh91}. In the cases of NGC 2146 and NGC 3239, we have not been able to break the degeneracy of the R23 relation for the majority of the HII regions, and we plot both upper-branch and lower branch R23 abundances. On the gradients of NGC 2146 and NGC 2782, we show oxygen abundance values for a few HII regions derived from the PP04 calibration. We use different symbols for these abundances and larger error bars (0.4 dex), and caution against a direct comparison with the R23 abundances since the different calibrations are systematically offset from each other. We show these few points merely as a rough reference point where no other data are available. 

Examining dispersion in measurements relative to sample size,  \cite{zaritsky94} find that 5 HII regions per galaxy should serve as a minimum on which to base reliable measurements of radial abundance gradients. For reference, we measure at least 5 HII regions per galaxy in all but NGC 5775 and UGC 9562. We measure the oxygen abundances of only 4 HII regions in these galaxies.  The range of  projected radial distances over which we measure oxygen abundance gradients is generally between 0.4 and 1.5 R/R$_{25}$, though several gradients extend beyond 2 R/R$_{25}$. Tables \ref{tab:prop2146} through \ref{tab:prop9562} provide all of the relevant data on the HII regions for which we determine abundances in every galaxy: RA and declination,  R/R$_{25}$, the projected distance from the nearest galaxy center in kpc, H$\alpha$ fluxes and R$-$band magnitudes from MDM narrow-band images, oxygen emission-line fluxes relative to H$\beta$ = 100, and the oxygen abundance (upper and lower branch values, where appropriate). 

\begin{itemize}
\item{{\bf{NGC 2146}} is classified in the HI Rogues catalog as a minor merger and has HI streams extending to the north and south out to 6 Holmberg radii \citep{taramopoulos01}.  These streams are thought to be due to a tidal interaction between NGC 2146 and an unseen LSB companion approximately one Gyr ago.   The outlying HII regions in this system are embedded within the HI  of this system and are within the $3-10 \times 10^{19}$ \cm~column density contours.  The total HI mass of NGC 2146 is $2.1 \times 10^{10}$ \Msun, with $4.2 \times 10^9$ \Msun~coming from within R$_{25}$ and $1.7 \times 10^{10}$ \Msun~coming from the extended component. All 9 HII regions in NGC 2146 appear to lie on a single extended spiral arm that begins just to the SE of the central burst of star formation, turns north, and then ends at the outlying HII regions to the SW of the galaxy center.  Regions 12, 13, and 14, all centrally-located, lie at the edge of the CCD in the GMOS observations, and we therefore measure the oxygen abundance of these HII regions using only the PP04 N2 index calibration (shown as filled circles in Figure 2). The dual-valued oxygen abundance for this galaxy complicates the interpretation of the gradient. If the outlying HII regions lie on the upper branch, then we find little to no abundance gradient ($-$0.16 dex per R/R$_{25}$)  for this galaxy. However, if the outlying HII regions have lower-branch oxygen abundances, then we find a very steeply declining gradient ($-$0.79 dex per R/R$_{25}$) at large galactocentric radii.   }


\begin{table*} \centering \scriptsize
\begin{tabular*}{1.0\textwidth}{@{\extracolsep{\fill}}lcccrccccc}
\hline
\footnotesize
ID & RA  & dec & R/R$_{25}$ & R$_{proj}$ & F$_{H\alpha}$ & m$_{R}$ & [OII] & [OIII] & 12 $+$ Log(O/H) \\
\hline
(1) & (2) & (3) & (4) & (5) & (6) & (7) & (8) & (9) & (10) \\
\hline
\hline
  5 &  06 18 14.9 &  +78 19 48.4 &  1.5  &13.12 &  4.13$\pm$.49 & 21.49$\pm$.08 &  226 &  147 &  7.76/8.80 \\
  4 &  06 18 21.8  & +78 19 38.2 &  1.4  &12.94 &  1.11$\pm$.13 & 22.93$\pm$.19 &  243 &  466 &  8.04/8.56 \\
  7 &  06 18 13.4  & +78 20 18.1 &  1.3  &11.02 &  1.94$\pm$.23 & 22.02$\pm$.10 &  447 &  130 &  8.19/8.57 \\
  8 &  06 18 09.1 &  +78 21 02.4 &  1.1  & 9.91 &  2.98$\pm$.35 & 22.14$\pm$.11 &  328 &  137 &  7.97/8.70 \\
  9 &  06 18 04.4  & +78 21 23.3 &  1.1  &11.13 &  1.66$\pm$.19 & 20.64$\pm$.07 &  267 &   64 &  7.84/8.82 \\
 14 &  06 19 00.5  & +78 20 44.1 &  0.5  & 8.68 &  3.57$\pm$.42 & 21.22$\pm$.08 &    *** &    *** &  8.61* \\
 13 &  06 18 51.0  & +78 21 31.7 &  0.4  & 4.28 &  8.89$\pm$.06 & 19.48$\pm$.07 &    *** &    *** &  8.61* \\
 12 &  06 18 44.7  & +78 21 50.4 &  0.4  & 3.40 & 28.14$\pm$.37 & 19.71$\pm$.07 &    *** &    *** &  8.54* \\
 15 &  06 18 52.5  & +78 20 37.8 &  0.3  & 7.09 &  4.44$\pm$.53 & 22.94$\pm$.19 &  161 &   79 &  8.92 \\
\hline
\end{tabular*}
\caption[HII Region Properties for NGC 2146]{HII Region properties for NGC 2146, sorted by R/R$_{25}$. Description of columns applies to tables 4 $-$ 16. (1) The slit mask ID for the HII region. (2) Right Ascension (J2000) in hours, minutes, seconds. (3) Declination (J2000) in degrees, minutes, seconds. (4) Projected distance from the 25$^{th}$ magnitude elliptical isophote. (5) Projected distance in kpc from the center of the galaxy. (6) H$\alpha$ flux and associated error from MDM 2.4-m images in units of 10$^{-15}$ergs s$^{-1}$ cm$^{-2}$ (7) R$-$band apparent magnitude and associated error  from MDM 2.4-m images in AB magnitudes. We use the 5$\sigma$ limiting magnitude for a point source in each image when an HII region is undetected in the R$-$band(8) and (9) The reddening-corrected [OII] $\lambda\lambda3727$ and [OIII] $\lambda4959$ and $\lambda5007$ summed emission line strengths, relative to H$\beta$ of 100.  Reddening correction was done for case B recombination with T = 10000 K and n$_{e}$ = 100 cm$^{-3}$, where H$\alpha$/H$\beta$ = 2.86 and H$\gamma$/H$\beta$ = 0.459. (10) The strong-line oxygen abundance, 12$+$Log(O/H), from the M91 calibration of the R23 relation described in Section \ref{sec:rogues:anal}. In cases where there is one value, we have used the [NII]/H$\alpha$ to break the R23 degeneracy. In cases where there are two values, the [NII]/H$\alpha$ either falls in the ambiguous range (0.05 $-$ 0.08) or was not available. The first value is the lower-branch R23 oxygen abundance, the second is the upper branch R23 oxygen abundance. The error associated with strong-line abundances is notoriously large, ranging from 0.15 to 0.2 dex , as discussed in Section \ref{sec:rogues:anal}. In a few cases  (for just NGC 2146 and NGC 2782), values for 12$+$Log(O/H) marked with an asterisk indicate that the GMOS multi-slit spectra were red enough to obtain H$\alpha$ and [NII] $\lambda6583$, but not the blue lines.  In these few cases, we used the [NII]/H$\alpha$ method from PP04 to determine the oxygen abundance. These values cannot be compared directly with R23 values because of systematic offsets of the calibration methods. \label{tab:prop2146}}
\end{table*}

\item{{\bf{NGC 2782}} is classified as a minor merger in the HI Rogues catalog, with a one-sided HI tail on the opposite side of an extension of optical emission.  The outlying HII regions are located within this HI tail, with column densities between $10^{19-20}$ \cm.  The tail contains $3.4\times10^9$ \Msun~ or $\sim$80\% of the galaxy's HI mass and is noted to consist of $\sim$10 dwarf galaxy-sized clumps of $\sim10^8$\Msun.  The basic properties of NGC 2782 can be reproduced with the collision of two disk galaxies with a 1:4 mass ratio \citep{smith94}.   The three HII regions in the extended tidal tail of NGC 2782 (regions 1, 2, and 3) have the same oxygen abundance as the two HII regions closer to the center of the galaxy (regions 4 and 6), of 12 + Log (O/H) $\sim$8.6. Regions 4 and 6 lie on the edge of the GMOS CCD, and we are able to use only the N2 index of PP04 to get rough estimates of their oxygen abundances, which are roughly consistent with the R23 values of \cite{moustakas06} from the central nebular SDSS spectrum that give 12 + Log (O/H) = 8.80. }



\begin{table*}[h!]\centering \scriptsize
\begin{tabular*}{1.0\textwidth}{@{\extracolsep{\fill}}lcccrccccc}
\hline
\footnotesize
ID & RA  & dec & R/R$_{25}$ & R$_{proj}$ & F$_{H\alpha}$ & m$_{R}$ & [OII] & [OIII] & 12 $+$ Log(O/H) \\
\hline
(1) & (2) & (3) & (4) & (5) & (6) & (7) & (8) & (9) & (10) \\
\hline
\hline
 1 &  09 13 50.9  & +40 08 16.4 &  2.6  &33.51 &  0.44$\pm$.05 & 23.07$\pm$.30 &  391 &   80 &  8.66* \\
  2 &  09 13 51.2  & +40 08 06.8 &  2.5  &32.14 &  1.71$\pm$.20 & 22.46$\pm$.19 &  338 &   84 &  8.72* \\
  3 &  09 13 52.3 & +40 07 46.2 &  2.2  &28.59 &  0.44$\pm$.05 & $>$23.72$\pm$.36 &  462 &  112 &  8.56 \\
  4 &  09 13 58.8 & +40 06 08.8 &  1.2  &15.00 &  0.28$\pm$.03 & $>$23.72$\pm$.36 &    *** &    *** &  8.55 \\
  6 &  09 14 01.4 & +40 06 36.3 &  0.6  & 7.99 &  1.40$\pm$.16 & 21.96$\pm$.13 &    ***&    *** &  8.47 \\

 \hline
\end{tabular*}
\caption[HII Region Properties: NGC2782]{HII Region Properties: NGC 2782. Same as Table \ref{tab:prop2146}, for NGC 2782}
\end{table*}

\item{{\bf{NGC 3226/7}}  is classified as an interacting double system in which only one of the galaxies is gas-rich in the HI rogues catalog.  NGC 3227 is the gas-rich galaxy and also a nearby Seyfert \citep{mundell01}.  HI tails extend 30 kpc to the south and 70 kpc to the north, beyond the region shown in our optical images. The total HI mass of the system is $2.0 \times 10^9$ \Msun, with $\sim1.0 \times 10^9$ \Msun~ of this mass located beyond R$_{25}$. HII regions 24 and 25 are found within a cloud of HI that is thought to be a dwarf galaxy, while the outermost HII regions are found in outer lower density gas with column densities of a few $\times 10^{19}$ \cm.

HII regions 10 and 12 are associated with the early-type galaxy NGC 3226, while the rest of the HII regions in this system appear to be associated with NGC 3227. R$_{25}$ for this particular system was very difficult to obtain, given the overlap of the isophotes of the two merging galaxies. Instead of fitting R$_{25}$ for each galaxy, we fit R$_{25}$ for both galaxies together, which somewhat downplays the extended star formation in this system. In a by-eye fit of  R$_{25}$ to NGC 3227 alone, Region 5 appears to lie at R/R$_{25}$ closer to 1.5, and regions 24 and 25 are probably both further outside of R$_{25}$. Nonetheless, within the uncertainties, the outlying HII regions of this system appear to be about as oxygen-rich as those closer to the center of the system. }


 
\begin{table*}[htp!]
\scriptsize
\begin{tabular*}{1.0\textwidth}{@{\extracolsep{\fill}}lcccrccccc}
\hline
\footnotesize
ID & RA  & dec & R/R$_{25}$ & R$_{proj}$ & F$_{H\alpha}$ & m$_{R}$ & [OII] & [OIII] & 12 $+$ Log(O/H) \\
\hline
(1) & (2) & (3) & (4) & (5) & (6) & (7) & (8) & (9) & (10) \\
\hline
\hline
 5 &  10 23 35.1 & +19 53 07.9 &  1.1  & 9.08 &  0.57$\pm$.06 & 22.75$\pm$.16 &  402 &  115 &  8.63 \\
 24 &  10 23 25.2 & +19 51 44.4 &  1.0  & 7.95 &  7.21$\pm$.86 & 20.61$\pm$.07 &  330 &  141 &  8.69 \\
 25 &  10 23 26.0 & +19 51 40.8 &  0.9  & 7.57 &  1.28$\pm$.15 & 22.13$\pm$.10 &  484 &  127 &  8.53 \\
 10 &  10 23 26.4 & +19 54 14.5 &  0.5  &10.04 &  0.87$\pm$.10 & 20.93$\pm$.07 &  461 &   85 &  8.58 \\
 12 &  10 23 26.1  & +19 54 01.9 &  0.5  & 9.03 &  0.93$\pm$.11 & 19.45$\pm$.07 &  446 &  124 &  8.57 \\
 22 &  10 23 32.9  & +19 51 54.3 &  0.4  & 7.39 &  1.42$\pm$.17 & 21.15$\pm$.07 &  234 &   50 &  8.87 \\
 15 &  10 23 26.4  & +19 52 54.2 &  0.3  & 3.83 &  1.94$\pm$.23 & 22.12$\pm$.10 &  182 &   33 &  8.94 \\

 \hline
\end{tabular*}
\caption[HII Region Properties: NGC3227]{HII Region Properties: NGC 3227. Same as Table \ref{tab:prop2146}, for NGC 3227.}
\end{table*}

\item{{\bf{NGC 3239}} is classified as an interacting double in which both of the galaxies are HI rich with two HI tails according to the HI rogues catalog \citep{iyer01}.   In the direction of the extended arm of HII regions shown in our optical images, the gas does not extend to a significantly larger radii, though there is a large HI feature on the opposite side of the galaxy.  The outermost HII regions lie in gas that has a column density of  10$^{20}$ cm$^{-2}$. The HI mass of this galaxy is $1.2 \times 10^{9}$ \Msun~ \citep{huchtmeier89}, about 50\% of which lies beyond R$_{25}$. Because the N2 indices for this galaxy lie in the ambiguous region (i.e. putting their resultant oxygen abundances near the turnover region of R23) we are unable to constrain the true abundance gradient. If we make the assumption that all outlying HII regions are either upper or lower branch, then the resultant abundance gradient is flat regardless of which branch is chosen. However, one can imagine a mixture of upper and lower branch oxygen abundances that would result in either increasing or decreasing gradients as well. }



\begin{table*}[htp]\centering \scriptsize
\begin{tabular*}{1.0\textwidth}{@{\extracolsep{\fill}}lcccrccccc}
\hline
\footnotesize
ID & RA  & dec & R/R$_{25}$ & R$_{proj}$ & F$_{H\alpha}$ & m$_{R}$ & [OII] & [OIII] & 12 $+$ Log (O/H) \\
\hline
(1) & (2) & (3) & (4) & (5) & (6) & (7) & (8) & (9) & (10) \\
\hline
\hline
 4 &  10 25 09.4  & +17 07 19.9 &  1.3  & 5.23 &  1.10$\pm$.13 & 21.69$\pm$.08 &  360 &  172 &  8.03/8.64 \\
  5 &  10 24 56.2 & +17 10 13.3 &  1.1  & 5.17 &  0.27$\pm$.03 & 23.54$\pm$.21 &  363 &   47 &  8.08/8.72 \\
  8 &  10 24 57.1  & +17 10 20.4 &  1.0  & 4.88 &  0.88$\pm$.10 & 23.20$\pm$.17 &  474 &  296 &  8.27/8.44 \\
 13 &  10 25 07.3  & +17 07 41.1 &  1.0  & 3.92 &  0.89$\pm$.10 & 22.48$\pm$.10 &  280 &  400 &  8.06/8.56 \\
 24 &  10 25 03.0  & +17 07 52.6 &  0.8  & 3.19 &  2.65$\pm$.31 & 21.77$\pm$.08 &  262 &  324 &  7.97/8.63\\
 14 &  10 25 03.3 & +17 08 05.8 &  0.7  & 2.65 &  0.35$\pm$.04 & 23.15$\pm$.16 &  404 &  166 &  8.10/8.59 \\
 18 &  10 24 57.0 & +17 09 05.1 &  0.7  & 4.13 &  0.59$\pm$.07 & 22.97$\pm$.14 &  394 &   43 &  8.15/8.68 \\
 15 &  10 25 00.8 & +17 08 23.9 &  0.6  & 2.73 &  3.58$\pm$.42 & 20.77$\pm$.07 &  361 &  389 &  8.18/8.49 \\
 22 &  10 25 02.7 & +17 10 00.8 &  0.6  & 2.13 &  2.53$\pm$.30 & 21.07$\pm$.07 &  204 &  231 &  7.78/8.75\\
 17 &  10 24 58.8 & +17 09 29.3 &  0.6  & 3.15 &  2.63$\pm$.31 & 20.94$\pm$.07 &  348 &  331 &  8.12/8.54\\
 21 &  10 25 06.8 & +17 09 57.5 &  0.5  & 2.29 &  3.77$\pm$.45 & 20.84$\pm$.07 &  455 &  189 &  8.20\\
 19 &  10 25 00.2  &+17 09 38.7 &  0.5  & 2.53 &  3.01$\pm$.36 & 20.75$\pm$.07 &  446 &  215 &  8.21 \\

 \hline
\end{tabular*}
\caption[HII Region Properties: NGC3239]{HII Region Properties: NGC 3239. Same as Table \ref{tab:prop2146}, for NGC 3239}
\end{table*}

\item{{\bf{NGC 3310}} is classified as a minor merger in the HI rogues catalog.  The gas disk of the main galaxy shows signs of perturbation from its high velocity dispersion and tails that extend 23 kpc to the north and 51 kpc to the south of the galaxy \citep{kregel01}.  The HI mass of the entire system is $5.2 \times 10^9$ \Msun, with 4.2 and 5.0 $\times 10^8$ \Msun~ in the northern and southern tails respectively.  Only the northern tail was imaged in H$\alpha$ and contains the HII regions presented here.   The peak column densities in the northern tail are $10^{20-21}$ \cm. Most of the 11 HII regions in this galaxy lie in an extended arm-like structure that extends to the northwest of NGC 3310, and all have the same oxygen enrichment of 12 + Log (O/H) $\sim$ 8.6. }
\label{sec:rogues:3310}

\begin{table*}[htp]\centering \scriptsize
\begin{tabular*}{1.0\textwidth}{@{\extracolsep{\fill}}lcccrccccc}
\hline
\footnotesize
ID & RA  & dec & R/R$_{25}$ & R$_{proj}$ & F$_{H\alpha}$ & m$_{R}$ & [OII] & [OIII] & 12 $+$ Log(O/H) \\
\hline
(1) & (2) & (3) & (4) & (5) & (6) & (7) & (8) & (9) & (10) \\
\hline
\hline
 11 &  10 38 27.2 & +53 31 58.6 &  1.8  &16.54 &  0.06$\pm$.00 & $>$23.79$\pm$.35 &  319 &   90 &  8.74 \\
  7 &  10 38 31.0 & +53 31 46.5 &  1.5  &13.54 &  0.33$\pm$.03 & 23.01$\pm$.19 &  302 &   88 &  8.76 \\
 16 &  10 38 35.1 & +53 32 03.9 &  1.4  &12.46 &  0.51$\pm$.06 & 22.83$\pm$.17 &  388 &  429 &  8.44 \\
 12 &  10 38 32.4 & +53 31 25.7 &  1.3  &11.54 &  1.40$\pm$.16 & 21.54$\pm$.08 &  421 &  184 &  8.56 \\
 14 &  10 38 33.8 & +53 31 16.7 &  1.1  &10.19 &  1.43$\pm$.17 & 22.39$\pm$.12 &  411 &  220 &  8.55 \\
 15 &  10 38 34.6 & +53 31 12.3 &  1.0  & 9.49 &  2.40$\pm$.28 & 21.74$\pm$.09 &  341 &  111 &  8.70 \\
 18 &  10 38 36.8 & +53 31 06.1 &  0.9  & 7.80 &  1.30$\pm$.15 & 21.90$\pm$.09 &  464 &  173 &  8.52 \\
 21 &  10 38 39.6 & +53 30 55.4 &  0.6  & 5.53 &  2.18$\pm$.26 & 20.84$\pm$.07 &  275 &  219 &  8.69 \\
 24 &  10 38 41.3 & +53 31 01.7 &  0.6  & 5.27 &  6.52$\pm$.78 & 21.06$\pm$.07 &  127 &  664 &  8.57 \\
 26 &  10 38 44.9 & +53 30 57.2 &  0.5  & 4.34 &  5.85$\pm$.70 & 21.15$\pm$.07 &  288 &  497 &  8.49 \\
 25 &  10 38 43.1 & +53 30 43.8 &  0.4  & 3.27 &  8.34$\pm$.00 & 20.75$\pm$.07 &  140 &  540 &  8.62 \\
  \hline
\end{tabular*}
\caption[HII Region Properties: NGC3310]{HII Region Properties: NGC 3310. Same as Table \ref{tab:prop2146}, for NGC 3310}
\end{table*}

\item{{\bf{NGC 3359}} is categorized as a galaxy with a detached HI cloud in the HI rogues catalog, but from the deeper contours it appears that the cloud may be a further extension of the extended HI spiral arm to the north of the galaxy \citep{boonyasait01}.  The HII regions we study here are located within the extended gaseous envelope of the galaxy, but not this far extended arm.  NGC3359 contains $4.5 \times 10^9$ \Msun~ in neutral hydrogen gas, and we estimate that about half of it is contained within R$_{25}$. 

The 16 HII regions we consider in NGC 3359 all have similar oxygen abundances near 12 + Log (O/H) $\sim$ 8.6. \cite{martin95} report oxygen abundances for 77 HII regions within R$_{25}$ for this galaxy, and find a steeply decreasing oxygen gradient in the centermost part of the galaxy, out to $\sim$0.35 R$_{25}$, and then a flat oxygen abundance gradient that levels off at 12 + Log (O/H) = 8.6 from 0.35 to 0.9 R$_{25}$. Our results are consistent with this flat gradient, and indicate that it extends to even larger radii, out to 1.5 R$_{25}$.}



\begin{table*} [htp]\centering \scriptsize
\begin{tabular*}{1.0\textwidth}{@{\extracolsep{\fill}}lcccrccccc}
\hline
\footnotesize
ID & RA  & dec & R/R$_{25}$ & R$_{proj}$ & F$_{H\alpha}$ & m$_{R}$ & [OII] & [OIII] & 12 $+$ Log(O/H) \\
\hline
(1) & (2) & (3) & (4) & (5) & (6) & (7) & (8) & (9) & (10) \\
\hline
\hline
  6 &  10 46 10.6 & +63 14 26.8 &  1.5  &16.65 &  0.26$\pm$.02 &  $>$23.16$\pm$.36 &  269 &  209 &  8.71 \\
 12 &  10 46 30.9 & +63 16 49.3 &  1.4  &18.05 &  0.55$\pm$.06 & 22.32$\pm$.19 &  434 &  164 &  8.56 \\
 11 &  10 46 26.6 & +63 16 28.4 &  1.4  &17.08 &  0.95$\pm$.11 & 22.43$\pm$.21 &  206 &  320 &  8.69 \\
  7 &  10 46 20.1 & +63 15 24.0 &  1.2  &14.48 &  0.45$\pm$.05 & $>$23.16$\pm$.36 &  291 &   51 &  8.80 \\
  8 &  10 46 24.0 & +63 15 48.2 &  1.2  &14.70 &  0.41$\pm$.04 & $>$23.16$\pm$.36 &  230 &  227 &  8.73 \\
 13 &  10 46 37.9 & +63 16 20.2 &  1.2  &15.09 &  1.85$\pm$.22 & 21.54$\pm$.11 &  454 &  182 &  8.53 \\
 14 &  10 46 39.6 & +63 16 10.3 &  1.1  &14.28 &  7.59$\pm$.91 & 20.75$\pm$.08 &  158 &  555 &  8.59 \\
 17 &  10 46 26.7 & +63 15 05.4 &  0.9  &10.74 &  3.99$\pm$.47 & 21.18$\pm$.09 &  425 &  213 &  8.54 \\
 16 &  10 46 37.6 & +63 15 32.4 &  0.8  &10.96 &  6.95$\pm$.83 & 20.94$\pm$.08 &  221 &  525 &  8.54 \\
 19 &  10 46 51.1 & +63 14 13.9 &  0.7  & 8.86 &   0.78$\pm$0.08 & $>$23.16$\pm$.36 &  210 &  244 &  8.74 \\
 28 &  10 46 28.6 & +63 12 06.9 &  0.7  & 8.60 & 12.08$\pm$.44 & 19.74$\pm$.07 &  576 &  299 &  8.40 \\
 20 &  10 46 44.8 & +63 14 38.1 &  0.6  & 7.51 & 13.90$\pm$.66 & 20.19$\pm$.07 &  247 &  867 &  8.37 \\
 27 &  10 46 31.1 & +63 12 32.7 &  0.5  & 5.94 &  1.92$\pm$.22 & $>$23.16$\pm$.36 &  276 &  145 &  8.74 \\
 23 &  10 46 28.6 & +63 13 38.1 &  0.5  & 5.40 & 13.39$\pm$.60 & 20.27$\pm$.07 &  253 &  277 &  8.67 \\
 24 &  10 46 30.7 & +63 13 22.2 &  0.4  & 4.06 &  9.76$\pm$.17 & 20.53$\pm$.07 &  207 &  323 &  8.69 \\
 25 &  10 46 42.8 & +63 13 16.7 &  0.3  & 3.07 &  4.81$\pm$.57 & 20.78$\pm$.08 &  423 &  111 &  8.61 \\

  \hline
\end{tabular*}
\caption[HII Region Properties: NGC3359]{HII Region Properties: NGC 3359. Same as Table \ref{tab:prop2146}, for NGC 3359}
\end{table*}

\item{{\bf{NGC 3432}} is noted as a minor merger in the HI rogues catalog.  It has a strong warp and a companion to the southwest corresponding to the side with the most dominant warp \citep{swaters02}.  The total HI mass of the system is $6.0 \times 10^9$ \Msun, with $2.58 \times 10^9$ \Msun~ in the outer regions beyond R$_{25}$.  The HII regions studied here are in the northern part of the galaxy's gaseous disk, with some of the outermost HII regions lying in gas at 4 $\times$ 10$^{20}$ \cm. NGC 3432 appears to have a flat oxygen abundance gradient (within the large error) with a median value of 12 + Log (O/H) $\sim$ 8.5. }



\begin{table*}[htp]\centering \scriptsize
\begin{tabular*}{1.0\textwidth}{@{\extracolsep{\fill}}lcccrccccc}
\hline
\footnotesize
ID & RA  & dec & R/R$_{25}$ & R$_{proj}$ & F$_{H\alpha}$ & m$_{R}$ & [OII] & [OIII] & 12 $+$ Log(O/H) \\
\hline
(1) & (2) & (3) & (4) & (5) & (6) & (7) & (8) & (9) & (10) \\
\hline
\hline
  7 &  10 52 47.2 & +36 39 38.9 &  1.4  &15.46 &  0.56$\pm$.06 & 23.77$\pm$.25 &  277 &  433 &  8.54 \\
  5 &  10 52 43.4 & +36 40 34.7 &  1.3  &16.01 &  0.35$\pm$.04 & 23.47$\pm$.20 &  314 &  383 &  8.54 \\
  6 &  10 52 43.6 & +36 40 23.8 &  1.3  &15.55 &  0.15$\pm$.02 &  23.81$\pm$.23 &  291 &   63 &  8.79 \\
 12 &  10 52 42.1 & +36 39 51.3 &  1.1  &13.26 &  0.11$\pm$.01 & $>$24.20$\pm$.36 &  269 &  514 &  8.50 \\
 14 &  10 52 40.5 & +36 39 52.2 &  1.0  &12.56 &  0.78$\pm$.09 & 22.36$\pm$.10 &  111 &  297 &  8.80 \\
 17 &  10 52 38.1 & +36 38 46.9 &  0.7  & 8.21 &  0.36$\pm$.04 & 23.16$\pm$.16 &  288 &  695 &  8.30 \\

   \hline
\end{tabular*}
\caption[HII Region Properties: NGC 3432]{HII Region Properties: NGC 3432. Same as Table \ref{tab:prop2146}, for NGC 3432}
\end{table*}

\item{{\bf{NGC 3718}} is a galaxy located in the outer regions of the Ursa Major Cluster with a two-sided extreme HI warp \citep{verheijen01}.  It has
a total mass of $10^{10}$ \Msun~ in neutral hydrogen and a FWHM HI linewidth of 465 km s$^{-1}$.  The HII regions studied here are generally not located in the highest column density gas along the warp, although that is where much of the new star formation appears to be concentrated. All 9 HII regions presented here have abundances consistent with the mean of 12 + Log (O/H) $\sim$ 8.6. }


\begin{table*}[htp] \centering \scriptsize
\begin{tabular*}{1.0\textwidth}{@{\extracolsep{\fill}}lcccrccccc}
\hline
\footnotesize
ID & RA  & dec & R/R$_{25}$ & R$_{proj}$ & F$_{H\alpha}$ & m$_{R}$ & [OII] & [OIII] & 12 $+$ Log(O/H) \\
\hline
(1) & (2) & (3) & (4) & (5) & (6) & (7) & (8) & (9) & (10) \\
\hline
\hline
  4 &  11 32 48.3  & +53 06 55.5 &  1.7  &17.16 &  0.17$\pm$.02 & $>$23.60$\pm$.36 &  393 &  239 &  8.55 \\
  6 &  11 32 48.3 & +53 05 56.8 &  1.4  &13.46 &  0.31$\pm$.03 & $>$23.60$\pm$.36 &  327 &   77 &  8.74 \\
 15 &  11 32 51.5 & +53 02 06.3 &  1.4  &15.65 &  2.50$\pm$.30 & 21.58$\pm$.11 &  394 &  162 &  8.60 \\
 12 &  11 32 45.7 & +53 02 33.7 &  1.0  &10.96 &  1.16$\pm$.13 & 22.72$\pm$.25 &  378 &  303 &  8.53 \\
 11 &  11 32 47.2 & +53 03 22.4 &  1.0  & 9.72 &  0.55$\pm$.06 & $>$23.60$\pm$.36 &  506 &  114 &  8.51 \\
  8 &  11 32 45.2 & +53 04 07.1 &  0.8  & 7.58 &  0.28$\pm$.03 & 23.01$\pm$.32 &  529 &  267 &  8.40 \\
  7 &  11 32 38.9 & +53 05 43.0 &  0.8  & 8.56 &  0.29$\pm$.03 & $>$23.60$\pm$.36 &  322 &   71 &  8.75 \\
 19 &  11 32 41.9 & +53 02 54.6 &  0.7  & 7.73 &  1.81$\pm$.21 & 21.54$\pm$.11 &  266 &   74 &  8.81 \\
 20 &  11 32 40.0 & +53 03 12.0 &  0.5  & 5.71 &  0.96$\pm$.11 & $>$23.60$\pm$.36 &  291 &  145 &  8.73 \\

   \hline
\end{tabular*}
\caption[HII Region Properties: NGC3718]{HII Region Properties: NGC 3718. Same as Table \ref{tab:prop2146}, for NGC 3718}
\end{table*}

\item{{\bf{NGC 3893}} is classified as an M51-type minor merger and is located in the Ursa Major Cluster \citep{verheijen01}.  The interacting companion (NGC3896) and NGC3893 appear to be embedded in a common envelope of HI.  The HII regions studied here are embedded in high column density gas in the inner part of the large envelope and appear to trace an extended, low-surface brightness spiral arm.   The HI mass of this system is $6 \times 10^9$ \Msun, about half of which is located within R$_{25}$. All 8 of the HII regions we study in NGC 3893 have similar oxygen abundances, near the mean of $\sim$8.75. }


\begin{table*}[htp]\centering \scriptsize
\begin{tabular*}{1.0\textwidth}{@{\extracolsep{\fill}}lcccrccccc}
\hline
\footnotesize
ID & RA  & dec & R/R$_{25}$ & R$_{proj}$ & F$_{H\alpha}$ & m$_{R}$ & [OII] & [OIII] & 12 $+$ Log(O/H) \\
\hline
(1) & (2) & (3) & (4) & (5) & (6) & (7) & (8) & (9) & (10) \\
\hline
\hline
  7 &  11 48 41.2  & +48 45 48.0 &  1.4  &16.61 &  0.93$\pm$.11 & 22.73$\pm$.16 &  135 &  521 &  8.64 \\
 10 &  11 48 46.3 & +48 44 38.2 &  1.2  &12.30 &  1.04$\pm$.12 & 22.65$\pm$.15 &  209 &  110 &  8.85 \\
 13 &  11 48 48.1 & +48 44 04.8 &  1.2  &11.05 &  0.50$\pm$.06 & 23.30$\pm$.24 &  315 &  142 &  8.71 \\
 15 &  11 48 48.5 & +48 43 44.7 &  1.1  &10.31 &  1.13$\pm$.13 & 22.72$\pm$.15 &  301 &   77 &  8.77 \\
 19 &  11 48 49.8 & +48 42 58.5 &  1.1  & 9.83 &  0.51$\pm$.06 & 23.68$\pm$.32 &  361 &  168 &  8.64 \\
 21 &  11 48 41.5 & +48 43 50.6 &  0.6  & 6.62 & 17.42$\pm$.09 & 19.79$\pm$.07 &  250 &  169 &  8.75 \\
 22 &  11 48 43.0 & +48 43 12.4 &  0.5  & 4.73 & 11.47$\pm$.37 & 19.95$\pm$.07 &  260 &  214 &  8.71 \\
 25 &  11 48 41.9 & +48 42 58.5 &  0.4  & 3.24 &  8.07$\pm$.96 & 20.10$\pm$.07 &  234 &   64 &  8.86 \\

    \hline
\end{tabular*}
\caption[HII Region Properties: NGC3893]{HII Region Properties: NGC 3893. Same as Table \ref{tab:prop2146}, for NGC 3893}
\end{table*}

\item{{\bf{NGC 5774/5}} is classified in the HI rogues catalog as an interacting double with no tails and an HI bridge joining the two gas-rich galaxies. 
HII regions in both galaxies and in the HI bridge were targeted in this study.  The NGC 5774/5 system belongs to a small group of at least 6 galaxies and the gas in the bridge between the galaxies is thought to be transferring gas from NGC 5774 to NGC 5775 \citep{irwin94}.  The HI masses of the galaxies are 5.4 and 9.1 $\times 10^9$ \Msun~ respectively. The oxygen abundances of the HII regions in this system are all close to an average value of 12 + Log (O/H) $\sim$ 8.7. }


\begin{table*}[htp] \centering \scriptsize
\begin{tabular*}{1.0\textwidth}{@{\extracolsep{\fill}}lcccrccccc}
\hline
\footnotesize
ID & RA  & dec & R/R$_{25}$ & R$_{proj}$ & F$_{H\alpha}$ & m$_{R}$ & [OII] & [OIII] & 12 $+$ Log(O/H) \\
\hline
(1) & (2) & (3) & (4) & (5) & (6) & (7) & (8) & (9) & (10) \\
\hline
\hline
&&&&&NGC 5774&\\
\hline
7 &  14 53 48.6  & +03 34 14.1 &  1.1  &13.11 &  0.04$\pm$.00 & 20.20$\pm$.07 &  320 &  300 &  8.59 \\
11 &  14 53 43.4 & +03 36 09.6 &  1.0  & 9.66 &  2.11$\pm$.25 & 22.10$\pm$.15 &  469 &  262 &  8.46 \\
12 &  14 53 44.6 & +03 35 58.0 &  1.0  & 8.94 &  5.74$\pm$.68 & 19.85$\pm$.07 &  189 &  616 &  8.53 \\
23 &  14 53 41.1 & +03 35 43.7 &  0.6  & 6.65 & 10.10$\pm$.21 & 20.06$\pm$.07 &  343 &  276 &  8.58 \\
10 &  14 53 43.1 & +03 35 16.0 &  0.3  & 2.78 &  1.91$\pm$.22 & 21.92$\pm$.13 &  373 &  293 &  8.54 \\
9 &  14 53 41.6 & +03 35 03.2 &  0.1  & 1.79 &  2.30$\pm$.27 & 20.52$\pm$.07 &  361 &  126 &  8.67 \\
 \hline
&&&&& NGC 5775&\\
 \hline
    4 &  14 53 51.0 &  +03 35 55.9 &  1.5  &27.20 &  0.40$\pm$.04 & $>$23.19$\pm$.34 &  439 &  230 &  8.51 \\
  1 &  14 53 51.9  &+03 35 27.0 &  1.3  &23.05 &  0.54$\pm$.06 & 21.67$\pm$.11 &  396 &  263 &  8.54 \\
 15 &  14 53 54.8 & +03 33 50.7 &  0.5  & 9.29 &  4.22$\pm$.50 & 22.42$\pm$.19 &  320 &  154 &  8.69 \\
 17 &  14 53 54.5 & +03 33 34.4 &  0.4  & 7.98 &  4.80$\pm$.57 & 22.26$\pm$.17 &  270 &   87 &  8.80 \\

    \hline
\end{tabular*}
\caption[HII Region Properties: NGC5774/5]{HII Region Properties: NGC 5774/5. Same as Table \ref{tab:prop2146}, for NGC 5774 and NGC 5775}
\end{table*}

\item{{\bf{NGC 6239}} is  a minor merger in the HI rogues catalog.  Both galaxies appear to be gas rich, although we imaged only the primary galaxy in H$\alpha$.  The one outlying HII region in this galaxy is in the northern HI tail with an HI column density of $5  \times 10^{19}$ \cm.  The HI mass of this galaxy is $7 \times 10^9$ \Msun~ \citep{springob05}.  The radial oxygen abundance gradient is consistent with being flat over $\sim$ 15 kpc. }



\begin{table*}[ht] \centering \scriptsize
\begin{tabular*}{1.0\textwidth}{@{\extracolsep{\fill}}lcccrccccc}
\hline
\footnotesize
ID & RA  & dec & R/R$_{25}$ & R$_{proj}$ & F$_{H\alpha}$ & m$_{R}$ & [OII] & [OIII] & 12 $+$ Log(O/H) \\
\hline
(1) & (2) & (3) & (4) & (5) & (6) & (7) & (8) & (9) & (10) \\
\hline
\hline
  5 &  16 49 53.6 & +42 45 28.2 &  1.9  &16.42 &  0.46$\pm$.05 & $>$23.15$\pm$.35 &  480 &  269 &  8.42 \\
  8 &  16 50 00.9 & +42 44 30.6 &  0.7  & 6.07 &  0.53$\pm$.06 & $>$23.15$\pm$.35 &  256 &  150 &  8.76 \\
  9 &  16 50 01.3 & +42 44 23.3 &  0.7  & 5.42 &  2.05$\pm$.24 & 21.80$\pm$.13 &  382 &  313 &  8.52 \\
 11 &  16 50 06.6 & +42 44 40.1 &  0.7  & 2.45 &  3.29$\pm$.39 & 21.01$\pm$.08 &  261 &  170 &  8.74 \\
 17 &  16 50 03.5 & +42 44 28.7 &  0.3  & 3.04 & 73.58$\pm$.82 & 18.36$\pm$.07 &  122 &  698 &  8.55 \\
 16 &  16 50 05.5 & +42 44 10.0 &  0.3  & 1.06 &  2.21$\pm$.26 & 21.62$\pm$.11 &  275 &  321 &  8.62 \\

    \hline
\end{tabular*}
\caption[HII Region Properties: NGC6239]{HII Region Properties: NGC 6239. Same as Table \ref{tab:prop2146}, for NGC 6239}
\end{table*}

\begin{table*}[ht] \centering \scriptsize
\begin{tabular*}{1.0\textwidth}{@{\extracolsep{\fill}}lcccrccccc}
\hline
\footnotesize
ID & RA  & dec & R/R$_{25}$ & R$_{proj}$ & F$_{H\alpha}$ & m$_{R}$ & [OII] & [OIII] & 12 $+$ Log(O/H) \\
\hline
(1) & (2) & (3) & (4) & (5) & (6) & (7) & (8) & (9) & (10) \\
\hline
\hline
   4 &  09 51 10.3 & +07 49 17.4 &  2.3  & 3.07 &  0.19$\pm$.02 & $>$24.57$\pm$.34 &  323 &  336 &  8.08 \\
  1 &  09 51 15.2 & +07 50 35.9 &  1.4  & 1.93 &  0.22$\pm$.02 & $>$24.57$\pm$.34 &  286 &  403 &  8.08 \\
 12 &  09 51 15.2 & +07 49 42.0 &  0.7  & 0.88 &  0.95$\pm$.11 & 23.41$\pm$.19 &  441 &  198 &  8.18 \\
 11 &  09 51 15.9 & +07 49 46.0 &  0.5  & 0.64 &  1.41$\pm$.16 & 21.93$\pm$.08 &  336 &   54 &  8.01 \\
 10 &  09 51 16.6 & +07 49 50.7 &  0.4  & 0.48 & 25.57$\pm$.06 & 19.31$\pm$.07 &  270 &  461 &  8.09 \\

    \hline
\end{tabular*}
\caption[HII Region Properties: UGC5288]{HII Region Properties: UGC 5288. Same as Table \ref{tab:prop2146}, for UGC 5288}
\end{table*}

\begin{table*}[ht] \centering \scriptsize
\begin{tabular*}{1.0\textwidth}{@{\extracolsep{\fill}}lcccrccccc}
\hline
\footnotesize
ID & RA  & dec & R/R$_{25}$ & R$_{proj}$ & F$_{H\alpha}$ & m$_{R}$ & [OII] & [OIII] & 12 $+$ Log(O/H) \\
\hline
(1) & (2) & (3) & (4) & (5) & (6) & (7) & (8) & (9) & (10) \\
\hline
\hline
 7 &  14 51 12.7  & +35 31 58.9 &  1.2  & 4.55 &  0.74$\pm$.08 & 22.97$\pm$.31 &  297 &  529 &  8.19 \\
 12 &  14 51 13.5 & +35 31 58.2 &  1.1  & 4.12 &  0.44$\pm$.05 & $>$23.17$\pm$.36 &  448 &  278 &  8.24 \\
 17 &  14 51 15.2 & +35 32 58.0 &  0.8  & 3.18 &  1.24$\pm$.14 & 21.86$\pm$.13 &  396 &  200 &  8.11 \\
 14 &  14 51 13.9 & +35 32 17.5 &  0.5  & 1.80 &  9.13$\pm$.09 & 19.99$\pm$.07 &  572 &  276 &  8.35 \\

     \hline
\end{tabular*}
\caption[HII Region Properties: UGC9562]{HII Region Properties: UGC 9562. Same as Table \ref{tab:prop2146}, for UGC 9562. \label{tab:prop9562}}
\end{table*}

\item{{\bf{UGC 5288}} is a blue compact dwarf galaxy with an extended HI envelope and no sign of interaction in the gas disk.  Its optical and HI properties are qualitatively similar to those of NGC 2915 (see Werk et al. 2010a\nocite{werk10b}). The HII regions probed in the outer gaseous regions of this galaxy are found in gas with column densities as low as 2 $\times$ $10^{20}$ \cm.  The total HI mass of this galaxy is $2 \times 10^8$ \Msun~ \citep{wong06}, only 10\% of which is within R$_{25}$. The radial oxygen abundance gradient for UGC 5288 is flat.}

\item{{\bf{UGC 9562}}, (also known as II Zw 71) is classified as an interacting double (the companion is UGC 9560/II Zw 70), or a polar ring galaxy.
Both galaxies are gas rich, with a bridge joining the two galaxies and a tail on the opposite side of the bridge linked to UGC 9562 \citep{cox01}.   All of the HII regions we study here are close to UGC 9562 which contains $8.2 \times 10^8$ \Msun~ in HI. The oxygen abundance gradient in UGC 9562 is flat, with an average oxygen abundance near the turnover of R23.  }
\end{itemize}

Excluding the two ambiguous cases (NGC 2146 and NGC 3239), our derived abundance gradients range from $+$0.25 dex/R$_{25}$ to $-$0.25 dex/R$_{25}$. We note that our data generally do not apply to nuclear oxygen abundance gradients between 0 and $\sim$0.4 R/R$_{25}$, which may very well be declining at different rates. Furthermore, we derived these gradients using projected galactocentric distances. Given uniformly shallow, if not flat, radial abundance gradients, correcting for inclination angle would have little impact on the overall results. The error-weighted mean of the radial abundance gradients (excluding the two ambiguous cases), normalized to R/R$_{25}$, is $-$0.02 $\pm$ 0.18. These galaxies have radial oxygen abundance gradients  (beyond 0.4 R/R$_{25}$) consistent with being flat within 1$\sigma$ over an average of $\sim$15 kpc, where typical $\sigma_{slope}$ = 0.15 dex. Additionally, NGC 3239 is consistent with these results, regardless of branch choice. If the outlying HII regions of NGC 2146 lie on the upper branch as they do in the vast majority of our sample of galaxies, then NGC 2146, too, is consistent with having a flat abundance gradient. Alternatively, NGC 2146 may be the only galaxy in our sample that has a significantly declining radial abundance gradient of $-$0.79 dex/R$_{25}$, corresponding to $-$0.035 dex/kpc. However, this gradient is still somewhat shallow compared to those of typical spiral galaxies, $-$0.07 dex/kpc (Zaritsky et al. 1994; see Section 6.1).

\subsection{Gas Densities in the Outer Disks: Low-Density Star Formation}
\label{sec:rogues:schmidt}
The HI gas densities at the locations of the outlying HII regions range from 3 $-$ 500 $\times$ 10$^{19}$ HI atoms \cm, corresponding to a range in gas surface density of 0.25 $-$ 80 M$_{\odot}$ pc$^{-2}$. The median value, which is typical of most of the gas column densities at the outlying HII region locations, is 4 $\times$ 10$^{20}$ \cm, or $\sim3$ M$_{\odot}$ pc$^{-2}$. This gas surface density is at the traditional ``threshold" level for star formation, usually cited between 2 and 8 M$_{\odot}$ pc$^{-2}$ \citep{schaye04}. Thus, we are truly examining star formation at low average gas surface densities, in the same outer-disk regime explored by \cite{bigiel10}. The outlying HII regions presented here cover a range of H$\alpha$ luminosities, from 10$^{35.5}$ $-$ 10$^{37.5}$ergs s$^{-1}$, corresponding to SFRs that range from 2 $\times$ 10$^{-4}$ to 2$\times$ 10$^{-6}$ M$_{\odot}$ yr$^{-1}$. Most of the outlying HII regions are isolated within surface areas of roughly a kpc$^2$, so these SFRs can be viewed roughly as star formation rate surface densities in units of M$_{\odot}$ yr$^{-1}$ kpc$^{-2}$. These values of SFRs and HI gas surface densities place the outlying HII regions in the same, or lower, ``downturn" portion of the Kennicutt-Schmidt law \citep{wyder09}, as expected. One caveat to keep in mind is that these stars may be forming in places where the local gas density is higher than the average surrounding gas densities, on a size scale which would be unresolved in the HI synthesis maps. We will present a full analysis of the relationship between the atomic hydrogen content and star formation activity in the gaseous outskirts of HI Rogues in a future paper.

\section{Discussion: The Lack of Significant Abundance Gradients at Large Galactocentric Radii}
\label{sec:rogues:disc}
	
	The 13 HI rogue galaxies in our sample are diverse, in terms of both their HI and optical properties. Their total baryonic masses range from dwarf galaxies with M $<$ 10$^{9}$ \Msun~(UGC 5288) to massive spiral galaxies with M $\sim$ 10$^{11}$ \Msun (NGC 2146). Their HI morphologies range from very extended, undisturbed HI envelopes, to warped HI distributions, to large tidal tails, representing a wide range of galaxy interactions;  and, their SFRs range from 0.006 \Msun~yr$^{-1}$ to 3 \Msun~yr$^{-1}$. Our results show that the outer gas of disturbed, interacting, or extended gas disk galaxies is oxygen-enriched.  Predominantly flat radial oxygen abundance gradients in this wide variety of galaxies with varying degrees of interaction and star formation provide compelling evidence of uniform metal distribution across extended gaseous features independent of large-scale galaxy properties and individual accretion histories.    In this section, we present an analysis of the oxygen yields of all the rogue galaxies, their positions on the mass-metallicity relation, and discuss the implications of flat oxygen abundance gradients out to large galactocentric radii. 

	 
\subsection{Comparison with Previous Work}

First, we discuss previous work on the subject of radial abundance distributions and attempt to reconcile it with the flat gradients observed here. Declining radial abundance gradients, often out to, and/or slightly beyond R$_{25}$, are commonly observed in spiral galaxies \citep{oey93, zaritsky94, kennicutt03a}.  The broadest study to date is that of \cite{zaritsky94}. They present metallicity gradients between 0.1 and 1.0 R/R$_{25}$ using at least 5 HII regions in each of 39 local spiral galaxies with a wide range of morphologies (Sab - Sm; 7 galaxies have bars) and luminosities ( $-21 <$ M$_{B}$ $< -$17).  Because the scatter in oxygen abundance at a given galactic radius (0.4 dex at most)  for each galaxy is much smaller the the mean abundance range for their entire sample (8.34 - 9.31), they conclude that these gradients are determined globally, rather than set by local conditions (i.e. the local SFR and gas fraction).  \cite{zaritsky94} find a wide range of slopes in the 39 radial oxygen distributions, from flat to very steeply declining, at $-$1.45 dex/R$_{25}$ corresponding to $-$0.231 dex/kpc. Their average slope is $-$0.59 dex/R$_{25}$ or $-$0.07 dex/kpc. Therefore, we do not see declining abundance gradients similar to those of \cite{zaritsky94} where our average gradient is  -0.02 $\pm$ 0.18 dex/R$_{25}$, or $-$0.001 dex/kpc. Furthermore, our data rule out decreasing radial oxygen abundance gradients steeper than $-$0.30 dex over 15 kpc ($-$0.02 dex kpc$^{-1}$) in these outer regions.

Several studies of metallicity gradients in the outer parts of relatively quiescent spiral galaxies have found a flattening beyond 0.5 R/R$_{25}$ \citep{martin95,bresolin09a}. Similarly, studies of low surface brightness disk galaxies \citep{deblok98} and dwarf galaxies \citep{croxall09} are suggestive of flat oxygen abundance gradients out to R$_{25}$. And, the metallicity gradients of interacting pairs appear to be consistently flatter than those of spiral galaxies \citep{kewley10}. Yet, this study is the first to measure radial oxygen abundance gradients in the outer regions of a modest sample of galaxies beyond R$_{25}$.  The majority of our galaxies are interacting, which may partially explain a flattening in the abundance gradients \citep{rupke10, kewley10}. Given the wide range of systems (including minor mergers, and warps) over which we measure flat gradients, our results suggest that galaxy interactions are not the sole explanation.

\subsection{Relationship to the Mass-Metallicity Relation}

There is a fundamental, global relation between galaxy mass and metallicity in the sense that lower mass galaxies have lower overall heavy-element content. \cite{larson74} first predicted that the average stellar metal abundance in a galaxy would depend on its mass owing to more significant gas loss from the energy supplied by supernovae in lower-mass galaxies. His notion turned out to fit well with observations of nebular oxygen abundances of irregular galaxies several years later by \cite{lequeux79}, and has become known as the mass-metallicity relation for galaxies in the 36 years since \citep{skillman89,tremonti04}. This relation is often parameterized in terms of effective yields and and total (gas $+$ stellar) mass to elucidate its origin, where effective yield measures how much a galaxy's metallicity deviates from its ``closed-box" value, given a certain gas mass fraction (Dalcanton 2007 provides a more detailed, clear explanation).  Systematically low effective yields of dwarf galaxies have been taken as evidence that low mass galaxies have evolved less as ``closed-boxes" than their more massive counterparts  \citep{tremonti04}. 

Although supernova-driven metal expulsion is the most popular and classic explanation for the low effective yields of dwarf galaxies \citep{larson74, tremonti04}, there are several issues that complicate the picture. \cite{dalcanton07} shows, for instance, that metal loss itself does not have a strong dependence on galaxy mass, and that any subsequent star formation following an episode of mass loss quickly pushes the effective yield up to its closed-box value. In the calculations presented by \cite{dalcanton07}, low star formation efficiencies in gas-rich dwarfs between episodes of mass-loss are what keep their effective yields low. Dilution by infalling low-metallicity gas may also play some role, but alone cannot explain the mass-metallicity relation \citep{dalcanton07}. An explanation offered by \cite{tassis08} is that the metals are not expelled via supernova-driven winds, but rather efficiently mixed via gravitational processes into the hot gas, where they are unobservable by standard ground-based methods. Coupled with low star formation efficiencies of low-mass, gas-rich dwarfs, \cite{tassis08} can reproduce the observed mass-metallicity relation without including winds from supernova in their model. Finally, if low-mass galaxies preferentially do not make as many O-type stars as more massive galaxies \citep{meurer09, lee09}, as in the IGIMF theory, then their effective yields will appear to be low \citep{koppen07}.

We calculate the effective oxygen yield for each galaxy, Z/ln($\mu^{-1}$), where $\mu$ is the gas fraction and Z is the metallicity by mass, using the mean HII region oxygen abundance as a proxy for total metal content, the stellar masses listed in Table \ref{tab:c5t1}, and total gas masses computed from the HI masses given in Table \ref{tab:c5t1} plus a 36\% contribution from helium. We do not include molecular gas in these calculations, although it most certainly is present. Although the molecular-to-neutral gas content varies widely within and among galaxies \citep{bigiel08,leroy08}, the total galaxy molecular gas mass is rarely, if ever, larger than its neutral gas mass.  If we assume that, on galaxy-wide scales, the molecular gas content scales to first order with the neutral gas content, then the effect on the overall mass-metallicity relation would be negligible. 

\begin{figure*}[htp] \centering
\subfigure {{\centering \rotatebox{0}{\includegraphics[width=0.68\linewidth]{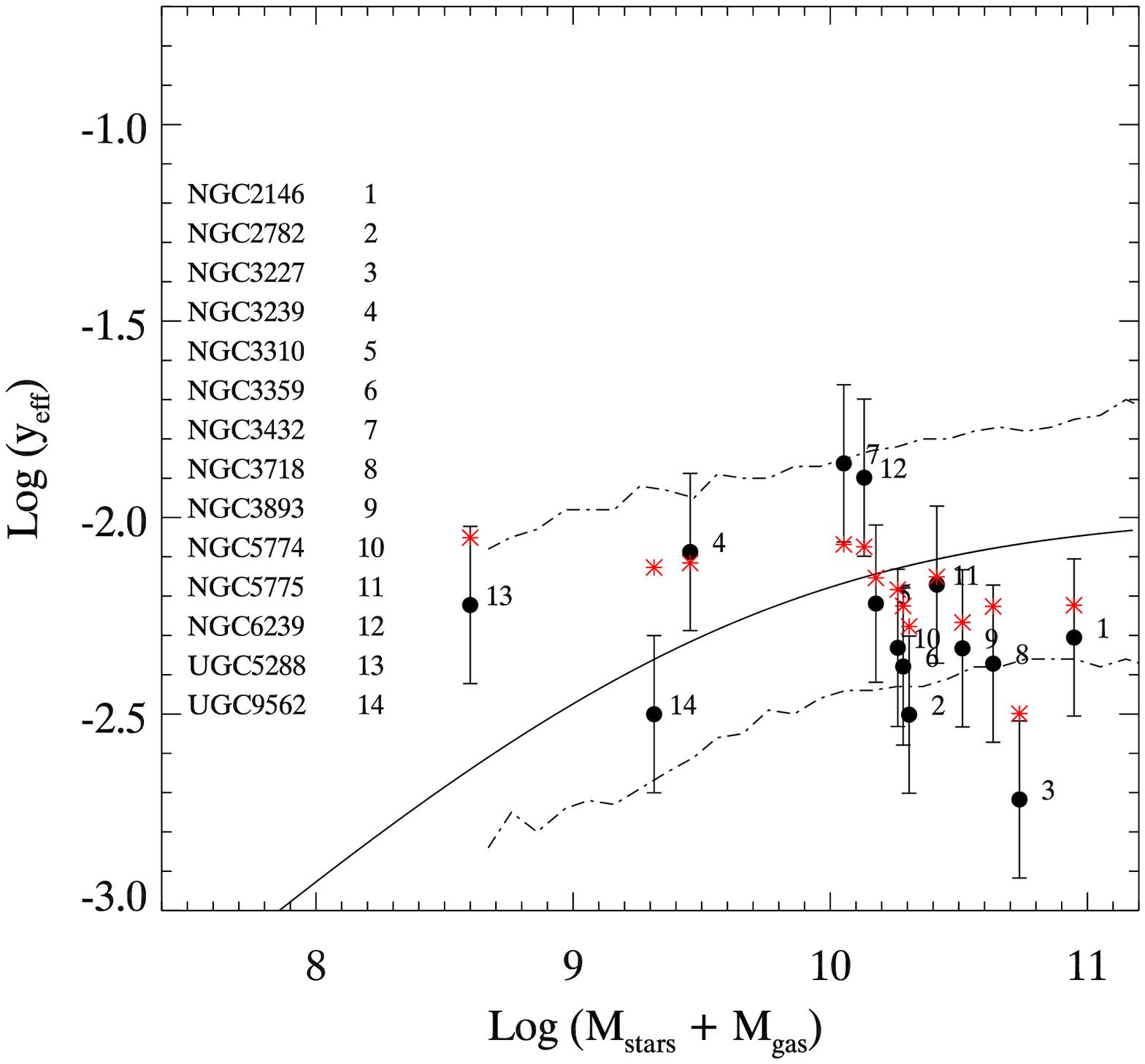}}}\label{fig:yields}}
\subfigure {{\centering \rotatebox{0}{\includegraphics[width=0.68\linewidth]{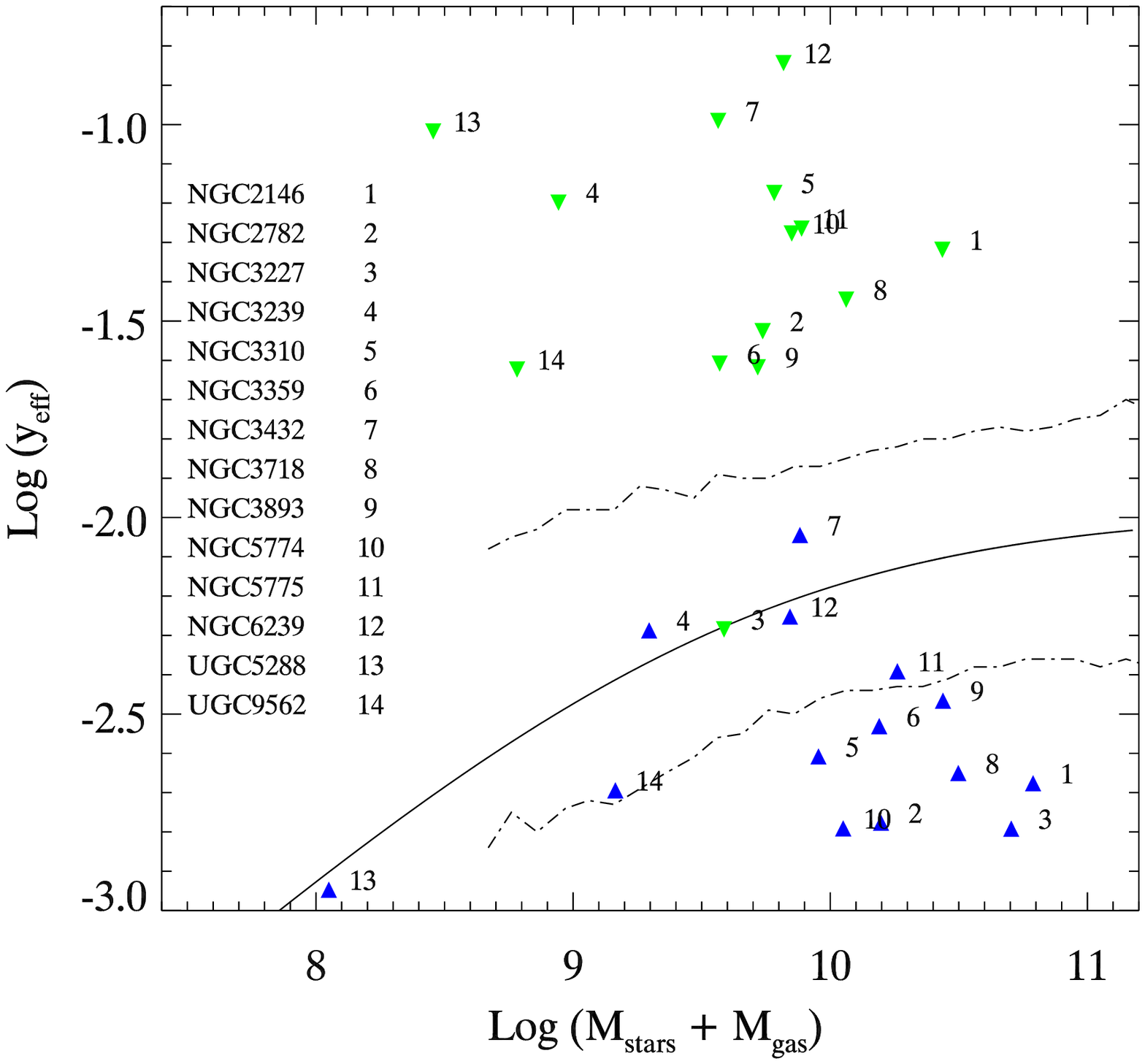}}}
\label{fig:yields_inout}
}
\caption[The effective yield versus the total baryonic (stellar plus gas) mass for our sample of galaxies, shown as filled circles.]{Top: The effective yield versus the total baryonic (stellar plus gas) mass for the HII regions in every galaxy, shown as filled circles. We plot ``expected" oxygen yields, determined from total stellar masses and a constant net oxygen yield, for each galaxy as red asterisks.  Bottom: The ``Inner" and ``outer" galaxy have been separated. The blue triangles show the inner (R $<$ R$_{25}$) galaxies, and the green triangles represent the outer (R $>$ R$_{25}$) regions. Each point is marked with a number that corresponds to a galaxy listed in the key on the left hand side of the figure. The empirical relation in \cite{tremonti04}, their equation 6, is shown for reference (solid line), along with contours (dashed-dotted lines) that enclose 95\% of the SDSS data presented in \cite{tremonti04}. }
\end{figure*}

Figure \ref{fig:yields} plots the effective oxygen yield versus total baryonic mass for our galaxies (filled circles; NGC 5774 and NGC 5775 we consider separately, though they are technically part of the same HI envelope) along with the empirical relation of \cite{tremonti04}.  To include both NGC 2146 and NGC 3239 in this figure, we have assumed upper branch metallicities for the HII regions. While this assumption may not be valid, the points in the upper panel of Figure 3 do not change appreciably with branch assumption. With the exception of NGC 3227, every galaxy in our sample appears to lie along the empirical  \cite{tremonti04} relation, within the 95\% contours (dashed-dotted lines). \cite{tremonti04} explain the decrease in metal yields with decreasing galaxy baryonic mass by invoking  metal-rich supernova-driven winds that preferentially escape from smaller potential wells of less-massive galaxies. 


The red asterisks on Figure \ref{fig:yields} directly above or below the filled-circle galaxy points mark an ``expected" oxygen yield determined by using the galaxy's total stellar mass and assuming a net oxygen yield of 0.01 (average value from Maeder 1992; \nocite{maeder92} the mass of oxygen ejected by all stars per unit mass of matter locked up in stars). All galaxies, with the exception of UGC 9562, have yields consistent with those predicted from a rough estimate that simply accounts for all metals using the existing stellar population. 

As was the case with NGC 2915, considered in \cite{werk10b}, there is only a low-level of ongoing and previous star formation at the locations of the outermost HII regions.  In this sense, the lack of radial oxygen abundance gradients beyond the  optical ``edges" of galaxy disks seems to necessarily imply some type of efficient metal transport.  In the context of the mass-metallicity relationship, we can consider the yields for the outer and inner gaseous regions separately. We show this separation in Figure \ref{fig:yields_inout} with the assumption that 95\% of the stellar mass is contained within R$_{25}$, and calculating the HI mass within R$_{25}$ from the HI 21-cm synthesis data, when it was available to us. In the four cases that it was not, we estimated by eye the total HI mass within and outside of R25 from the HI column density maps available from the online the HI Rogues Gallery. The percentage of HI mass contained in R$_{25}$ varies from 10\% to 60\% for the galaxies. 

 First, keeping metallicity constant while lowering the stellar mass contribution to the effective yields relative to the gas mass contribution (outer HII regions) shifts the black points in Figure \ref{fig:yields} to the left and up, shown as downward-facing green triangles in Figure \ref{fig:yields_inout}. Note that the green triangles for NGC 2146 and NGC 3239 move down by 0.7 dex and 0.4 dex, respectively,  if we instead assume lower-branch metallciities for their outer HII regions.  This figure shows that the effective yields for the outer disk lie above their ``closed-box" values. Conversely, the general trend of lowering the gas mass contribution relative to the stellar mass contribution while keeping metallicity constant (inner HII regions) has the effect of moving the points to the left and down, shown as blue triangles in Figure \ref{fig:yields_inout}. For the most part, the effective yields for the inner disk fall below their ``closed box" values. Including molecular gas in this analysis would only make the effect more striking, since, in general,  the fraction of molecular gas decreases with galactocentric radius \citep{leroy08}. Outer-galaxy oxygen abundances are generally more metal-enriched than their gas fractions otherwise suggest, while inner-galaxy oxygen abundances lie below what is expected based on these simple stellar-population based assumptions. This duality appears to underlie the flat radial abundance gradients of gas-rich, interacting galaxies and apparently isolated extended gaseous disk dwarf galaxies (UGC 5288 and NGC 2915; Werk et al. 2010b).

\subsection{Methods of Metal Transport}
\label{sec:rogues:flat}

As with the flat radial oxygen abundance gradient of NGC 2915, a number of physical mechanisms could be at work in various combinations to distribute metals in these HI Rogue galaxies. Very generally, our results indicate that there is efficient metal mixing out to large galactocentric radii, facilitated by galaxy interactions. Nonetheless, there are a few galaxies in our sample that show no evidence of a recent merger or interaction, yet are still well-mixed out to large radii. The rest of our sample represents a myriad of interaction types.  If indeed interactions are the primary driver of the metal mixing seen here, then the mechanism by which it acts is independent of the interaction details. 


 The accretion of  gas following a merger event or tidal interaction results in tidal torques that drive the gas to the system center \citep{barnes96} and subsequently mix metals into the outer regions \citep{rupke10}.  Other than tidal torques, metal mixing can be driven by a number of processes: magnetorotational instabilities \citep{sellwood99}, infall of gas clouds \citep{santillan07}, self-gravity in conjunction with differential rotation \citep{wada99}, quasi-stationary spiral structure interacting with a central bar \citep{minchev09}, viscous flows generated by gravitational instability or  cloud-cloud collisions \citep{ferguson01}, thermal instability triggered self-gravitational angular momentum transport \citep{mcnally09}.  In general, the sound speed, $\sim$10 km/s, is a robust upper limit to the speed at which mixing occurs in cold, neutral gas. Measurements of velocity dispersions in the outer HI disks of galaxies are roughly consistent with this number, generally no less than 8 km/s \citep{tamburro09}.  In some strongly interacting galaxies, the velocity dispersions may be higher due to these gravitational perturbations, up to 30 km/s in some rare cases (e.g. NGC 1533; Ryan-Weber et al. 2003\nocite{emmaconf}). Furthermore, in the extended HI disk of NGC 2915, \cite{elson10b} find evidence for radial flows of 5 $-$ 17 km s$^{-1}$ that may be driving central material outward into the extended disk. They note that these flows are consistent with the finding of higher-than-expected oxygen abundances in the outer gaseous disk, as observed by \citep{werk10b}. 

A simple calculation of mixing timescales, assuming there is mixing at the sound speed over 15 kpc, yields 1.5 Gyr as the lower limit to the timescale over which mixing occurs in our extended HI features. Yet, the metals generated by massive stars are generally returned to the ISM on much shorter timescales, on the order of 100 Myr \citep{tenoriotagle96}.  Even if the velocity dispersion in the cold gas is three times higher (30 km/s; the most extreme end) due to interactions, the lower limit to the cold-gas mixing timescale is 500 Myr.  By assuming a net oxygen yield of 0.01 (Maeder 1992; \nocite{maeder92} the mass of oxygen ejected by all stars per unit mass of matter locked up in stars), we can calculate this required average outward metal ``momentum" that would reproduce a flat oxygen distribution.  Taking an average inner-R$_{25}$ SFR of 0.5 M$_{\odot}$ yr$^{-1}$, and an average separation between inner and outer SF of 15 kpc,  we find that there needs to be a constant outward movement of  0.0025 M$_{\odot}$ of oxygen at 150 km/s. We can repeat this calculation for the most compact galaxies in our sample (NGC 3239, UGC 5288, and UGC 9562), now taking  an average inner-R$_{25}$ SFR of 0.15 M$_{\odot}$ yr$^{-1}$, and an average separation between inner and outer SF of 4 kpc. The corresponding ``metal momentum" needed to wipe out any abundance gradient is 0.0007 M$_{\odot}$ of oxygen at 40 km/s.  While metal mixing in cold neutral gas probably happens, it most likely does not happen on fast enough timescales to wipe out a metallicity gradient that would arise in a typical star-forming spiral galaxy from a radial dependence of the star formation rate (i.e. star formation is concentrated in the central regions, therefore metals are concentrated in central regions). With the advent of high-resolution radio telescopes such as the EVLA and ALMA, future work will be able to address the kinematics and motions of cold gas in considerably more detail. 

Given the fast mixing speeds required, we consider that the metal transport may be occurring predominantly in a hot gas component as proposed by \cite{tassis08} and others.  Although the dominant driver of the mixing of metals in the hot-phase gas is unclear, suggested sources include diffusion, gravitational instabilities, cold accretion flows,  and galaxy interactions \citep{keres05, tassis08}. The recent finding of OVI absorption over the entire Milky-Way disk, irrespective of circumstellar environments and spiral arms \citep{bowen08} may provide some support for the idea that there can be efficient mixing in the hot gas phase of the ISM. Furthermore, the hot X-ray gas generated by metal-enriched supernova-driven blowouts (i.e. winds) is able to propagate to the outer 100 kpc of a galaxy's halo \citep{strickland04}. Finally, the three-dimensional adaptive mesh refinement code of \cite{deavillez02} finds that inhomogeneities in the ISM can be erased over a couple hundred Myrs (dependent on supernova rate) when gas cycles between the disk and the hot halo in a galactic fountain process. Whether or not enough metals can be transported in the hot gas and quickly cooled at large enough radii to reproduce the flat oxygen distribution seen in these 13 galaxies is unclear. However, our data generally support a scenario in which galaxy interactions induce efficient metal mixing in hot gas over the full scale of the galaxy. 


Efficient metal mixing has important implications for the origin of the mass-metallicity relationship. Large-scale and universal metal-outflow may not be the dominant cause of the low effective yields in some {\emph{gas-rich}} dwarf galaxies as advanced by \cite{tremonti04} and others. Instead, these low-mass gas-rich galaxies on the whole would have fewer metals than expected from the simple nucleosynthetic yield because they are stirring those centrally-generated metals throughout their gaseous outskirts. Coupled with the idea that star formation is increasingly inefficient in these low-mass, gas-rich galaxies \citep{dalcanton07}, mixing could account for lower effective yields in lower mass galaxies with no real need for large-scale metal blow out. 

\section{Summary and Conclusions}
\label{sec:rogues:sum}

Using narrow-band H$\alpha$ and continuum R$-$band images from the MDM 2.4-m telescope, we have identified 13 ``HI rogues" with massive star formation occurring beyond their optical radii. GMOS multi-slit optical spectroscopy targeting both inner and outlying HII regions in these 13 rogue galaxies allowed for the measurement of strong-line oxygen abundances of $\sim$100 HII regions with projected galactocentric distances  ranging from 0.3 $-$ 2.5 R/R$_{25}$. Despite the diversity of these 13 systems in terms of their optical and HI morphologies, star-forming properties, and level of disruption, we find that all of them have flat radial oxygen abundance gradients from their central optical bodies to their outermost regions. In addition to the interacting systems with flat abundance gradients, we find several non-interacting, gas-rich galaxies that have flat oxygen abundance gradients out to large projected radii. 

 We find that efficient metal-mixing or pre-enrichment and subsequent mixing from recent interactions are the two most likely potential causes for the enrichment of the gas beyond R$_{25}$.  Large scale metal blowout then may not be the primary driver of the mass-metallicity relation for gas-rich or interacting galaxies. That metals are distributed throughout the extended gaseous regions of both interacting and non-interacting, strongly star-forming and quiescent, massive and dwarf galaxies has important implications for the origin of the mass-metallicity relation, and galaxy chemical evolution in general. 


 \section{Acknowledgements}
   
   JKW would like to thank members of her dissertation committee at the University of Michigan for valuable insights and suggestions that improved this final manucript: Fred Adams, Lee Hartmann, Mario Mateo, and Sally Oey. Discussions with Eric Pelligrini and the entire University of Michigan FANG (Feedback Activity in Nearby Galaxies) research group were also very helpful for understanding the results of this work. JKW gratefully acknowledges the generosity of the COS-halos group, specifically Xavier Prochaska, for permitting the Keck LRIS observations of several HI Rogues that were instrumental in breaking the R23 degeneracy.  Observations at Lick went considerably more smoothly because of the aid of Michele Fumagalli and Robert da Silva $-$ many thanks to both! The following individuals kindly provided their reduced HI data cubes for the interpretation of our results:  Judith Irwin (NGC 5774/5), Carole Mundell (NGC 3227), Beverly Smith (NGC 2782), David Hogg and Mort Roberts (NGC 6239), Caroline Simpson (NGC 3239), Rob Swaters (NGC 3432), Marcel Clemens (NGC 3395; not used), and Marc Verheijen (NGC 3718 and NGC 3893). Finally, JKW wishes to thank the very helpful support staff, both at MDM and Gemini Observatories. MEP and JKW acknowledge support for this work through NSF CAREER AST-0904059, the Research Corporation, and support from the Luce Foundation.  GRM was partially supported by NASA LTSA grant NAG5-13083 for the work presented here.


    {{\it Facilities:} \facility{Gemini: GMOS-N}, \facility{MDM},\facility{Keck: LRIS}, \facility{VLA}, \facility{WSRT}}
\bibliography{thesisrefs}
\bibliographystyle{apj}

 \clearpage

\end{document}